\RequirePackage{etoolbox}
\csdef{input@path}%
{%
 {sty/},
 {img/},
}
\documentclass{elsevierbook}

\usepackage{url}
\usepackage{chngcntr}
\counterwithout{figure}{chapter}
\counterwithout{equation}{chapter}
\renewcommand{\thesection}{\arabic{section}}

\usepackage{upgreek}
\usepackage{amsfonts, bm}
\usepackage[ruled]{algorithm2e}
\usepackage{xcolor}
\usepackage[nopostdot,style=super,section=section,nonumberlist,toc,nogroupskip]{glossaries}
\DeclareMathOperator*{\argmax}{arg\,max}
\DeclareMathOperator*{\argmin}{arg\,min}

\newglossary{symbols}{sym}{sbl}{List of Abbreviations and Symbols}
\makeglossaries

\newglossaryentry{gls:forward}{name={forward function},description={a function that predicts data for any particular values of model parameters},sort=aa}
\newglossaryentry{gls:inverse}{name={inversion},description={the process that infers the value of model parameters from measurements or observations},sort=ab}
\newglossaryentry{gls:bayes}{name={Bayesian inference},description={a method that uses Bayes' theorem to infer the posterior probability distribution of model parameters given the observed data},sort=ag}
\newglossaryentry{gls:prior}{name={prior pdf},description={a probability density function of model parameters which describes information that is independent of the data},sort=ac}
\newglossaryentry{gls:likelihood}{name={likelihood function},description={a probability density function that defines the probability of observing certain data give a specific set of model parameters},sort=ad}
\newglossaryentry{gls:posterior}{name={posterior pdf},description={a probability density function which describes the uncertainty of model parameters by combining the prior information and the information from the data},sort=ae}
\newglossaryentry{gls:evidence}{name={evidence},description={the probability distribution of observed data marginalized over the model parameters},sort=af}
\newglossaryentry{gls:variational}{name={variational inference},description={a method that uses optimization to solve Bayesian inference problem},sort=ba}
\newglossaryentry{gls:kl}{name={KL-divergence},description={the Kullback-Leibler divergence is a measure of difference between two probability distributions},sort=bb}
\newglossaryentry{gls:variationalfamily}{name={variational family},description={a family of probability density functions from which one seeks an optimal approximation to the posterior probability density function},sort=bc}
\newglossaryentry{gls:elbo}{name={ELBO},description={a lower bound for the evidence},sort=afa}
\newglossaryentry{gls:meanfield}{name={mean field approximation},description={probability density functions that assume mutually independent parameters},sort=bd}
\newglossaryentry{gls:advi}{name={ADVI},description={automatic differential variational inference, a method that seeks an optimal Gaussian distribution to approximate the posterior probability distribution},sort=be}
\newglossaryentry{gls:flow}{name={normalizing flow},description={an invertible transform which transforms an initial distribution to a target distribution},sort=bf}
\newglossaryentry{gls:svgd}{name={SVGD},description={Stein variational gradient descent, a method that optimizes a set of model samples to approximate the posterior probability distribution},sort=bg}

\newglossaryentry{stein}{type=symbols,name={\ensuremath{\mathcal{A}_{p}}},description={Stein operator defined on probability distribution $p$},sort=stein}
\newglossaryentry{m}{type=symbols,name={\ensuremath{\mathbf{m}}},sort=m,
	description={vector of model parameters}}
\newglossaryentry{dobs}{type=symbols,name={\ensuremath{\mathbf{d}_{\mathrm{obs}}}},sort=dobs,
	description={observed data vector}}
\newglossaryentry{pdf}{type=symbols,name={pdf},sort=pdf,description={probability density function}}
\newglossaryentry{pm}{type=symbols,name={\ensuremath{p(\mathbf{m})}},sort=pm,description={prior probability density function}}
\newglossaryentry{plike}{type=symbols,name={\ensuremath{p(\mathbf{d}_{\mathrm{obs}}|\mathbf{m})}},sort=pdm,description={likelihood function}}
\newglossaryentry{pjoint}{type=symbols,name={\ensuremath{p(\mathbf{m},\mathbf{d}_{\mathrm{obs}})}},description={joint probability density function of $\mathbf{m}$ and $\mathbf{d}_{\mathrm{obs}}$},sort=pjoint}
\newglossaryentry{posterior}{type=symbols,name={\ensuremath{p(\mathbf{m}|\mathbf{d}_{\mathrm{obs}})}},sort=posterior,description={posterior probability density function}}
\newglossaryentry{posteriorparam}{type=symbols,name={\ensuremath{p(\mathbf{m}|\mathbf{d}_{\mathrm{obs}},\Theta)}},sort=posteriorparam,description={posterior probability density function parameterized with $\Theta$}}
\newglossaryentry{logevidence}{type=symbols,name={\ensuremath{L(\Theta;\mathbf{d}_{\mathrm{obs}})}},description={logarithmic evidence defined as a function of parameters $\Theta$},sort=logevidence}
\newglossaryentry{evidenceparam}{type=symbols,name={\ensuremath{F(q;\Theta)}},description={evidence lower bound of probability distribution $q$ defined as a function of parameters $\Theta$},sort=evidenceparam}
\newglossaryentry{pevidence}{type=symbols,name={\ensuremath{p(\mathbf{d}_{\mathrm{obs}})}},sort=pdobs,description={normalization factor in Bayes' theorem (evidence)}}
\newglossaryentry{mc}{type=symbols,name={MC},description={Monte Carlo},sort=MC}
\newglossaryentry{mcmc}{type=symbols,name={McMC},description={Markov chain Monte Carlo},sort=McMC}
\newglossaryentry{kl}{type=symbols,name={KL},description={Kullback-Leibler divergence},sort=kl}
\newglossaryentry{q}{type=symbols,name={\ensuremath{q}},description={a pobability density function used to approximate the posterior pdf},sort=q}
\newglossaryentry{Q}{type=symbols,name={\ensuremath{Q}},description={$q \in Q$},sort=qQ}
\newglossaryentry{elbo}{type=symbols,name={ELBO},description={evidence lower bound},sort=elbo}
\newglossaryentry{psi}{type=symbols,name={\ensuremath{\psi,\phi}},description={scalar functions},sort=psi}
\newglossaryentry{phi}{type=symbols,name={\ensuremath{\bm \upphi}},description={a smooth vector function},sort=pphi}
\newglossaryentry{EM}{type=symbols,name={EM},description={Expectation-Maximization},sort=em}
\newglossaryentry{jaco}{type=symbols,name={\ensuremath{\mathbf{J}}},description={Jacobian matrix},sort=jaco}
\newglossaryentry{det}{type=symbols,name={\ensuremath{det}},description={determinant},sort=det}
\newglossaryentry{mu}{type=symbols,name={\ensuremath{\bm \upmu}},description={mean of a Gaussian distribution},sort=gmu}
\newglossaryentry{sigma}{type=symbols,name={\ensuremath{\bm\Sigma}},description={covariance matrix of a Gaussian distribution},sort=gsigma}
\newglossaryentry{T}{type=symbols,name={\ensuremath{T}},description={an invertible transform},sort=transform}
\newglossaryentry{ADVI}{type=symbols,name={ADVI},description={automatic differential variational inference},sort=advi}
\newglossaryentry{SVGD}{type=symbols,name={SVGD},description={Stein variational gradient descent},sort=svgd}
\newglossaryentry{ftheta}{type=symbols,name={\ensuremath{\mathbf{F}_{\uptheta}}},description={a normalizing flow parameterized by $\uptheta$},sort=ftheta}
\newglossaryentry{rbf}{type=symbols,name={RBF},description={radial basis function},sort=rbf}
\newglossaryentry{map}{type=symbols,name={MAP},description={maximum a posterior},sort=MAP}
\newglossaryentry{mhmcmc}{type=symbols,name={MH-McMC},description={Metropolis-Hastings Markov chain Monte Carlo},sort=MHMcMC}
\newglossaryentry{hmc}{type=symbols,name={HMC},description={Hamiltonian Monte Carlo},sort=hmc}
\newglossaryentry{fwi}{type=symbols,name={FWI},description={full-waveform inversion},sort=fwi}
\newglossaryentry{vfwi}{type=symbols,name={VFWI},description={variational full-waveform inversion},sort=vfwi}
\newglossaryentry{mrf}{type=symbols,name={MRF},description={Markov random field},sort=MRF}
\newglossaryentry{kappa}{type=symbols,name={\ensuremath{\bm \kappa}},description={geological facies},sort=rockface}
\newglossaryentry{gamma}{type=symbols,name={\ensuremath{\bm \gamma}},description={geological rock properites},sort=rockproperty}
\newglossaryentry{gm}{type=symbols,name={GM},description={Gaussian mixture},sort=gm}
\newglossaryentry{clique}{type=symbols,name={\ensuremath{c}},description={a subset of variables (clique)},sort=clique}
\newglossaryentry{Clique}{type=symbols,name={\ensuremath{\mathcal{C}}},description={a set of cliques, i.e. $c \in \mathcal{C}$},sort=cliques}
\newglossaryentry{lower}{type=symbols,name={\ensuremath{\mathbf{L}}},description={a lower-triangular matrix},sort=lowermatrix}
\newglossaryentry{normal}{type=symbols,name={\ensuremath{N(\cdot|\mathbf{0},\mathbf{I})}},description={a standard normal distribution with zero mean and an identity covariace matrix},sort=norm}
\newglossaryentry{expectation}{type=symbols,name={\ensuremath{\mathrm{E}_{q}}},description={expectation with respect to probability distribution $q$},sort=expectation}
\newglossaryentry{nn}{type=symbols,name={\ensuremath{NN}},description={a neural network function},sort=neuralnetwork}
\newglossaryentry{kernel}{type=symbols,name={\ensuremath{k(\cdot,\cdot)}},description={a scalar kernel function},sort=kernel}
\newglossaryentry{kernelm}{type=symbols,name={\ensuremath{\mathbf{K}(\cdot,\cdot)}},description={a matrix-valued kernel function},sort=kernelm}

\renewcommand*{\glossarymark}[1]{}

\begin{document}

\Frontmatter
\begin{titlepage}%
  \begin{titlepagerule}
    \title{An Introduction to Variational inference in Geophysical Inverse Problems}
  \end{titlepagerule}



  \author{Xin Zhang, Muhammad Atif Nawaz, Xuebin Zhao, Andrew Curtis}
  \address{School of Geosciences, University of Edinburgh, Edinburgh, EH9 3FE, United Kingdom}

\vspace{2cm}
\noindent \textit{This article is a preprint version of the manuscript. The final version is published in Advances in Geophysics:} \url{https://doi.org/10.1016/bs.agph.2021.06.003}
\end{titlepage}

\Mainmatter
  \begin{frontmatter}

\chapter{An Introduction to Variational Inference in Geophysical Inverse Problems}
%
\begin{aug}
	\author[addressrefs={ad1}]%
	{\fnm{Xin}   \snm{Zhang}\footnote{x.zhang2@ed.ac.uk}}%
	\author[addressrefs={ad1}]%
	{\fnm{Muhammad Atif} \snm{Nawaz}}%
	
	\author[addressrefs={ad1}]
	{\fnm{Xuebin} \snm{Zhao}}%
	
	\author[addressrefs={ad1}]%
	{\fnm{Andrew} \snm{Curtis}}
	
	\address[id=ad1]%
	{School of Geosciences, University of Edinburgh, Edinburgh, EH9 3FE, United Kingdom}%
\end{aug}

\begin{abstract}
 In a variety of scientific applications we wish to characterize a physical system using measurements or observations. This often requires us to solve an inverse problem, which usually has non-unique solutions so uncertainty must be quantified in order to define the family of all possible solutions. Bayesian inference provides a powerful theoretical framework which defines the set of solutions to inverse problems, and variational inference is a method to solve Bayesian inference problems using optimization while still producing fully probabilistic solutions. This chapter provides an introduction to variational inference, and reviews its applications to a range of geophysical problems, including petrophysical inversion, travel time tomography and full-waveform inversion. We demonstrate that variational inference is an efficient and scalable method which can be deployed in many practical scenarios.
\end{abstract}

\begin{keywords}
	\kwd{Inverse problem}
	\kwd{Bayesian inference}
	\kwd{Variational inference}
	\kwd{Uncertainty quantification}
	\kwd{Petrophysical inversion}
	\kwd{Seismic tomography}
	\kwd{Full-waveform inversion}
\end{keywords}

\newpage
\glsaddall
\printglossary[type=symbols,title=Summary of notation,nonumberlist]

\end{frontmatter}

\newpage
\section{Introduction}
In a variety of scientific applications scientists often wish to characterize a physical system using measurements or observations which do not represent the system directly. A simplified model of the system is defined which includes a physical relation that predicts measurements or observations for any particular values of the model parameters. One then seeks parameter values that match the measurements or observations. This process is called inversion, and the physical relation that predicts observations that would be made if any particular set of parameter values were true is called the \textit{forward} function. In this article we focus on Geophysics. Geophysicists often need to characterize properties of the Earth's interior using measurements such as seismic, gravitational or electromagnetic data. Subsurface properties are usually parameterized such that one can construct a forward function that predicts corresponding data, and the inverse problem is therefore a parameter estimation problem \citep{aki1976determination, tarantola2005inverse}.

Due to nonlinearity of the physical relation, insufficient data coverage and noise in the data, the inverse problem almost always has non-unique solutions, as infinitely many sets of parameter values fit the observed data to within their measurement uncertainties. This family of values defines uncertainty in the inverse problem solution. In order to reduce this uncertainty, any available \textit{prior} information about parameters (information known independently of the geophysical data) is usually imposed on the solution, and remaining uncertainties in the estimated parameters must be described \citep{tarantola2005inverse}.

Inverse problems are often solved using optimisation methods by seeking parameter values that minimize misfits between observed data and the data predicted by the forward function. Since most inverse problems are under-determined, some form of regularization is often imposed on the model. This process is well-established for linear problems in which the system reduces to solving a set of linear simultaneous equations \citep{aster2018parameter}. This approach can also be applied to nonlinear problems by linearising (approximating) the nonlinear physics around a reference model and solving that linearised problem for the parameter values. The process of linearising and solving the problem is iterated until the misfit or update to the values is sufficiently small \citep{aki1976determination, tarantola1982generalized, dziewonski1987global, constable1987occam, iyer1993seismic, tarantola2005inverse, aster2018parameter}. However, since the regularization is often ad-hoc in the sense that it does not correspond to genuine prior information, the results can be biased and valuable information can be concealed in the process \citep{zhdanov2002geophysical}. In addition this method cannot provide accurate uncertainty estimate for nonlinear problems, nor even for linear problems with complex data uncertainty distributions.
 
Bayesian inference provides a different way to solve inverse problems and quantify uncertainties. In Bayesian inference the prior information is represented by a probability density function (pdf) and is updated with new information from the data to produce a probability density function that describes all information post inversion, called a \textit{posterior} pdf. According to Bayes' theorem, the posterior pdf can be expressed as:
\begin{equation}
	p(\mathbf{m}|\mathbf{d}_{\mathrm{obs}}) = \frac{p(\mathbf{m})p(\mathbf{d}_{\mathrm{obs}}|\mathbf{m})}{p(\mathbf{d}_{\mathrm{obs}})}
	\label{eq:bayes}
\end{equation}
where $\mathbf{m}$ is a vector of model parameter values, $\mathbf{d}_{\mathrm{obs}}$ is the observed data, and $p(\mathbf{m}|\mathbf{d}_{\mathrm{obs}})$ is the posterior pdf; $p(\mathbf{m})$ represents the prior pdf which describes information independent of data, $p(\mathbf{d}_{\mathrm{obs}}|\mathbf{m})$ is called the \textit{likelihood} which represents the probability of observing data $\mathbf{d}_{\mathrm{obs}}$ given parameters $\mathbf{m}$ which in turn depends on the forward function, and $p(\mathbf{d}_{\mathrm{obs}})$ is a normalization factor called the \textit{evidence}. The term \textit{inference} indicates that the prior information is combined with uncertainties in the measured data and forward function to infer the posterior pdf.

A common way to solve Bayesian inference problems is to use Markov chain Monte Carlo (McMC). In McMC one constructs a set (chain) of successive samples of $\mathbf{m}$ drawn from the posterior pdf by taking a structured random walk through a parameter space \citep{brooks2011handbook}; those samples can thereafter be used to calculate useful statistics of that pdf, e.g. the mean and standard deviation. The Metropolis-Hastings algorithm is one such method \citep{metropolis1949monte, hastings1970monte} and has been applied to a range of geophysical applications \citep{mosegaard1995monte, malinverno2000monte, andersen2001bayesian, malinverno2002parsimonious, sambridge2002monte, oh2001geostatistical, mosegaard2002monte, ramirez2005stochastic, gallagher2009markov}. The method has been generalized to trans-dimensional inversion called \textit{reversible-jump} McMC, in which the number of parameters (the dimensionality of parameter space) can vary in the inversion and consequently the parameterization itself can be adapted to the data and the prior information \citep{green1995reversible, green2009reversible}. Reversible-jump McMC has been applied to various geophysical applications, including vertical seismic profile inversion \citep{malinverno2000monte}, electrical resistivity inversion \citep{malinverno2002parsimonious, galetti2018transdimensional}, electromagnetic inversion \citep{minsley2011trans, ray2013robust}, surface wave dispersion inversion \citep{bodin2012transdimensional, young2013transdimensional,  shen2012joint}, travel time tomography \citep{bodin2009seismic, piana2015local, hawkins2015geophysical, galetti2015uncertainty, galetti2017transdimensional, zhang20183, zhang20201d, zhang2020imaging} and full-waveform inversion \citep{ray2016frequency, sen2017transdimensional, ray2017low, guo2020bayesian}. However, due to its random-walk behaviour the method becomes inefficient in high dimensional space (e.g., > 1,000). Other more advanced McMC methods have been introduced to geophysics to solve high dimensional problems, such as Hamiltonian Monte Carlo \citep{duane1987hybrid, sen2017transdimensional, fichtner2018hamiltonian, gebraad2020bayesian, kotsi2020time}, Langevin Monte Carlo \citep{roberts1996exponential, siahkoohi2020uncertainty}, stochastic Newton McMC \citep{martin2012stochastic, zhao2019gradient}, and parallel tempering \citep{hukushima1996exchange, dosso2012parallel, sambridge2013parallel}. Nevertheless, these methods remain intractable for large datasets and high dimensionality because of their extremely high computational cost.

Variational inference solves Bayesian inference problems in a different way: one seeks an optimal approximation to the posterior pdf within a predefined family of (simplified) probability distributions. This is achieved by minimizing a measure of the difference between the posterior pdf and the approximating pdf, for example the Kullback-Leibler (KL) divergence \citep{kullback1951information}. Since the method uses optimization rather than random sampling, it can be computationally more efficient than McMC and provide better scaling to high dimensionality. The methods can also be applied to large datasets by dividing the dataset into minibatches and using stochastic optimization techniques \citep{robbins1951stochastic, kubrusly1973stochastic}. By contrast, stochastic optimisation cannot be applied to McMC because it breaks the detailed balance which is required by most McMC methods.

In variational inference the choice of the variational family determines the accuracy of the approximation and the complexity of the optimization problem. A good choice should be rich enough to approximate complex distributions and simple enough such that the optimization problem can be efficiently solved. A common choice is to use a \textit{mean-field} approximation in which the parameters are assumed to be mutually independent \citep{parisi1988statistical, bishop2006pattern, blei2017variational, zhang2018advances}. The optimisation problem can then be solved efficiently using a coordinate ascent algorithm \citep{bishop2006pattern, blei2017variational} which has been applied in geophysics to invert for spatial distributions of geological facies using seismic data \citep{nawaz2018variational, nawaz2019rapid, nawaz2020variational}.

Despite its wide application in practice, the mean-field method ignores correlations between parameters and requires tedious model-specific mathematical derivations and implementation. This restricts the method to a narrow range of inverse problems for which the derivations can be performed. To make variational inference applicable to general inverse problems, a variety of "black box" methods have been proposed based on different variational families, for example, the mean-field approximation \citep{ranganath2014black, ranganath2016hierarchical}, Gaussian distributions \citep{kucukelbir2017automatic} and probability transforms \citep{rezende2015variational, tran2015variational, liu2016stein, marzouk2016introduction}. These methods are quite general and can be applied to a wide range of applications, for example in geophysics they have been applied to travel time tomography \citep{zhang2020seismic, zhao2020bayesian}, full-waveform inversion \citep{zhang2020variational} and seismic image denoising \citep{siahkoohi2020faster}.

This chapter aims to give a brief introduction to variational inference and its applications in geophysics. In the following sections we first introduce the concepts of variational inference, and then describe four different variational methods: mean-field variational inference, automatic differential variational inference (ADVI), normalizing flows and Stein variational gradient descent (SVGD). The first of these shows how the structure of some inference problems can be exploited to obtain highly efficient variational methods of solution, whereas the latter three methods make few assumptions about the problem structure. In section 3 we demonstrate how these methods have been applied to a range of different applications, including petrophysical inversion, travel time tomography and full-waveform inversion. We conclude the chapter by discussing some limitations and possible improvements to the variational methodology.

\section{Variational Inference}
Variational inference uses optimization to solve Bayesian inference problems. First a family of known probability distributions $Q=\{q(\mathbf{m})\}$ is defined. For example, $Q$ could be the family of all Gaussians, or sums of Gaussians. The variational method then seeks the best approximation to the posterior pdf $p(\mathbf{m}|\mathbf{d}_{\mathrm{obs}})$ within that family by minimizing the KL-divergence between $q(\mathbf{m})$ and $p(\mathbf{m}|\mathbf{d}_{\mathrm{obs}})$:
\begin{equation}
q^{*}(\mathbf{m}) = \argmin_{q \in Q} \mathrm{KL}[q(\mathbf{m})||p(\mathbf{m}|\mathbf{d}_{\mathrm{obs}})]
\label{eq:argminKL} 
\end{equation}
$q^{*}(\mathbf{m})$ is then used as an approximation to the posterior pdf. The KL-divergence is a measure of difference between two pdfs, and can be expressed as:
\begin{equation}
\mathrm{KL}[q(\mathbf{m})||p(\mathbf{m}|\mathbf{d}_{\mathrm{obs}})] = \mathrm{E}_{q}[\mathrm{log}q(\mathbf{m})] - \mathrm{E}_{q}[\mathrm{log}p(\mathbf{m}|\mathbf{d}_{\mathrm{obs}})]
\end{equation} 
where the expectations are taken with respect to the known pdf $q(\mathbf{m})$. The KL-divergence is nonnegative and only equals zero when $q=p$ \citep{kullback1951information}. Expanding the posterior pdf $p(\mathbf{m}|\mathbf{d}_{\mathrm{obs}})$ using Bayes' theorem,
\begin{equation}
\mathrm{KL}[q(\mathbf{m})||p(\mathbf{m}|\mathbf{d}_{\mathrm{obs}})] = \mathrm{E}_{q}[\mathrm{log}q(\mathbf{m})] - \mathrm{E}_{q}[\mathrm{log}p(\mathbf{m},\mathbf{d}_{\mathrm{obs}})] + \mathrm{log}p(\mathbf{d}_{\mathrm{obs}})
\label{eq:KLdiv}
\end{equation}
The evidence term $\mathrm{log}p(\mathbf{d}_{\mathrm{obs}})$ is computationally intractable in many problems: it is the marginal pdf over $\mathbf{d}_{\mathrm{obs}}$ of the joint distribution $p(\mathbf{m},\mathbf{d}_{\mathrm{obs}})$, so the evidence calculation requires an integral of the forward function over the full prior pdf on $\mathbf{m}$ to be evaluated. This is often impossible. Therefore we move the evidence term to the left-hand side and reverse the sign of the equation:
\begin{equation}
\mathrm{log}p(\mathbf{d}_{\mathrm{obs}}) - \mathrm{KL}[q(\mathbf{m})||p(\mathbf{m}|\mathbf{d}_{\mathrm{obs}})]  = \mathrm{E}_{q}[\mathrm{log}p(\mathbf{m},\mathbf{d}_{\mathrm{obs}})] - \mathrm{E}_{q}[\mathrm{log}q(\mathbf{m})]  
\end{equation}
Given that the KL-divergence is nonnegative, the left-hand side defines a lower bound for the evidence, called the evidence lower bound (ELBO):
\begin{equation}
\begin{aligned}
\mathrm{ELBO}[q] &= \mathrm{log}p(\mathbf{d}_{\mathrm{obs}}) - \mathrm{KL}[q(\mathbf{m})||p(\mathbf{m}|\mathbf{d}_{\mathrm{obs}})] \\
&= \mathrm{E}_{q}[\mathrm{log}p(\mathbf{m},\mathbf{d}_{\mathrm{obs}})] - \mathrm{E}_{q}[\mathrm{log}q(\mathbf{m})]
\end{aligned}
\label{eq:ELBO}
\end{equation}
Since the second line of equation \ref{eq:ELBO} does not involve the intractable evidence term, it can be computed in practice by analytical or numerical methods. In addition because the evidence term $\mathrm{log}p(\mathbf{d}_{\mathrm{obs}})$ is a constant for a given problem, minimizing the KL-divergence is equivalent to maximizing the ELBO. Variational inference in equation \ref{eq:argminKL} can therefore be expressed as:
\begin{equation}
q^{*}(\mathbf{m}) = \argmax_{q \in Q} \mathrm{ELBO}[q(\mathbf{m})]
\label{eq:argmaxELBO} 
\end{equation}

In variational inference the choice of family $Q$ is important because it determines the accuracy of the approximation and the complexity of the optimization. A good choice should be flexible enough to approximate the posterior pdf accurately, but simple enough for efficient optimization. Depending on different choices of the family, different variational methods have been proposed. In the following sections we describe several such methods.

\subsection{Mean field approximation} \label{ssec:mfvi}

For problems that have particular types of structures, extremely efficient variational methods can be derived to find solutions. In this section we look at problems that have known, structured probabilistic relationships amongst the variable.

Exact Bayesian inference requires evaluation of the evidence – the denominator in Bayes theorem (equation \ref{eq:bayes}). As the model dimensionality increases, the cost of this calculation escalates exponentially. Thus, exact inference becomes infeasible for many-parameter models and for all practical purposes one needs to resort to approximate inference.

\begin{figure}
	\includegraphics[width=1.\linewidth]{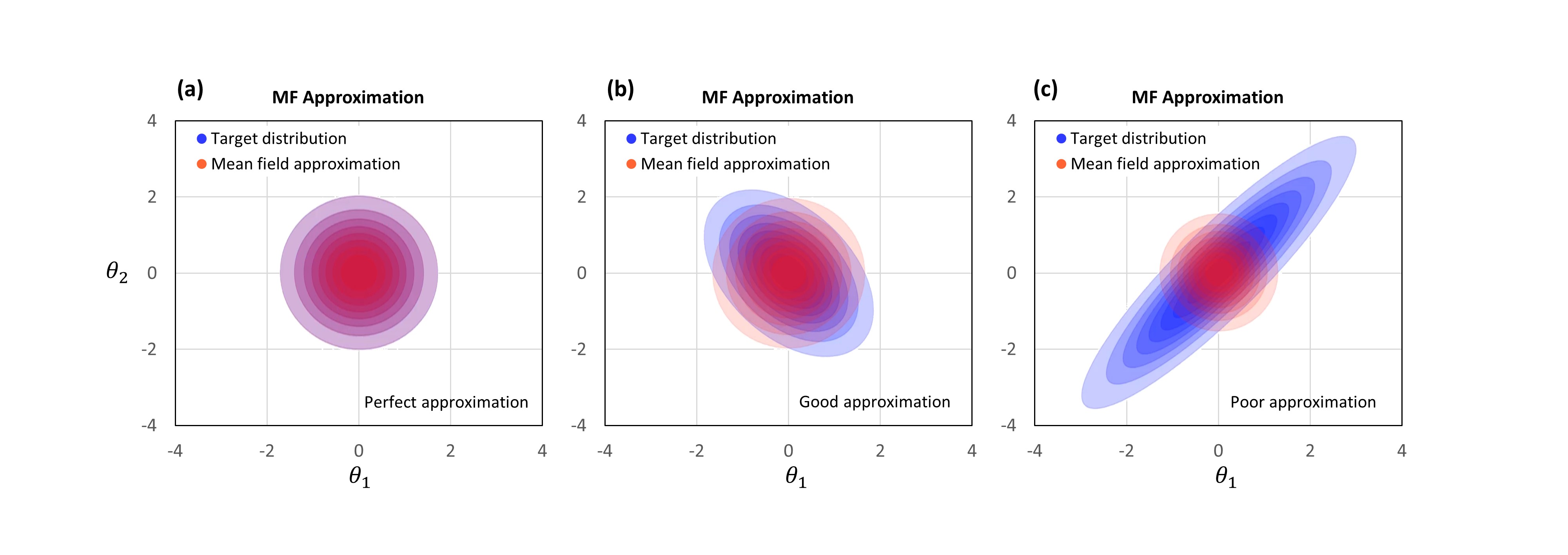}
	\caption{Examples of mean field approximation to bivariate Gaussian distributions with \textbf{(a)} zero correlation, \textbf{(b)} weak correlation and \textbf{(c)} strong correlation between the two parameters.}
	\label{fig:MF_approximation}
\end{figure}

Stochastic sampling-based inference, such as the commonly used Markov-chain Monte Carlo (McMC) method, is only asymptotically exact, i.e., sampling distributions in high-dimensional (henceforth, simply large) models converge to the true distribution only theoretically as sampling continues to infinite time. Instead of approximating the true distribution by a finite number of samples, one may consider other approximation schemes such as limiting the dimensionality of probabilistic dependence among variables \citep{nawaz2016bayesian}. One such scheme is the mean field approximation which provides an efficient method to model probabilistic dependence in high dimensional problems by exploiting structure in the probabilistic dependence among various variables, and replacing at least some probabilistic dependence in the model by an effective random field that is defined by a set of scalar potential functions $\psi_{i}$, each of which is defined over just a few variables. Thus, the intractable joint posterior distribution $p(\mathbf{m}|\mathbf{d}_{\mathrm{obs}})$ over all of the variables under the mean field approximation assumes a factorized form 
\begin{equation}
	p(\mathbf{m}|\mathbf{d}_{\mathrm{obs}}) \cong q(\mathbf{m}) =  \frac{1}{Z} \prod_{i} \psi_{i} (m_{i})
	\label{eq:meanfield}
\end{equation}
where $Z$ is the normalization constant and $q(\mathbf{m})$ is the factorized approximation of $p(\mathbf{m}|\mathbf{d}_{\mathrm{obs}})$. Such a factorized approximation allows computationally efficient inference in large models. In its simplest form, each random variable is regarded as independent of the others, and the only source of mutual interaction (or correlation) among several variables is a random field - a structured set of probabilistic relationships among various parameters of interest at multiple locations. Figure \ref{fig:MF_approximation} shows examples of the mean field approximation to different bivariate Gaussian distributions. While the method can provide accurate approximations to distributions that have zero or weak correlation between parameters (Figure \ref{fig:MF_approximation}a and b), it fails to produce accurate estimates of distributions with strong correlations (e.g., Figure \ref{fig:MF_approximation}c). Thus, a naive implementation of the mean field approximation cannot be used to infer posterior distributions with strong correlations.

A more common approach is to capture correlations among pairs of variables which results in the so called \textit{Ising} or \textit{Potts} model depending on whether each variable can take two or more possible states, respectively. Modelling of pairwise dependence among variables imposes smoothness constraints that can be described by second order statistics, e.g. using covariance matrix. These models, however, ignore higher-order dependence structure beyond pairs of variables, e.g. multiple-point statistics. \cite{nawaz2019rapid} introduced higher-order mean field inference method that makes use of structure of dependence among variables to capture most of the significant higher-order correlations among them while still allowing computationally efficient inference. Factorization of joint distribution in this case takes the form
\begin{equation}
	q(\mathbf{m})= \frac{1}{Z} \prod_{c \in \mathcal{C}} \psi_{c} (\mathbf{m}_{c})
	\label{eq:factorized_mf}
\end{equation}
where $\psi_{c}$ represents potential functions (called \textit{clique} potentials) defined over some subsets $c$ of variables called a \textit{clique} and $\mathcal{C}$ represents the set of cliques in the graph. Full probabilistic dependence among variables $\mathbf{m}_{c}$ within a clique $c$ is honoured, however, it is assumed that a variable in $c$ may interact with other variables outside $c$ only through an effective field defined by the functional form of clique potentials $\psi_{c}$. According to the Hammersley-Clifford theorem \citep{hammersley1971markov,besag1974spatial}, the joint distribution $q(\mathbf{m})$ over all $\mathbf{m}$ may be expressed as a Gibbs distribution which takes the form
\begin{equation}
	q(\mathbf{m})= \frac{1}{Z} \mathrm{exp} \left\{-\frac{1}{r} \sum_{c \in \mathcal{C}} E_{c}(\mathbf{m}_{c}) \right\}
	\label{eq:gibbs}
\end{equation}
where $E_{c}(\mathbf{m}_{c})$ represents the energy function that associates low energy states correspond to high probability configurations of $\mathbf{m}_{c}$, and $r$ is a constant. The clique potentials $\psi_{c}$, therefore, take the function form
\begin{equation}
	\psi_{c} (\mathbf{m}_{c}) =  \mathrm{exp} \left\{ -\frac{E_{c}(\mathbf{m}_{c})}{r}
	\label{eq:clique_potential} \right\}
\end{equation}
A factorized distribution that takes the form of the Gibbs distribution is commonly known as a Markov random field. The quality of the mean field approximation can be determined using some distance measure between the true (unknown) posterior distribution $p(\mathbf{m}|\mathbf{d}_{\mathrm{obs}})$ and its factorized approximation $q(\mathbf{m})$. This may be achieved by using the KL divergence (equation \ref{eq:KLdiv}) which we then minimize, showing that mean field inference is a special form of variational inference where the approximating distribution takes a factorized form. Mean field inference commonly employs iterative optimization methods to perform probabilistic inference in an optimization framework without stochastic sampling while still providing full probabilistic results, as described below.

In order to estimate the intractable constant $p(\mathbf{d}_{\mathrm{obs}})$ in Bayes’ theorem under the above simplified model, we denote its logarithm as a function of parameters $\Theta$ as $L(\Theta;\mathbf{d}_{\mathrm{obs}})$ and refer to it as the log evidence. Any choice of the auxiliary distribution $q$ defines a lower bound $F(q;\Theta)$ (the ELBO in equation \ref{eq:ELBO}) on the log-evidence $L(\Theta;\mathbf{d}_{\mathrm{obs}})$ \citep{neal1998view, beal2003variational, nawaz2018variational}, such that
\begin{equation}
	L(\Theta;\mathbf{d}_{\mathrm{obs}}) = F(q;\Theta) + \mathrm{KL}(q(\mathbf{m}) || p(\mathbf{m}|\mathbf{d}_{\mathrm{obs}},\Theta))
\end{equation}
where the lower bound $F(q;\Theta)$ is also called \textit{variational free energy} or simply \textit{free energy}. It has its origin in statistical physics where it corresponds to the negative of Gibbs free energy \citep{feynman1972statistical}, and $ \mathrm{KL}(q(\mathbf{m}) || p(\mathbf{m}|\mathbf{d}_{\mathrm{obs}},\Theta)) \ge 0$ is the KL divergence (also called relative-entropy) between $q(\mathbf{m})$ and $p(\mathbf{m}|\mathbf{d}_{\mathrm{obs}},\Theta)$ as defined above. For a  factorisable distribution $q$, $F(q; \Theta)$ assumes a closed-form expression in terms of marginal distributions of $q$ \protect \citep{nawaz2018variational}.

Although $L(\Theta;\mathbf{d}_{\mathrm{obs}})$ is intractable, its lower bound $F(q;\Theta)$ may be estimated for a suitably chosen family of approximate pdf's $q$. This suggests that an iterative scheme may be devised to estimate $L(\Theta;\mathbf{d}_{\mathrm{obs}})$ by successively updating $q$ and $\Theta$ in each iteration. For example, a variational form of the Expectation-Maximization (EM) algorithm \citep{dempster1977maximum} may be used to approximate $L(\Theta;\mathbf{d}_{\mathrm{obs}})$ in an iterative fashion such that its lower bound $F(q;\Theta)$ is increased which effectively decreases $ \mathrm{KL}(q(\mathbf{m}) || p(\mathbf{m}|\mathbf{d}_{\mathrm{obs}},\Theta))$ for a given set of parameters $\Theta$ in each iteration. 

\protect
	\begin{algorithm}[H] \label{al:meanfield}
		\SetAlgoNoLine
		
		\KwIn{An initial distribution $q^{0}$ from an approximating family $q(\mathbf{m})$; An initial set of parameters $\Theta^0$ of the (unkonwn) posterior distribution $p(\mathbf{m}|\mathbf{d}_{\mathrm{obs}},\Theta)$.}
		\KwOut{Updated $\Theta$ of the true posterior pdf $p(\mathbf{m}|\mathbf{d}_{\mathrm{obs}},\Theta)$.}
		$l \gets 0$ \\
		\While{$l < N$}{
			Calculate $F(q^{l}; \Theta^{l})$ \protect \cite[e.g. see][]{nawaz2018variational}\\
			
			E-step: \\ 
			\qquad $q^{l+1} \gets \argmax_{q} F(q;\Theta^{l})$ \\
			M-step: \\
			\qquad $\Theta^{l+1} \gets \argmax_{\Theta} F(q^{l+1};\Theta)$ \\
			$\delta \gets \mathrm{abs}(\Theta^{l+1}-\Theta^{l}$) \\
			\uIf{$\delta$ is sufficiently small}{
				exit
			}
			\Else{
				$l \gets l + 1$
			}
		}
		return $\Theta^{N}$.
		\caption{Mean field approximation}
	\end{algorithm}

The E-step of the EM algorithm at any iteration $l$ updates the variational distribution $q(\mathbf{m})$ by maximizing the free-energy $F(q;\Theta)$ with respect to $q$ while keeping the parameters $\Theta^{l}$ fixed such that 
\begin{equation}
	q^{l+1} = \argmax_{q} F(q;\Theta^{l})
	\label{eq:e-step}
\end{equation}
where the bracketed superscripts refer to the iteration number. \cite{nawaz2018variational} showed that the E-step of the EM algorithm can be solved using a message passing algorithm, called belief propagation (BP) \citep{pearl1982reverend}, or its variant, the loopy belief propagation (LBP) \citep{mariethoz2014multiple, yedidia2003understanding}. The M-step of the EM algorithm at any iteration $l$ computes an updated set of parameters $\Theta^{l+1}$ by maximizing the free-energy $F(q;\Theta)$ with respect to $\Theta$ while keeping the variational distribution $q$ fixed at its value $q^{l+1}$ estimated during the E-step, such that
\begin{equation}
	\Theta^{l+1} = \argmax_{\Theta} F(q^{l+1};\Theta)
	\label{eq:m-step}
\end{equation} 

In summary, at the end of $(l+1)^{th}$ iteration the E-Step of the EM algorithm yields the free energy $F(q^{l+1},\Theta^{l})$ equal to $L(\Theta^{l};\mathbf{d}_{\mathrm{obs}})$ which is the upper bound of $F(q,\Theta^{l})$.  Therefore, the E-step improves the estimate of the posterior distribution $p(\mathbf{m}|\mathbf{d}_{\mathrm{obs}}, \Theta)$ while the M-step improves the estimate of parameters $\Theta$, such that the combined E-M steps are guaranteed not to decrease the estimate of log evidence $L(\Theta;\mathbf{d}_{\mathrm{obs}})$ during any iteration of the EM algorithm. On convergence, the EM algorithm yields the best mean field approximation $q(\mathbf{m})$ of the true intractable posterior distribution $p(\mathbf{m}|\mathbf{d}_{\mathrm{obs}})$. We summarize the method in Algorithm \ref{al:meanfield}.

\subsection{Automatic differential variational inference} \label{ssec:advi}
The mean-field approximation allows highly efficient variational methods to be derived, at the expense of losing full correlation information between parameters. Such methods require model-specific derivations and implementations, which restricts them to those types of problems for which approximation applies. In this section we describe a method called automatic differential variation inference (ADVI) which can be applied to a general class of inverse problems, and which is made efficient by introducing a different approximation \citep{kucukelbir2017automatic}.

The key idea behind ADVI is to use a Gaussian variational family. Gaussians are defined over the entire set of real numbers whereas in reality model parameters often have hard bound constrains (for example seismic velocity is greater than zero). To apply ADVI to constrained variables we first transform those variables into an unconstrained space using an invertible transform $T$: $\bm{\uptheta}=T(\mathbf{m})$. In this space the joint pdf $p(\mathbf{m},\mathbf{d}_{\mathrm{obs}})$ becomes:
\begin{equation}
p(\bm{\uptheta},\mathbf{d}_{\mathrm{obs}}) = p(\mathbf{m},\mathbf{d}_{\mathrm{obs}}) |det\mathbf{J}_{T^{-1}}(\bm{\uptheta})|
\end{equation}
where $\mathbf{J}_{T^{-1}}(\bm{\uptheta})$ is the Jacobian matrix of the inverse of transform $T$, and $|\cdot|$ denotes absolute value. Define a Gaussian family in this unconstrained space,
\begin{equation}
q(\bm{\uptheta};\bm{\zeta})=N(\bm{\uptheta}|\bm{\upmu},\bm{\Sigma})=N(\bm{\uptheta}|\bm{\upmu},\mathbf{L}\mathbf{L}^{\mathrm{T}})
\end{equation}
where $\bm{\zeta}$ represents variational parameters, that is the mean vector $\bm{\upmu}$ and the covariance matrix $\bm{\Sigma}$. To ensure the covariance matrix is positive semidefinite, we use a Cholesky factorization $\bm{\Sigma}=\mathbf{L}\mathbf{L}^{\mathrm{T}}$ where $\mathbf{L}$ is a lower-triangular matrix, to reparameterize the covariance matrix. If $\bm{\Sigma}$ is a diagonal matrix, $q$ reduces to a mean-field approximation as described in section \ref{ssec:mfvi}.

Within this Gaussian family the variational problem in equation \ref{eq:argmaxELBO} becomes:
\begin{equation}
\begin{aligned}
\bm{\zeta}^{*} &= \argmax_{\bm{\zeta}} \mathrm{ELBO}[q(\bm{\uptheta};\bm{\zeta})]\\
                &= \argmax_{\bm{\zeta}} \mathrm{E}_{q}[\mathrm{log}p\big(T^{-1}(\bm{\uptheta}),\mathbf{d}_{\mathrm{obs}}\big) + \mathrm{log}|det\mathbf{J}_{T^{-1}}(\bm{\uptheta})|] - \mathrm{E}_{q}[\mathrm{log}q(\bm{\uptheta};\bm{\zeta})] 
\label{eq:argmaxELBO_advi}
\end{aligned} 
\end{equation}
This optimisation problem can be solved by gradient-based optimisation methods, for example gradient ascent. In order to calculate the gradients of the ELBO with respect to variational parameters $\bm{\zeta}$, we first transform the Gaussian distribution $q(\bm{\uptheta};\bm{\zeta})$ to a standard Normal distribution $N(\bm{\upeta}|\mathbf{0},\mathbf{I})$ by using the transform $\bm{\upeta}=R(\bm{\uptheta})=\mathbf{L}^{-1}(\bm{\uptheta}-\bm{\upmu})$. The problem thereafter becomes:
\begin{multline}
\bm{\zeta}^{*} = \argmax_{\bm{\zeta}} \mathrm{ELBO}[q(\bm{\uptheta};\bm{\zeta})]\\
= \argmax_{\bm{\zeta}} \mathrm{E}_{N(\bm{\upeta}|\mathbf{0},\mathbf{I})}\big[\mathrm{log}p\big(T^{-1}R^{-1}(\bm{\upeta}),\mathbf{d}_{\mathrm{obs}}\big) + \mathrm{log}|det\mathbf{J}_{T^{-1}}\big(R^{-1}(\bm{\upeta})\big)|\big] \\
- \mathrm{E}_{q}[\mathrm{log}q(\bm{\uptheta};\bm{\zeta})] 
\label{eq:argmaxELBO_advi_standardized}
\end{multline} 
where the first expectation in equation \ref{eq:argmaxELBO_advi_standardized} is calculated with respect to a standard Normal distribution. There is no Jacobian term appearing in equation \ref{eq:argmaxELBO_advi_standardized} according to the rules of integration by substitution. For example for any function $h(\bm{\uptheta})$,
\begin{equation}
\begin{aligned}
	\mathrm{E}_{q}[h(\bm{\uptheta})]
	&= \int h(\bm{\uptheta}) q(\bm{\uptheta};\bm{\zeta}) d\bm{\uptheta} \\
	&= \int h\big(R^{-1}(\bm \upeta)\big) q\big(R^{-1}(\bm \upeta);\bm \zeta\big)|det\mathbf{J}_{R^{-1}}(\bm \upeta)|d\bm{\upeta} \\
	&= \int h\big(R^{-1}(\bm{\upeta})\big) N\big(\bm{\upeta}|\mathbf{0},\mathbf{I}\big) d\bm{\upeta} \\
	&= \mathrm{E}_{N(\bm{\upeta}|\mathbf{0},\mathbf{I})}[h(R^{-1}(\bm{\upeta}))]
\end{aligned}
\end{equation}
The second expectation in equation \ref{eq:argmaxELBO_advi_standardized} does not need to be transformed because the expectation has an analytic form. In fact this expectation is called the \textit{entropy} of $q$, written $H[q(\bm{\uptheta};\bm{\zeta})]$:
\begin{equation}
\begin{aligned}
H[q(\bm{\uptheta};\bm{\zeta})] 
&= - \mathrm{E}_{q}[\mathrm{log}q(\bm{\uptheta};\bm{\zeta})] \\
&= \frac{k}{2} + \frac{k}{2}\mathrm{log}(2\pi) + \frac{1}{2}\mathrm{log}|det\mathbf{L}\mathbf{L}^{\mathrm{T}}|
\label{eq:entropy}
\end{aligned}
\end{equation}
where $k$ is the dimension of vector $\bm{\uptheta}$.

The gradients of the ELBO with respect to variational parameters can be calculated by exchanging the derivative and expectation according to the dominant convergence theorem \citep{ccinlar2011probability} which allows the derivatives to be calculated inside the expectations, and by applying the chain rule:
\begin{equation}
\nabla_{\bm{\upmu}}\mathrm{ELBO}
=\mathrm{E}_{N(\bm{\upeta}|\mathbf{0},\mathbf{I})} 
    \big[
    \nabla_{\mathbf{m}}\mathrm{log}p(\mathbf{m},\mathbf{d}_{\mathrm{obs}}) 
    \nabla_{\bm{\uptheta}}T^{-1}(\bm{\uptheta})
    + \nabla_{\bm{\uptheta}} \mathrm{log}|det\mathbf{J}_{T^{-1}}(\bm{\uptheta})| 
    \big]
\label{eq:gradient_mu}
\end{equation}
The gradients of the ELBO with respect to $\mathbf{L}$ can be written similarly:
\begin{multline}
\nabla_{\mathbf{L}}\mathrm{ELBO}
=\mathrm{E}_{N(\bm{\upeta}|\mathbf{0},\mathbf{I})} 
\big[
 \big(
\nabla_{\mathbf{m}}\mathrm{log}p(\mathbf{m},\mathbf{d}_{\mathrm{obs}}) 
\nabla_{\bm{\uptheta}}T^{-1}(\bm{\uptheta})
+ \nabla_{\bm{\uptheta}} \mathrm{log}|det\mathbf{J}_{T^{-1}}(\bm{\uptheta})| 
 \big) \bm{\upeta}^{\mathrm{T}}
\big] \\
+ (\mathbf{L}^{-1})^\mathrm{T}
\label{eq:gradient_L}
\end{multline}
The expectations can be estimated using Monte Carlo (MC) integration which provides noisy, unbiased estimates of the expectations. The accuracy of MC integration increases with the number of samples, but in practice a low number or even a single sample can be sufficient at each iteration since optimisations are usually performed over many iterations so that statistically they will converge towards the solution \citep{kucukelbir2017automatic}. The variational problem in equation \ref{eq:argmaxELBO_advi_standardized} can therefore be solved by standard gradient-based optimisation methods, by gradient ascent. The final approximation $q(\mathbf{m})$ can then be obtained by transforming the solution $q^{*}(\bm{\uptheta})$ back to the constrained parameter space, either numerically or analytically depending on the form of transform $T$. By combining with the automatic differential technique \citep{wengert1964simple, baydin2018automatic} the whole process can be conducted automatically, hence the name "automatic differential". The procedure is summarized in Algorithm \ref{al:advi}.

\protect \begin{algorithm}[H] \label{al:advi}
	\SetAlgoNoLine
	
	\KwIn{The joint pdf $p(\mathbf{m},\mathbf{d}_{\mathrm{obs}})$ in a constrained space, which can be estimated for any particular value of $\mathbf{m}$; a transform $T: \bm \uptheta = T(\mathbf{m})$ which transforms $\mathbf{m}$ into an unconstrained variable $\bm \uptheta$ and an initial Gaussian distribution $q^0(\bm \uptheta; \bm \upmu^0, \mathbf{L}^0)$ in the unconstrained space.}
	\KwOut{A distribution $q^{N}(\mathbf{m})$ that approximates the posterior pdf.}
	\For{$l \gets 1$ to $N$}{
		Calculate gradients $\nabla_{\bm{\upmu}^{l-1}}\mathrm{ELBO}$ and $\nabla_{\mathbf{L}^{l-1}}\mathrm{ELBO}$ using equation \ref{eq:gradient_mu} and \ref{eq:gradient_L}. \\
		Update $\bm \upmu$ and $\mathbf{L}$
		\begin{equation}
			\begin{aligned}
			\bm \upmu^{l} = \bm \upmu^{l-1} + \epsilon^{l}\nabla_{\bm{\upmu}^{l-1}}\mathrm{ELBO} \\
			\mathbf{L}^{l} = \mathbf{L}^{l-1} + \epsilon^{l}\nabla_{\mathbf{L}^{l-1}}\mathrm{ELBO}
			\end{aligned}
		\end{equation} 
		where $\epsilon^{l}$ is the step size at the $l^{th}$ iteration.
	}
	Transform $q^N(\bm \uptheta;\bm \upmu^{N},\mathbf{L}^{N})$ to $q^N(\mathbf{m})$ using $T$.
	\caption{Automatic differential variational inference (ADVI)}
\end{algorithm}

Note that the final approximation is determined by the Gaussian distribution $q^{*}(\bm{\uptheta})$ in the unconstrained space and by the transform $T$. Unfortunately the optimal transform is difficult to determine because it depends on the unknown properties of the posterior distribution $p(\mathbf{m}|\mathbf{d}_{\mathrm{obs}})$. A commonly-used transform is:
\begin{equation}
\begin{aligned}
\theta_{i} &= T(m_{i}) = \mathrm{log}(m_{i}-a) - \mathrm{log}(b-m_{i}) \\
 m_{i} &= T^{-1}(\theta_{i}) = a_{i} + \frac{(b_{i}-a_{i})}{1+exp(-\theta_{i})}
\end{aligned}
\label{eq:transform}
\end{equation}
where $m_{i}$ represents the $i^{th}$ parameter in the original constrained space, $\theta_{i}$ is the transformed unconstrained variable, and $a_{i}$ and $b_{i}$ are the lower and upper bound on $m_{i}$ respectively \citep{stan2016stan}. The final approximation is then limited by the Gaussian distribution $q^{*}(\bm{\uptheta})$ and the transform in equation \ref{eq:transform}.

\begin{figure}
	\includegraphics[width=.9\linewidth]{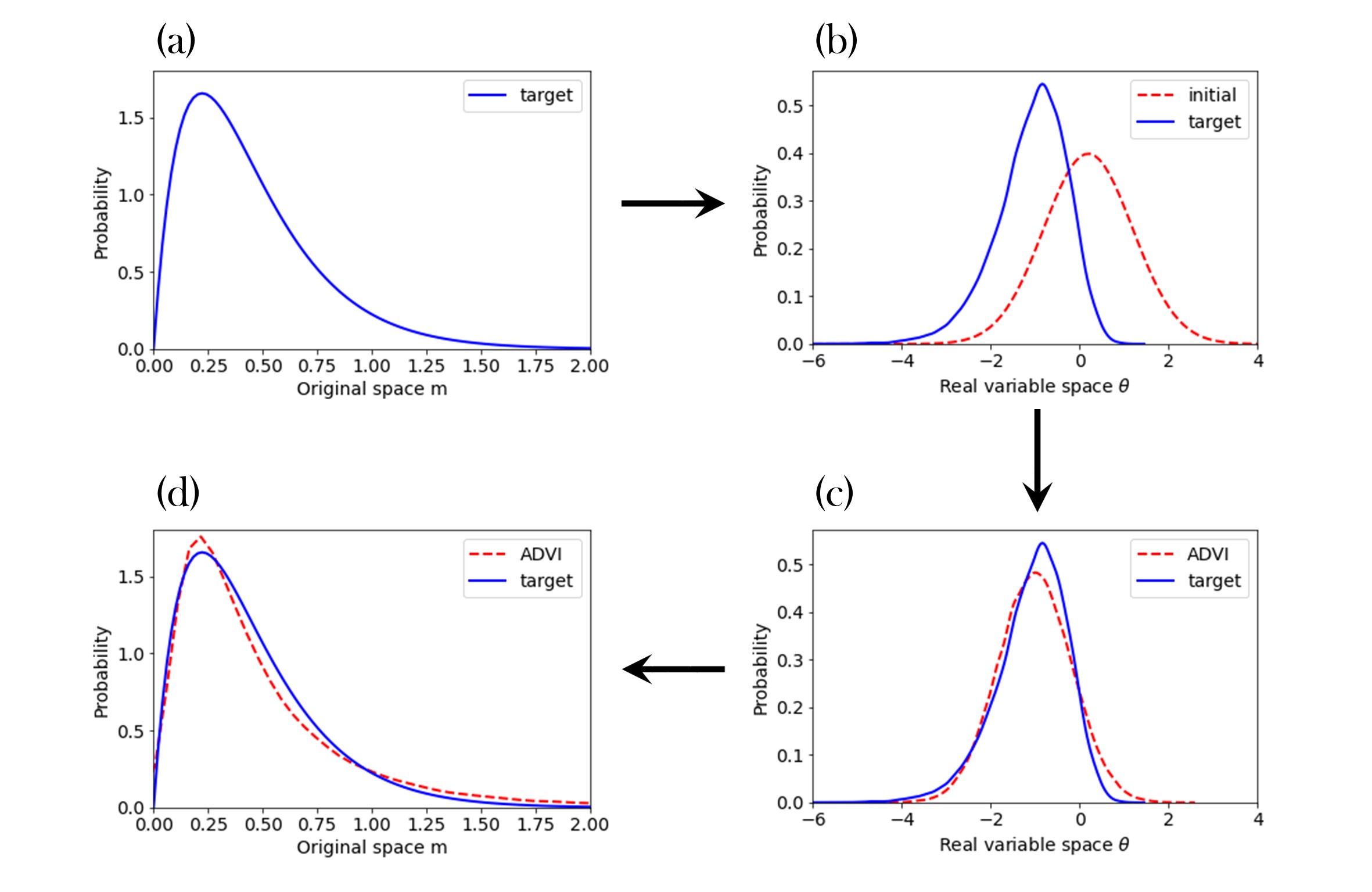}
	\caption{A 1D example of ADVI. \textbf{(a)} the target (posterior) pdf in the original positive half space. \textbf{(b)} The target pdf in the transformed unconstrained space (blue line) and an initial Gaussian approximation (red dashed line). \textbf{(c)} and \textbf{(d)} show the target pdf (blue line) and the approximation obtained using ADVI (red dashed line) in the unconstrained space and the original space, respectively.}
	\label{fig:advi_workflow}
\end{figure}
Figure \ref{fig:advi_workflow} shows a 1D example of the ADVI. The true target (posterior) pdf is defined in the positive half space (Figure \ref{fig:advi_workflow}a). An initial Gaussian distribution is defined in the transformed unconstrained space (Figure \ref{fig:advi_workflow}b) and updated using the gradient ascent method (Figure \ref{fig:advi_workflow}c). The final approximation is obtained by transforming the obtained Gaussian distribution back to the original space (Figure \ref{fig:advi_workflow}d). Since the true distribution is a non-Gaussian distribution in both original and transformed space, the obtained approximation is different to the true distribution. This indicates that ADVI can produce biased results for non-Gaussian posterior pdfs.

Note that in very high dimensional space ADVI may become inefficient because of the large size of the full covariance matrix (the number of variables is proportional to the square of dimensionality). In such cases if correlation between certain parameters can be ignored, a diagonal covariance matrix or a sparse covariance matrix may be used to reduce computational cost. Due to the Gaussian variational family, ADVI cannot provide accurate approximations to multimodal distributions. However, further improvements are made possible by using a mixture of Gaussian distributions \citep{zobay2014variational, arenz2018efficient}. 

\subsection{Normalizing flows}
The approximation to the posterior pdf obtained using ADVI is limited by a Gaussian distribution in the unconstrained space, and a fixed transform T that is used to transform that Gaussian distribution to the original parameter space. It should be possible to improve the approximation by finding a more suitable transform. This idea leads to a method called normalizing flows, in which a series of invertible and differential transforms (called flows) are applied to an initial known distribution (e.g. a Gaussian distribution); the flows are optimized to produce an improved approximation to the posterior pdf \citep{rezende2015variational}.

Let $\mathbf{m}_{0}$ be a random vector variable which has a simple and analytically known pdf $q_{0}(\mathbf{m}_{0})$, for example a Gaussian distribution, and apply an invertible transform $\mathbf{F}_{\bm{\uptheta}}$ (parameterized by $\bm{\uptheta}$) such that $\mathbf{m}_{1} = \mathbf{F}_{\bm{\uptheta}}(\mathbf{m}_{0})$. The pdf of transformed variable $\mathbf{m}_{1}$ can be expresses as:
\begin{equation}
q_{1}(\mathbf{m}_{1}) = q_{0}(\mathbf{m}_{0}) 
\big |det \frac{\partial \mathbf{F}_{\bm{\uptheta}}}{\partial \mathbf{m}_{0}}\big |^{-1}
\label{eq:flow}
\end{equation} 
where $\frac{\partial \mathbf{F}_{\bm{\uptheta}}}{\partial \mathbf{m}_{0}}$ is the Jacobian matrix of the transform $\mathbf{F}_{\bm{\uptheta}}$. The pdf $q_{0}$ is called an initial distribution and the transform $\mathbf{F}_{\bm{\uptheta}}$ is referred to as a normalizing flow which pushes the simple and known pdf $q_{0}$ to a target pdf $q_{1}$. Depending on the form of the normalizing flow, the initial pdf can be manipulated in different ways, for example it can be expanded, contracted, rotated or its location can be shifted to produce different target pdfs.

In Bayesian inference the goal is to estimate the posterior pdf, that is, to find a normalizing flow $\mathbf{F}_{\bm{\uptheta}}$ such that the pdf $q_{1}$ is close to the posterior pdf. However, in general it is difficult to construct a single flow that transforms a simple distribution to the posterior distribution given that real posterior pdfs often have complex forms (which a priori we do not know). Instead this ideal transform can be approximated by combining multiple simple flows and successively applying equation \ref{eq:flow}.

Assume we have $K$ flows, $\mathbf{F}(\bm{\uptheta}_{0}), \mathbf{F}(\bm{\uptheta}_{1}), ..., \mathbf{F}(\bm{\uptheta}_{K-1})$, and successively apply them to the initial variable $\mathbf{m}_{0}$:
\begin{equation}
\mathbf{m}_{K} = \mathbf{F}_{\bm{\uptheta}_{K-1}} \cdot \mathbf{F}_{\bm{\uptheta}_{K-2}} \cdots \mathbf{F}_{\bm{\uptheta}_{1}} \cdot \mathbf{F}_{\bm{\uptheta}_{0}}(\mathbf{m}_{0}) 
\end{equation}
where $\mathbf{m}_{K}$ is the variable after the combined transformation. The pdf of $\mathbf{m}_{K}$ can be obtained using equation \ref{eq:flow}:
\begin{equation}
	q_{K}(\mathbf{m}_{K}) = q_{0}(\mathbf{m}_{0})
	\prod_{i=0}^{K-1}\big |det \frac{\partial \mathbf{F}_{\bm{\uptheta}_{i}}}{\partial \mathbf{m}_{i}}\big |^{-1}
	\label{eq:multflows}
\end{equation} 
Hereafter for simplicity we use the notation $\Uptheta=(\bm{\uptheta}_{0}, \bm{\uptheta}_{1}, ..., \bm{\uptheta}_{K-1})$ and $\mathbf{F}_{\Uptheta}$ to represent the chain of transforms: $\mathbf{F}_{\Uptheta} = \mathbf{F}_{\bm{\uptheta}_{K-1}} \cdot \mathbf{F}_{\bm{\uptheta}_{K-2}} \cdots \mathbf{F}_{\bm{\uptheta}_{1}} \cdot \mathbf{F}_{\bm{\uptheta}_{0}}$, and use $| det \frac{\partial \mathbf{F}_{\Uptheta}}{\partial \mathbf{m}_{0}} | = 	\prod_{i=0}^{K-1} |det \frac{\partial \mathbf{F}_{\bm{\uptheta}_{i}}}{\partial \mathbf{m}_{i}} |$. By using a series of transforms equation \ref{eq:multflows} improves the expressibility of the combined transformation, so that more complex final distributions can be created. Note that if we use an analytically known initial distribution and construct transforms such that their Jacobian determinants are also analytically known, the final distribution is also analytic.

To approximate the posterior pdf using the distribution $q_{K}(\mathbf{m}_{K})$ obtained from normalizing flows, we optimize the flow parameters $\Uptheta$ by maximizing the ELBO as in equation \ref{eq:argmaxELBO}. This results in a variational problem:
\begin{equation}
	\begin{aligned}
		\Uptheta^{*} &= \argmax_{\Uptheta} \mathrm{ELBO}[q_{K}(\mathbf{m}_{K})]\\
		&= \argmax_{\Uptheta} \mathrm{E}_{q_{K}}[\mathrm{log}p(\mathbf{m}_{K},\mathbf{d}_{\mathrm{obs}}\big)
		-\mathrm{log}q_{K}(\mathbf{m}_{K})] 
		\label{eq:argmaxELBO_flows0}
	\end{aligned} 
\end{equation} 
According to the change of variables theorem, for any function $h(\mathbf{m}_{K})$ the expectation with respect to $q_{K}(\mathbf{m}_{K})$ can be expressed as:
\begin{equation}
	\int h(\mathbf{m}_{k}) q_{K}(\mathbf{m}_{K}) \,d\mathbf{m}_{K} = \int h(\mathbf{m}_{k}) q_{0}(\mathbf{m}_{0}) \,d\mathbf{m}_{0}
\label{eq:changevariables}
\end{equation}
Combining equation \ref{eq:multflows} and \ref{eq:changevariables} with equation \ref{eq:argmaxELBO_flows0} gives
\begin{equation}
\begin{aligned}
	\Uptheta^{*} &= \argmax_{\Uptheta} \mathrm{ELBO}[q_{K}(\mathbf{m}_{K})]\\
	&= \argmax_{\Uptheta} \mathrm{E}_{q_{0}}[\mathrm{log}p(\mathbf{m}_{K},\mathbf{d}_{\mathrm{obs}}) 
	- \mathrm{log}q_{0}(\mathbf{m}_{0})
	+ \mathrm{log}\big|det \frac{\partial \mathbf{F}_{\Uptheta}}{\partial \mathbf{m}_{0}}\big |
	] 
	\label{eq:argmaxELBO_flows1}
\end{aligned} 
\end{equation}
where the expectation is taken with respect to the initial distribution $q_{0}(\mathbf{m}_{0})$. This problem can be solved using standard gradient-based optimization methods, for example, gradient ascent. Similarly to the gradient computations in ADVI, the gradients of ELBO with respect to $\Uptheta$ can be obtained by exchanging the expectations and derivatives and by applying the chain rule:
\begin{equation}
	\nabla_{\Uptheta} \mathrm{ELBO} = \mathrm{E}_{q_{0}} 
	\big[ \nabla_{\mathbf{m}_{K}} \mathrm{log}p(\mathbf{m}_{K},\mathbf{d}_{\mathrm{obs}}) 
	\nabla_{\Uptheta} \mathbf{m}_{K} 
	+ \nabla_{\Uptheta} \mathrm{log} \big | det \frac{\partial \mathbf{F}_{\Uptheta}}{\partial \mathbf{m}_{0}} \big |
	\big]
	\label{eq:gradients_normalizing_flow}
\end{equation} 
As in ADVI the expectation can be calculated using MC integration over a small number of samples, and the resulting gradients can be used to solve the optimization problem using gradient ascent methods. The final approximation can be obtained using equation \ref{eq:multflows} with the optimal parameters $\Uptheta^{*}$. The procedure is summarized in Algorithm \ref{al:nfvi}.

\protect
\begin{algorithm}[H] \label{al:nfvi}
	\SetAlgoNoLine
	
	\KwIn{An initial pdf $q_{0}(\mathbf{m}_{0})$; the joint pdf $p(\mathbf{m},\mathbf{d}_{\mathrm{obs}})$; a series of flows $\mathbf{F}_{\Uptheta} = \mathbf{F}_{\bm{\uptheta}_{K-1}} \cdot \mathbf{F}_{\bm{\uptheta}_{K-2}} \cdots \mathbf{F}_{\bm{\uptheta}_{1}} \cdot \mathbf{F}_{\bm{\uptheta}_{0}}$ parameterized by $\Uptheta=(\bm{\uptheta}_{0}, \bm{\uptheta}_{1}, ..., \bm{\uptheta}_{K-1})$.}
	\KwOut{A distribution $q(\mathbf{m})$ that approximates the posterior pdf.}
	Initialize $\Uptheta$ with $\Uptheta^0$. \\
	\For{$l \gets 1$ to $N$}{
		Calculate gradients $\nabla_{\Uptheta^{l-1}} \mathrm{ELBO}$ using equation \ref{eq:gradients_normalizing_flow}. \\
		Update $\Uptheta$
		\begin{equation}
			\Uptheta^{l} = \Uptheta^{l-1} + \epsilon^{l}\nabla_{\Uptheta^{l-1}}\mathrm{ELBO}
		\end{equation} 
		where $\epsilon^{l}$ is the step size at the $l^{th}$ iteration.
	}
	Obtain final approximation $q(\mathbf{m}) = q_0(\mathbf{m}_0)| det \frac{\partial \mathbf{F}_{\Uptheta^N}}{\partial \mathbf{m}_{0}} |$.
	\caption{Normalizing flows}
\end{algorithm}

As described in section \ref{ssec:advi}, in ADVI we apply two transforms: one transforms constrained variables to unconstrained variables and the other transforms a Gaussian distribution to a standard Gaussian distribution. The first transform is fixed, while the parameters of the latter transform (the mean and covariance matrix of the Gaussian distribution) are optimized such that they maximise the ELBO between the Gaussian distribution and the posterior pdf in the unconstrained space (equation \ref{eq:argmaxELBO_advi_standardized}). Thus, ADVI is in fact a single normalizing flow.

\begin{figure}
	\includegraphics[width=.9\linewidth]{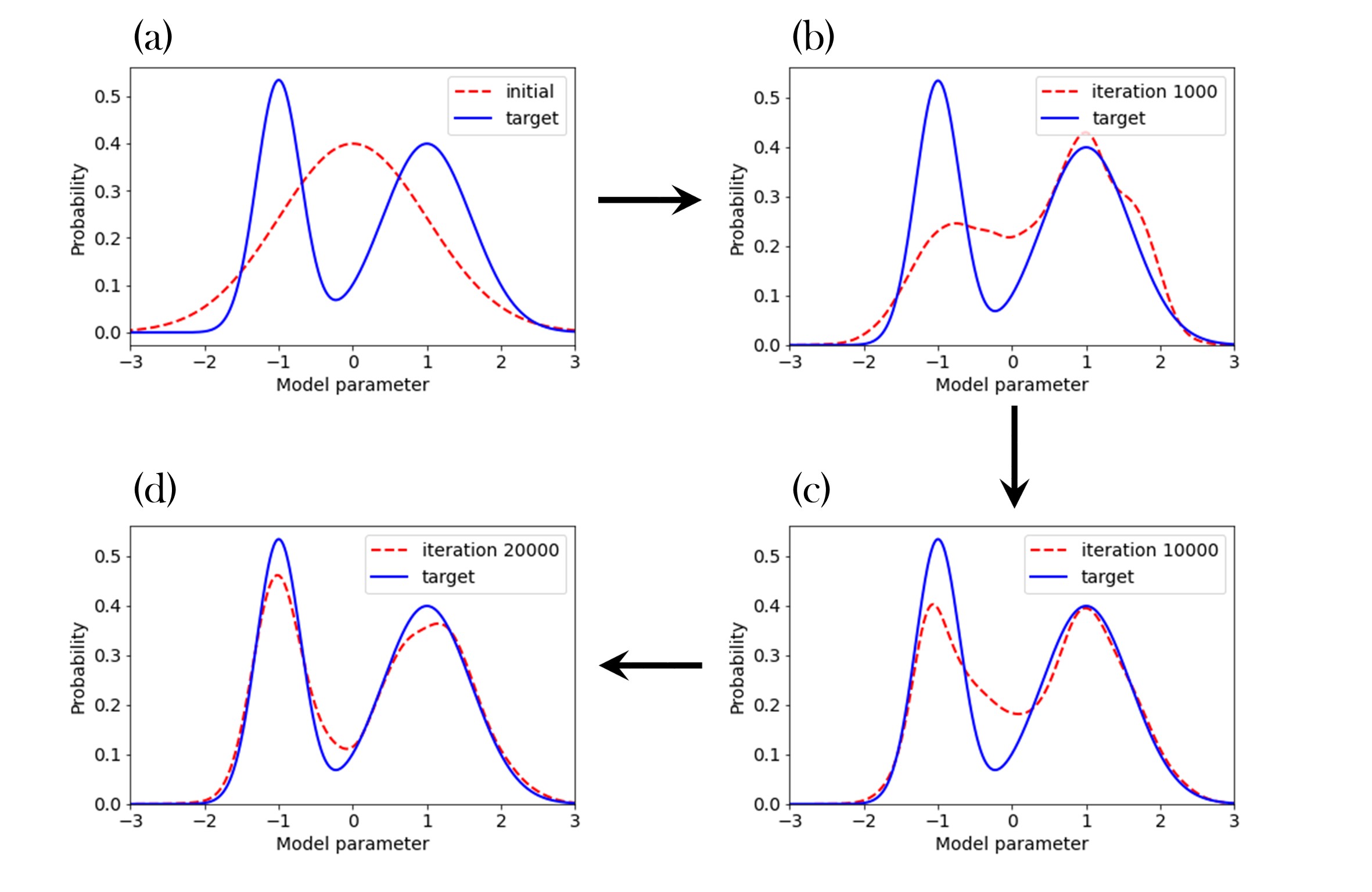}
	\caption{A 1D example using normalizing flows. \textbf{(a)} the true or target pdf (blue line) and the initial pdf (red dashed line). \textbf{(b)}, \textbf{(c)} and \textbf{(d)} show the estimated pdfs after 1000, 10000 and 20000 iterations of gradient ascent respectively.}
	\label{fig:nfvi_workflow}
\end{figure}

To construct a flexible normalizing flow for practical applications, several conditions are required: the flows must be 1) invertible, and 2) expressive enough to represent any desired pdf; 3) The forward and backward transform and associated Jacobian determinant must be able to be computed efficiently. A simple example of such flows is planar flow:
\begin{equation}
	\mathbf{m}_{i+1} = \mathbf{m}_{i} + \mathbf{u}h(\mathbf{w}^{\mathrm{T}}\mathbf{m}_{i}+b)
	\label{eq:planar_flow}
\end{equation}
where $\mathbf{u}$ and $\mathbf{w}$ are vectors, $b$ is a scalar and $h$ is a smooth  function \citep{rezende2015variational}: $h(x)=tanh(x)$ is usually used. The determinant of the Jacobian matrix of this flow is:
\begin{equation}
	det \frac{\partial \mathbf{m}_{i+1}}{\partial \mathbf{m}_{i}} = 1 + \mathbf{u}^{\mathrm{T}}h'(\mathbf{w}^{\mathrm{T}}\mathbf{m}_{i}+b)\mathbf{w}
\end{equation}
The planar flow essentially expands or contracts a distribution along the direction perpendicular to the hypeplane $\mathbf{w}^{\mathrm{T}}\mathbf{m}_{i}+b=0$, and can be interpreted as a neural network with one hidden layer and one hidden unit \citep{kingma2018glow}.

Figure \ref{fig:nfvi_workflow} shows a 1D example using planar flows. The true target (posterior) pdf is a multimodal distribution (blue line in Figure \ref{fig:nfvi_workflow}a). We use a standard normal distribution as the initial distribution and a normalizing flows model with 10 planar flows in equation \ref{eq:planar_flow}. The model parameters are updated using gradient ascent with gradients calculated using equation \ref{eq:gradients_normalizing_flow}. Figure \ref{fig:nfvi_workflow}b, c and d show the estimated pdfs after 1000, 10000 and 20000 iterations respectively. The initial pdf is gradually reshaped and finally produces an accurate approximation to the true pdf.

\begin{figure}
	\includegraphics[width=.9\linewidth]{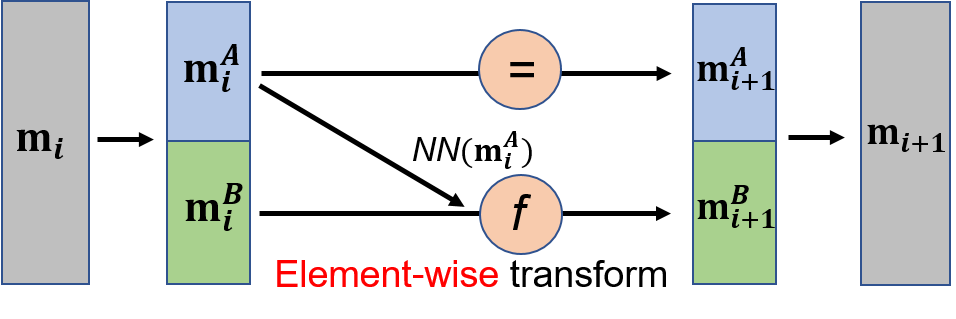}
	\caption{An illustration of a coupling flow. The input vector $\mathbf{m}_{i}$ is first divided into two halves $\mathbf{m}_{i}^{A}$ and $\mathbf{m}_{i}^{B}$. The former half $\mathbf{m}_{i}^{A}$ is copied as the first half $\mathbf{m}_{i+1}^{A}$ of the output. It is also input to a neural network which outputs hyperparameters of an element-wise function $f$; this function $f$ transforms the second half $\mathbf{m}_{i}^{B}$ to $\mathbf{m}_{i+1}^{B}$ which together with $\mathbf{m}_{i+1}^{A}$ forms the output $\mathbf{m}_{i+1}$.}
	\label{fig:coupling_flow}
\end{figure}

It becomes difficult to use planar flows to approximate complex posterior distributions in high dimensionality in the sense that each planar flow is a neural network with the necessarily simple structure of only one hidden layer and one hidden unit. To improve the expressiveness of the sequence of flows in equation \ref{eq:multflows}, many different forms of flow have been proposed \cite[an overview is given in][]{zhao2020bayesian}. One such flow is constructed from an invertible neural network with a specific design to enable invertibility and fast computation of the Jacobian determinant \citep{dinh2016density, kingma2018glow, behrmann2019invertible, greydanus2019hamiltonian}. In this study we describe an invertible neural network called a \textit{coupling flow} \citep{dinh2016density}. In a coupling flow an input vector $\mathbf{m}_{i}$ is divided into two half vectors $\mathbf{m}_{i}^{A}$ and $\mathbf{m}_{i}^{B}$, and the output halves $\mathbf{m}_{i+1}^{A}$ and $\mathbf{m}_{i+1}^{B}$ are obtained using (see Figure \ref{fig:coupling_flow}):
\begin{equation}
	\begin{aligned}
	\mathbf{m}_{i+1}^{A} &= \mathbf{m}_{i}^{A} \\
	\mathbf{m}_{i+1}^{B} &= f(\mathbf{m}_{i}^{B};NN(\mathbf{m}_{i}^{A}))
	\end{aligned}
\end{equation}
where $NN(\mathbf{m}_{i}^{A})$ represents any neural network which takes $\mathbf{m}_{i}^{A}$ as input, and $f$ transforms $\mathbf{m}_{i}^{B}$ to $\mathbf{m}_{i+1}^{B}$ and is an invertible, element-wise bijection function parameterized by the output of the neural network. The two halves $\mathbf{m}_{i+1}^{A}$ and $\mathbf{m}_{i+1}^{B}$ are combined to obtain the output vector $\mathbf{m}_{i+1}$. This transform can be easily inverted
\begin{equation}
	\begin{aligned}
		\mathbf{m}_{i}^{A} &= \mathbf{m}_{i+1}^{A} \\
		\mathbf{m}_{i}^{B} &= f^{-1}(\mathbf{m}_{i+1}^{B};NN(\mathbf{m}_{i}^{A}))
	\end{aligned}
\end{equation}
and the Jacobian determinant of the transform can also be calculated using
\begin{equation} 
	det \frac{\partial \mathbf{m}_{i+1}}{\partial \mathbf{m}_{i}} = det 
	\begin{bmatrix}
		\frac{\partial \mathbf{m}_{i+1}^{A}}{\partial \mathbf{m}_{i}^{A}} 
		& \frac{\partial \mathbf{m}_{i+1}^{A}}{\partial \mathbf{m}_{i}^{B}} \\
		\frac{\partial \mathbf{m}_{i+1}^{B}}{\partial \mathbf{m}_{i}^{A}} 
		& \frac{\partial \mathbf{m}_{i+1}^{B}}{\partial \mathbf{m}_{i}^{B}}
	\end{bmatrix} 
  = det \frac{\partial \mathbf{m}_{i+1}^{B}}{\partial \mathbf{m}_{i}^{B}}
\end{equation}
where we have used the fact that $\frac{\partial \mathbf{m}_{i+1}^{A}}{\partial \mathbf{m}_{i}^{A}}=\mathbf{I}$ and $\frac{\partial \mathbf{m}_{i+1}^{A}}{\partial \mathbf{m}_{i}^{B}} = \mathbf{0}$. Since the function $f$ is an element-wise function, the matrix $\frac{\partial \mathbf{m}_{i+1}^{B}}{\partial \mathbf{m}_{i}^{B}}$ is a diagonal matrix whose determinant can be calculated efficiently.

In practice a series of successive coupling flows are used to improve the expressiveness of the overall transform. To ensure that all elements in the input vector $\mathbf{m}_{i}$ are modified, the locations of the two outputs $\mathbf{m}_{i+1}^{A}$ and $\mathbf{m}_{i+1}^{b}$ are exchanged before feeding into the next flow. The function $f$ can be any element-wise functions which is invertible and differentiable, and many choices of $f$ can be used in practice \citep{dinh2014nice,  dinh2016density, kingma2018glow, durkan2019cubic, durkan2019neural, de2020block}. 

Note that instead of coupling flows, other designs of invertible neural networks can also be used in normalizing flows, for example invertible residual networks \citep{behrmann2019invertible}, neural ordinary differential equations \citep{chen2018neural, grathwohl2018ffjord} or Hamiltonian neural networks \citep{greydanus2019hamiltonian}. Further research that performs fair comparisons between these networks would be a useful contribution.
 
\subsection{Stein variational gradient descent}
In normalizing flows a series of analytical invertible transforms are applied to a simple initial distribution and are optimized by maximizing the ELBO between the final transformed distribution and the posterior distribution. In practice construction of effective analytic transforms can be a difficult task. Instead of using analytical transforms, Stein variational gradient descent (SVGD) uses a smooth transform whose analytical form remains unknown, and successively applies it to an initial probability distribution represented by a set of parameter-space samples which are referred to as particles \citep{liu2016stein}. Similarly to normalizing flows, the transforms are optimized to minimize the KL-divergence between the transformed distribution and the posterior distribution so that the final set of particles are distributed according to the posterior. 

In SVGD a smooth transform is used:
\begin{equation} 
T(\mathbf{m}) = \mathbf{m} + \epsilon\bm{\upphi}(\mathbf{m})
\label{eq:svgd_transform}
\end{equation}
where $\mathbf{m}$ is a $d$-dimensional vector, $\bm{\upphi}(\mathbf{m}) = [\phi_{1}, ... , \phi_{d}]$ is a smooth $d$-dimensional vector function which describes the perturbation direction and $\epsilon$ is the magnitude of the perturbation. When $\epsilon$ is sufficiently small, the transform $T$ is invertible as the Jacobian matrix is close to an identity matrix. Define $q$ as an initial distribution and $q_{T}$ as the transformed distribution, the gradient of KL-divergence between $q_{T}$ and the posterior pdf $p$ with respect to $\epsilon$ can be calculated as:
\begin{equation}
	\nabla_{\epsilon} \mathrm{KL}[q_{T}||p] |_{\epsilon=0} = - \mathrm{E}_{q}[trace (\mathcal{A}_{p} \bm{\upphi(\mathbf{m})})]
	\label{eq:grad_svgd}
\end{equation}
where $\mathcal{A}_{p}$ is the Stein operator such that $\mathcal{A}_{p} \bm{\upphi}(\mathbf{m}) = \nabla_{\mathbf{m}} \mathrm{log} p(\mathbf{m}|\mathbf{d}_{\mathrm{obs}}) \bm{\upphi} (\mathbf{m})^{\mathrm{T}} + \nabla_{ \mathbf{m} } \bm{\upphi} ( \mathbf{m} )$ \citep{liu2016stein}. This implies that by maximizing the right-hand side expectation we obtain the steepest direction of change in the KL-divergence; the KL-divergence can therefore be minimized by iteratively stepping a small distance in that direction.

The optimal $\bm{\upphi}^{*}$ which maximizes the expectation in equation \ref{eq:grad_svgd} can be found using kernels. Assume $x, y \in X$ and define a mapping $\varphi$ from $X$ to an inner product space; a \textit{kernel} is a function which satisfies $k(x,y) = \langle \varphi(x),\varphi(y) \rangle$ where $\langle,\rangle$ represents an inner product \citep{gretton2013introduction}. The optimal $\bm{\upphi}^{*}$ is found to be: 
\begin{equation}
	\bm{\upphi}^{*} \propto \mathrm{E}_{\{\mathbf{m'} \sim q\}} [\mathcal{A}_{p} k(\mathbf{m'},\mathbf{m})]
	\label{eq:phi_qp}
\end{equation}
where $k(\mathbf{m'},\mathbf{m})$ is a kernel function \citep{liu2016stein}.

Given equation \ref{eq:phi_qp}, the KL-divergence can be minimized by iteratively applying the transform in equation \ref{eq:svgd_transform} with the optimal $\bm{\upphi}^{*}$ to an initial distribution. For example, define an initial distribution $q_{0}$, and apply the transform $T_{0}(\mathbf{m}) = \mathbf{m} + \epsilon\bm{\upphi}_{0}^{*}(\mathbf{m})$ where $\bm{\upphi}_{0}^{*}(\mathbf{m})$ is given in equation \ref{eq:phi_qp}. This produces a new distribution $q_{[T_{0}]}$ which decreases the KL-divergence. This process is iterated to obtain an approximation to the posterior:
\begin{equation}
	\begin{aligned}
	T_{l}(\mathbf{m}) &= \mathbf{m} + \epsilon_{l}\bm{\upphi}_{l}^{*}(\mathbf{m}) \\
	q_{l+1} &= q_{l[T_{l}]}
	\end{aligned}
\end{equation}
where the subscript $l$ denotes the $l^{th}$ iteration. If the perturbation magnitude $\{\epsilon_{l}\}$ is sufficiently small, that is, the transform is invertible at each iteration, the process should eventually converge to the posterior distribution.

In practice since the posterior distribution $p(\mathbf{m}|\mathbf{d}_{\mathrm{obs}})$ and its gradient with respect to model $\mathbf{m}$ are analytically unknown  (and are needed in the definition of the Stein operator $\mathcal{A}_{p}$), we cannot obtain the analytical form of the optimal $\bm{\upphi}^{*}$ and consequently the optimal transform $T$. Fortunately the unnormalized posterior distribution can usually be estimated at a set of samples $\{\mathbf{m}_{1},...,\mathbf{m}_{n}\}$ distributed approximately according to the posterior pdf, which enables us to estimate the optimal $\bm{\upphi}^{*}$ numerically, for example using the mean value taken over the set of samples. Thus in SVGD we use a set of samples $\{ \mathbf{m}_{i}\}$ (the particles) to represent the approximate distribution $q$ and to approximate the optimal $\bm{\upphi}^{*}$ using the particles mean. Each particle is then updated using the estimated $\bm{\upphi}^{*}$. This results in Algorithm \ref{al:svgd}.

\begin{algorithm}[H] \label{al:svgd}
	\SetAlgoNoLine
	
	\KwIn{An initial pdf $q_{0}$; the posterior pdf $p(\mathbf{m}|\mathbf{d}_{\mathrm{obs}})$ which can be estimated up to a normalising constant for any particular value of $\mathbf{m}$.}
	\KwOut{A set of particles $\{ \mathbf{m}_{i}\}$ whose density approximates the posterior pdf.}
	Draw a set of particles $\{ \mathbf{m}_{i}^{0} \}_{i=1}^{n}$ from $q_{0}$; \\
	\For{$l \gets 1$ to $N$}{
		\begin{equation}
			\begin{aligned}
				\bm{\upphi}^{*}_{ q_{l}, p} (\mathbf{m}) &= \frac{1}{n} \sum_{j=1}^{n} \left[ k(\mathbf{m}_{j}^{l} , \mathbf{m}) \nabla_{\mathbf{m}_{j}^{l}} \mathrm{log} p(\mathbf{m}_{j}^{l}|\mathbf{d}_{\mathrm{obs}}) + \nabla_{\mathbf{m}_{j}^{l}} k(\mathbf{m}_{j}^{l}, \mathbf{m}) \right] \\
				\mathbf{m}_{i}^{l+1} &= \mathbf{m}_{i}^{l} + \epsilon^{l} \bm{\upphi}^{*}_{ q_{l}, p} (\mathbf{m}_{i}^{l})
			\end{aligned}
		\label{eq:phi_mean}
		\end{equation}
		where $\epsilon^{l}$ is the step size at the $l^{th}$ iteration.
	}
	\caption{Stein Variational gradient descent (SVGD)}
\end{algorithm}

Since SVGD uses particles to approximate the posterior pdf, the accuracy of the method increases with the number of particles. For sufficiently small $\{ \epsilon_{l}\}$ the method converges to the posterior distribution asymptotically with the number of particles. On the other hand for one single particle the method reduces to a standard gradient ascent method towards the model with maximum a posterior (MAP) pdf value if the gradient $\nabla_{\mathbf{m}} k(\mathbf{m}, \mathbf{m})$ vanishes (which is valid for many kernels, including the radial basis function kernel described below). This suggests that in practice we can start from a small number of particles and gradually increase the particles to produce more accurate results. In comparison to other particle-based methods, for example, sequential Monte Carlo \citep{smith2013sequential}, SVGD requires fewer samples to achieve the same accuracy which makes it more efficient \citep{liu2016stein}. It is also important to notice that sequential Monte Carlo is a stochastic sampling method, whereas SVGD is a deterministic sampling method.

The kernel function enables interactions between particles and strongly affects the efficiency of the method. We first describe a simple, commonly-used kernel function, the \textit{radial basis function} (RBF)
\begin{equation}
	k(\mathbf{m},\mathbf{m}') = \mathrm{exp} [- \frac{\Vert \mathbf{m}-\mathbf{m}' \Vert^{2}}{2\sigma^{2}}]
	\label{eq:rbf}
\end{equation}
where $\sigma$ is a scale factor which intuitively controls the interaction intensity between pairs of particles based on their distance apart.

With a RBF kernel the first term of $\bm{\upphi}^{*}$ in equation \ref{eq:phi_mean} is the weighted average of gradients of the posterior pdf from all particles, in which the weights are determined by particle distances and the scale factor $\sigma$. This term drives particles towards a local high probability area. The second term of $\bm{\upphi}^{*}$ becomes $\sum_{j} \frac{\mathbf{m}-\mathbf{m}_{j}}{\sigma^{2}}  k(\mathbf{m}_{j}, \mathbf{m})$ which pushes the particle $\mathbf{m}$ away from its neighbouring particles with high kernel values. The two terms therefore contribute in different ways to arrange particles to represent the posterior pdf: the first term drives particles towards a local high probability area, whereas the second term acts as a \textit{repulsive force} which prevents particles from collapsing to a single mode. These terms balance such that the limiting distribution is the posterior pdf provided that the derivative of the kernels (the second term of $\bm{\upphi}^{*}$ in equation \ref{eq:phi_mean}) does not vanish. Note that when $\sigma \to 0$, the method becomes independent gradient ascent for each particle as the kernel value and its derivative between any two particles vanish.

\begin{figure}
	\includegraphics[width=.9\linewidth]{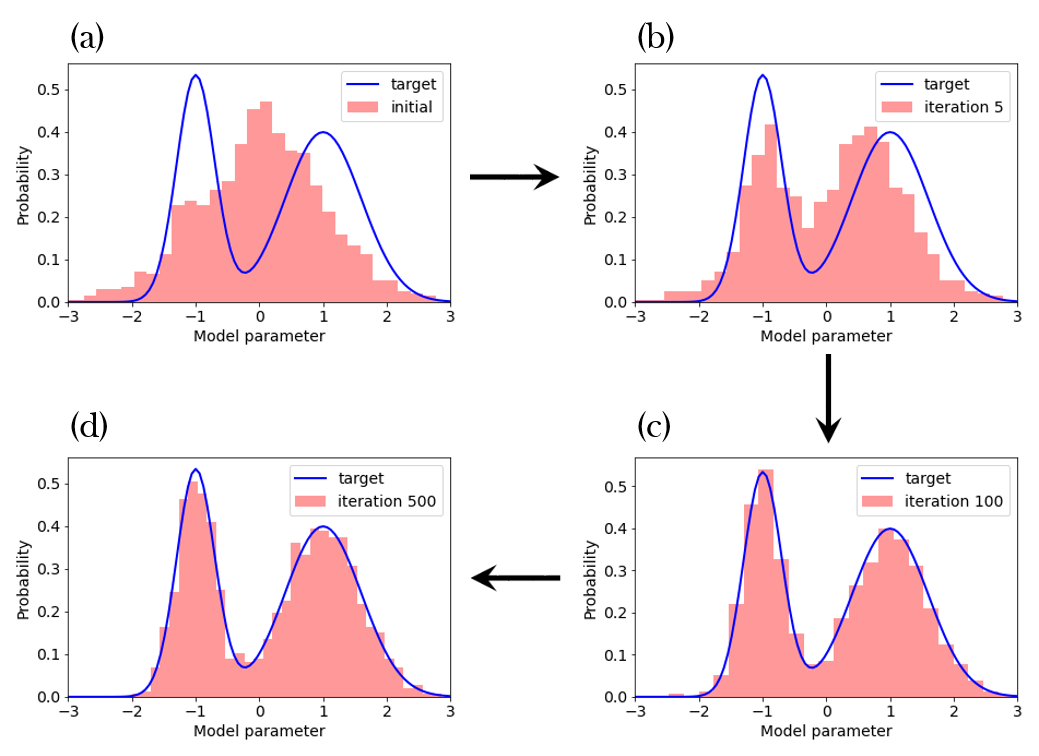}
	\caption{A 1D example using SVGD. \textbf{(a)} The true pdf (blue line) and the histogram of 1000 initial particles (red) which are generated from a Gaussian distribution. \textbf{(b)}, \textbf{(c)} and \textbf{(d)} show the histograms of the particles after 5, 100 and 500 iterations respectively.}
	\label{fig:svgd_workflow}
\end{figure}

Figure \ref{fig:svgd_workflow} shows a 1D example using SVGD with a RBF kernel. The target pdf is the same multimodal distribution in Figure \ref{fig:nfvi_workflow}a (blue line). We start from 1,000 particles generated from a standard Normal distribution (red histograms in Figure \ref{fig:svgd_workflow}a) and iteratively update them using equation \ref{eq:phi_mean}. Figure \ref{fig:svgd_workflow}b, c and d show the histograms of those particles after 5, 100 and 500 iterations respectively. After 100 iterations the method has almost converged to the true distribution. This example shows that SVGD arranges particles to represent the posterior pdf optimally.

Kernel functions can be generalized to matrix forms and used in SVGD instead of scalar kernel functions. By doing this one can inject information about correlations between the different parameters in $\mathbf{m}$ into the method. Assuming a matrix-valued kernel function $\mathbf{K}$, the $\bm{\upphi}^{*}$ in equation \ref{eq:phi_mean} becomes:
\begin{equation}
	\bm{\upphi}^{*}_{ q_{l}, p} (\mathbf{m}) = \frac{1}{n} \sum_{j=1}^{n} \left[ \mathbf{K}(\mathbf{m}_{j}^{l} , \mathbf{m}) \nabla_{\mathbf{m}_{j}^{l}} \mathrm{log} p(\mathbf{m}_{j}^{l}|\mathbf{d}_{\mathrm{obs}}) + \mathbf{K}(\mathbf{m}_{j}^{l}, \mathbf{m}) \nabla_{\mathbf{m}_{j}^{l}}  \right]
	\label{eq:phi_matrix_kernel}
\end{equation}
where $\mathbf{K}(\mathbf{m}_{j}^{l}, \mathbf{m}) \nabla_{\mathbf{m}_{j}^{l}}$ represents matrix multiplication \citep{wang2019stein}. A possible choice of a matrix-valued kernel is:
\begin{equation}
	\mathbf{K}(\mathbf{m'},\mathbf{m}) = \mathbf{Q}^{-1} \mathrm{exp}(-\frac{1}{2\sigma^{2}}||\mathbf{m}-\mathbf{m'}||^{2}_{\mathbf{Q}})
\end{equation}
where $\mathbf{Q}$ is a positive definite matrix,  $||\mathbf{m}-\mathbf{m'}||^{2}_{\mathbf{Q}}=(\mathbf{m}-\mathbf{m'})^{\mathrm{T}}\mathbf{Q}(\mathbf{m}-\mathbf{m'})$ and $\sigma$ is a scaling parameter. This kernel is essentially a RBF kernel with a preconditioning matrix $\mathbf{Q}$. \cite{wang2019stein} showed that by setting $\mathbf{Q}$ to be the average Hessian matrix of all particles, the method converges faster than a scalar RBF kernel. Other choice of $\mathbf{Q}$ include the inverse of the covariance matrix calculated from the particles, or the inverse of the diagonal covariance (variance) matrix.

\section{Applications}

\subsection{Petrophysical inversion}

In this section, we present an application of variational inference using the mean field approximation for joint estimation of geological facies $\bm{\kappa}$ and petrophysical rock properties $\bm{\gamma}$ using information derived from seismic data that are referred to as seismic attributes $\mathbf{d}$. These attributes represent elastic rock properties that may directly be inverted from seismic waveform data such as P- and S-wave impedances ($I_{p}$ and $I_{s}$), velocities ($V_{p}$ and $V_{s}$) and their ratios ($V_{p}⁄V_{s}$). Examples of petrophysical properties $\bm{\gamma}$ of interest include porosity ($\varrho$), clay volume ($V_{cl}$) and water saturations ($S_{w}$). Geological facies refer to well-defined discrete classes of lithology and fluid types that are in principle distinctively distinguishable from seismic and well data. Petrophysical rock properties and facies together represent the unknown model parameters, i.e. $\mathbf{m} \equiv \{\bm{\gamma},\bm{\kappa}\}$.

Estimation of rock properties from seismic attributes is a non-unique inverse problem. Usually the solution can be better constrained if the spatial distribution of geological facies is known \citep{nawaz2020variational}. For this reason, we would like to infer the rock properties $\bm{\gamma}$ and facies $\bm{\kappa}$ jointly from the seismic attributes $\mathbf{d}$ along with their associated uncertainty of prediction. In terms of probability theory, we seek the posterior distribution $p(\bm{\gamma},\bm{\kappa}|\mathbf{d})$ of unknown model parameters $\bm{\gamma}$ and $\bm{\kappa}$ conditioned on the attribute data $\mathbf{d}$. According to Bayes’ theorem
\begin{equation}
	p ( \bm{\gamma}, \bm{\kappa} | \mathbf{d} ) =\frac{p( \mathbf{d} | \bm{\gamma}, \bm  \kappa ) p( \bm{\gamma} | \bm  \kappa ) p(\bm \kappa) }{p(\mathbf{d}) }
	\label{eq:petro_bayes}
\end{equation}
where $p(\bm \kappa)$ represents the prior distribution of facies $\bm \kappa$, $p(\bm \gamma|\bm \kappa)$ represents the conditional prior distribution of the petrophysical properties $\bm \gamma$ given the facies $\bm \kappa$, $p(\mathbf{d}|\bm \gamma,\bm \kappa)$ represents the data likelihood given $\bm \gamma$ and $\bm \kappa$, and $p(\mathbf{d})$ represents the marginal distribution of data $\mathbf{d}$. Since the data $\mathbf{d}$ is observed,  $p(\mathbf{d})$ is a constant that normalizes the posterior distribution.

The joint distribution $p(\bm \kappa)$ of facies is modelled as a Markov random field (MRF) with pair-wise correlations, which according to equation (\ref{eq:factorized_mf}) is given by
\begin{equation}
	p(\bm \kappa)= \frac{1}{Z} \prod_{i,j} \psi_{ij} (\kappa_{i},\kappa_{j})
\end{equation}
where the potential functions $\psi_{ij} (\kappa_{i},\kappa_{j})$ define how probable it is to find the facies $\kappa_{i}$ and $\kappa_{j}$ in locations $i$ and $j$ in the model, and may be estimated by scanning a training image \citep{mariethoz2014multiple} and building histograms for various combinations of facies over various neighbouring locations. 

The conditional prior distribution $p(\bm \gamma|\bm \kappa)$ of $\bm \gamma$ given $\bm \kappa$ is usually modelled using well logs that have been up-scaled to the dominant seismic wavelength \citep{grana2010probabilistic}, and the likelihood $p(\mathbf{d}|\bm \gamma,\bm \kappa)$ is usually modelled using rock physics models \citep{grana2010probabilistic,grana2018joint} calibrated with the well data and local geological information. We adopt a different approach: we model both the conditional prior $p(\bm \gamma|\bm \kappa)$ and the likelihood $p(\mathbf{d}|\bm \gamma,\bm \kappa)$ jointly using up-scaled well-logs in the form of a joint distribution $p(\mathbf{d},\bm \gamma|\bm \kappa,\Theta)$ of elastic attributes $\mathbf{d}$ and petrophysical properties $\bm \gamma$ given the facies $\bm \kappa$, parameterized by $\Theta$. Equation (\ref{eq:petro_bayes}) may then be written as
\begin{equation}	
	p( \bm \gamma, \bm \kappa | \mathbf{d}, \Theta) = \frac{p( \mathbf{d}, \bm \gamma | \bm \kappa , \Theta) p (\bm \kappa) }{p(\mathbf{d}|\Theta) }
\end{equation}
Thus, we do not use a rock physics model explicitly. However, if only limited well data is available, rock physics models may be used to augment the existing data. 

We use a Gaussian mixture (GM) distribution to model $p(\mathbf{d},\bm \gamma|\bm \kappa,\Theta)$ that is defined as a linear combination of Gaussian kernels, usually referred to as the components of the mixture distribution. A GM distribution is a universal approximator of pdfs: given a sufficient number of Gaussian kernels with appropriate parameters, a GM can approximate any complex pdf to any desired non-zero accuracy \citep{mclachlan2004finite}. The GM distribution for rock properties $d_{i}$ and $\gamma_{i}$ given facies $\kappa_{i}$ at a location $i$ may be expressed as
\begin{equation}
	p ( d_{i},r_{i} | \kappa_{i}, \Theta ) = \sum _{t=1}^{T_{k}} \alpha_{t,k} g_{t,k} \left( d_{i},r_{i} \right) ,~  \forall i
\end{equation}
where $T_{k}$ is the number of mixture components (which may be different for each facies $k$), $\alpha_{t,k}$ is the component weight, and $g_{t,k} (d_{i},\gamma_{i})$ is the Gaussian kernel for the $t^{th}$ component given by
\begin{equation}
	g_{t,k} ( d_{i},r_{i} ) =
	N \left( \left[ 
	~\begin{matrix}
		\bm{\mu} _{d}\\
		\bm{\mu} _{r}\\
	\end{matrix}
	~ \right] _{t,k}, \left[ 
	\begin{matrix}
		\bm{\Sigma}  _{d,d}  &    \bm{\Sigma}  _{d,r}\\
		\bm{\Sigma}  _{r,d}  &    \bm{\Sigma}  _{r,r}\\
	\end{matrix}
	\right] _{t,k} \right) ,~  \forall i
\end{equation}
where $N$ represents the pdf of the Normal distribution, $\bm{\mu}$ and $\bm{\Sigma}$ are means and block covariance matrices of the kernel with subscripts indicating the components with respect to the data $\mathbf{d}$ and the petrophysical properties $\bm \gamma$.

\begin{figure}
	\includegraphics[width=1.0\linewidth]{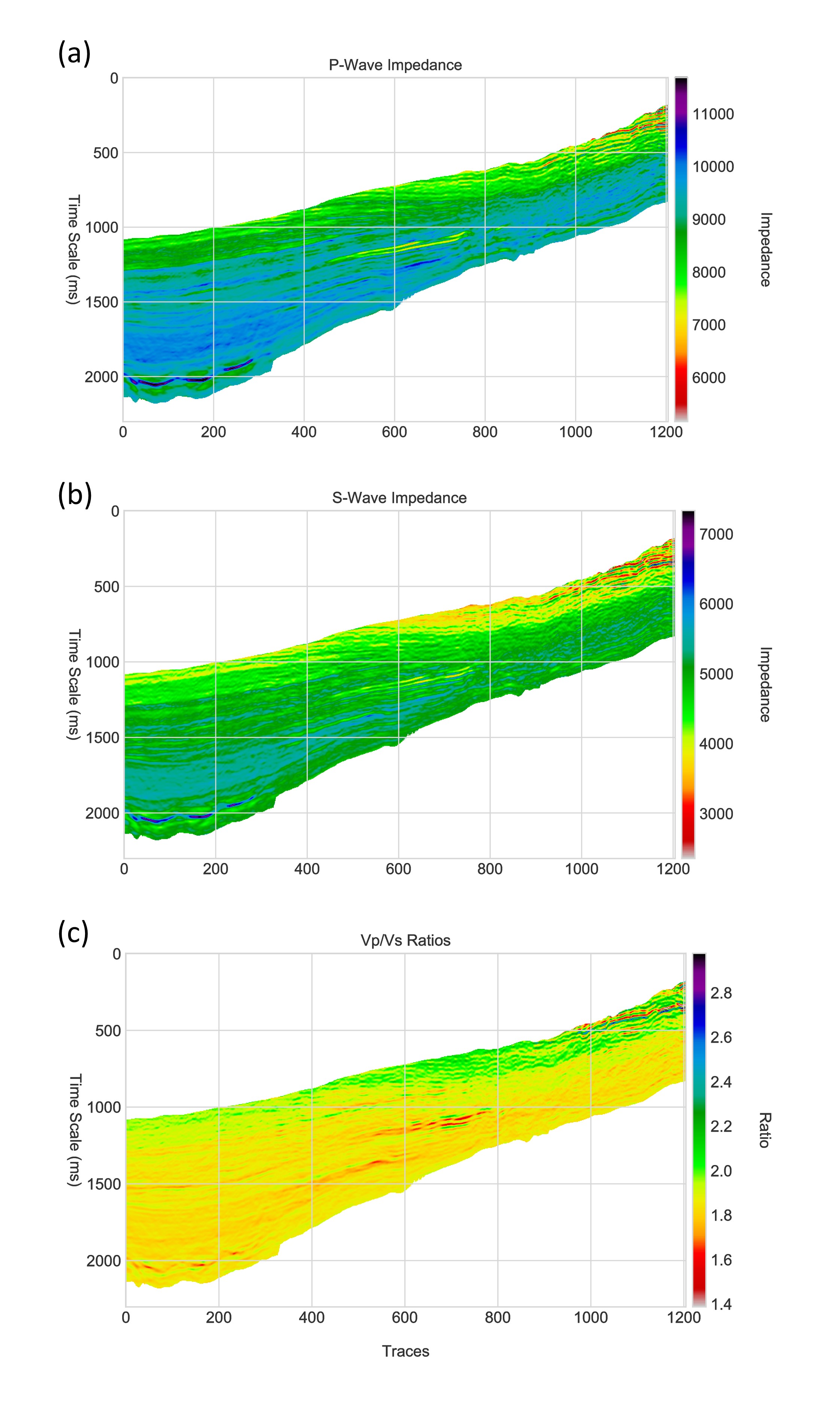}
	\caption{Seismic attributes \textbf{(a)} P-wave impedance, \textbf{(b)} S-wave impedance, and \textbf{(c)} Vp/Vs ratio, derived from a selected 2D section of waveform seismic data using a deterministic inversion method. These attributes are used as inputs to our method for the joint inversion of geological facies and petrophysical rock properties.}
	\label{fig:petro_seismic_attributes}
\end{figure}

Since the joint conditional distribution $p(\mathbf{d},\bm \gamma|\bm \kappa,\Theta)$ of seismic attributes $\mathbf{d}$ and rock properties $\bm \gamma$ given facies $\bm \kappa$ (and the distribution parameters $\Theta$) is modelled as a GM distribution, and the prior distribution of facies $p(\bm \kappa)$ is modelled as a MRF, the overall model of the joint distribution $p(\mathbf{d},\bm \gamma,\bm \kappa|\Theta)$ of the data $\mathbf{d}$ and unknown model parameters $\bm \gamma$ and $\bm \kappa$ represents a Gaussian mixture - Markov random field (GM-MRF) given by
\begin{equation}
	p\left( \bm \gamma, \bm \kappa | \mathbf{d}, \Theta  \right) 
	=\frac{p \left( \mathbf{d},\bm \gamma, \bm \kappa  |  \Theta  \right) }{p \left( \mathbf{d} |  \Theta  \right) } 
	\cong \frac{1}{Z'}~ \prod_{i}^{}p \left( d_{i},r_{i} | \kappa _{i}, \Theta  \right) ~ \prod_{ \left( i,j \right) }^{} \psi _{ij} \left(  \kappa _{i}, \kappa _{j} \right) 
	\label{eq:factorized_bayes}
\end{equation}
where $p(\mathbf{d}|\Theta)$ has been absorbed in the normalization constant $Z'$ on the right-hand side. This demonstrates that although we only assumed that the prior distribution $p(\bm \kappa)$ on facies $\bm \kappa$ is a MRF, the posterior distribution $p(\bm \gamma,\bm \kappa|\mathbf{d},\Theta)$ and the joint distribution $p(\mathbf{d},\bm \gamma,\bm \kappa|\Theta)$ then also turn out to be MRFs. This is a consequence of the conditional independence assumption on the rock properties $\mathbf{d}$ and $\bm \gamma$ that is invoked in the mean-field approximation. The factorization of the posterior distribution in equation (\ref{eq:factorized_bayes}) is instrumental in making inference tractable for real-scale models using, for example, the EM method of inference as described in section \ref{ssec:mfvi}.

\begin{figure}
	\includegraphics[width=1.0\linewidth]{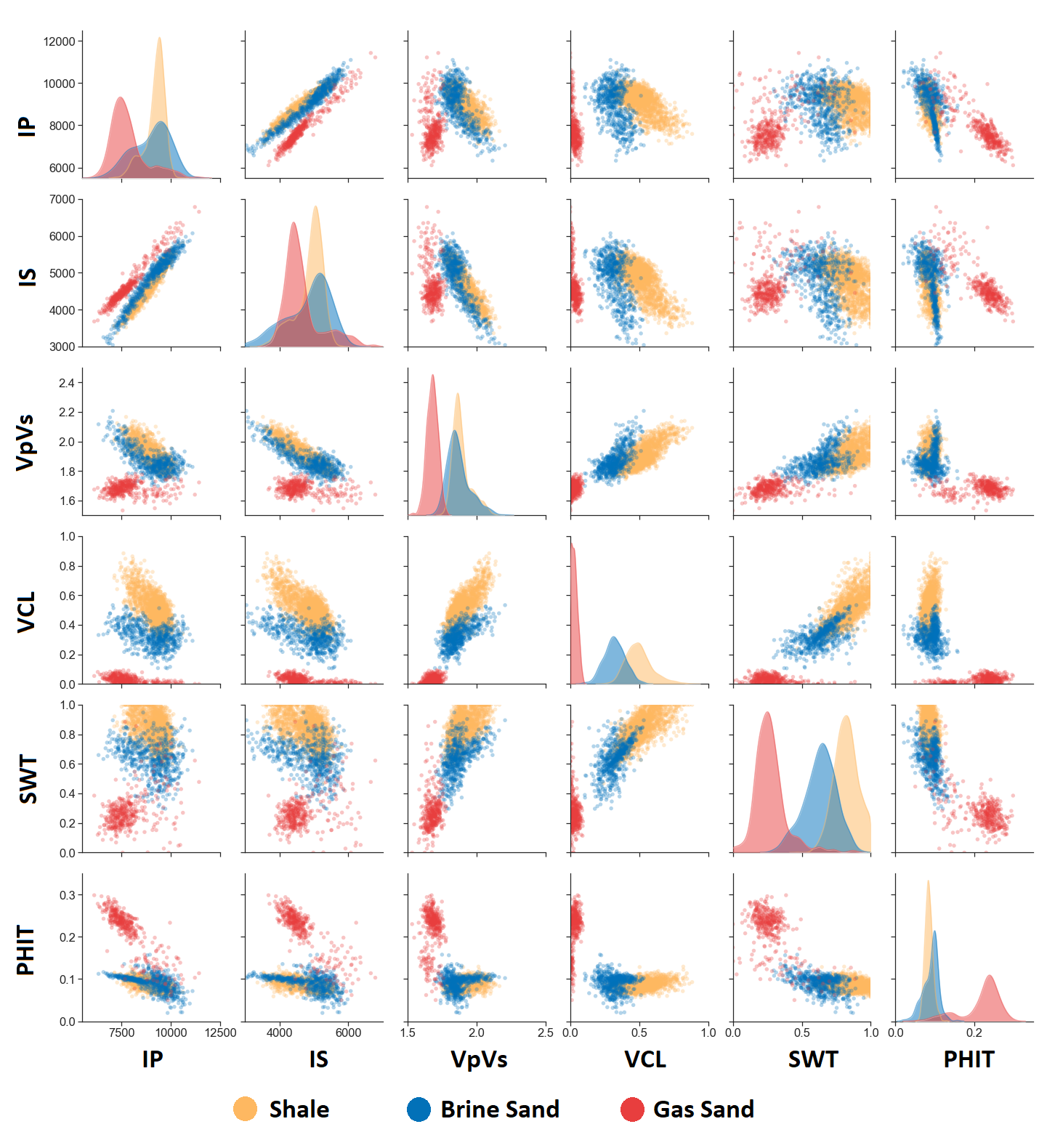}
	\caption{Seismic attributes P-wave impedance (IP), S-wave impedance (IS), and Vp/Vs ratio (VpVs), and petrophysical properties clay volume (VCL), water saturations (SWT) and porosity (PHIT) of three geological facies: Shale, Brine Sand and Gas Sand obtained from the well log data.}
	\label{fig:petro_matrix_plot}
\end{figure}

\begin{figure}
	\includegraphics[width=1.0\linewidth]{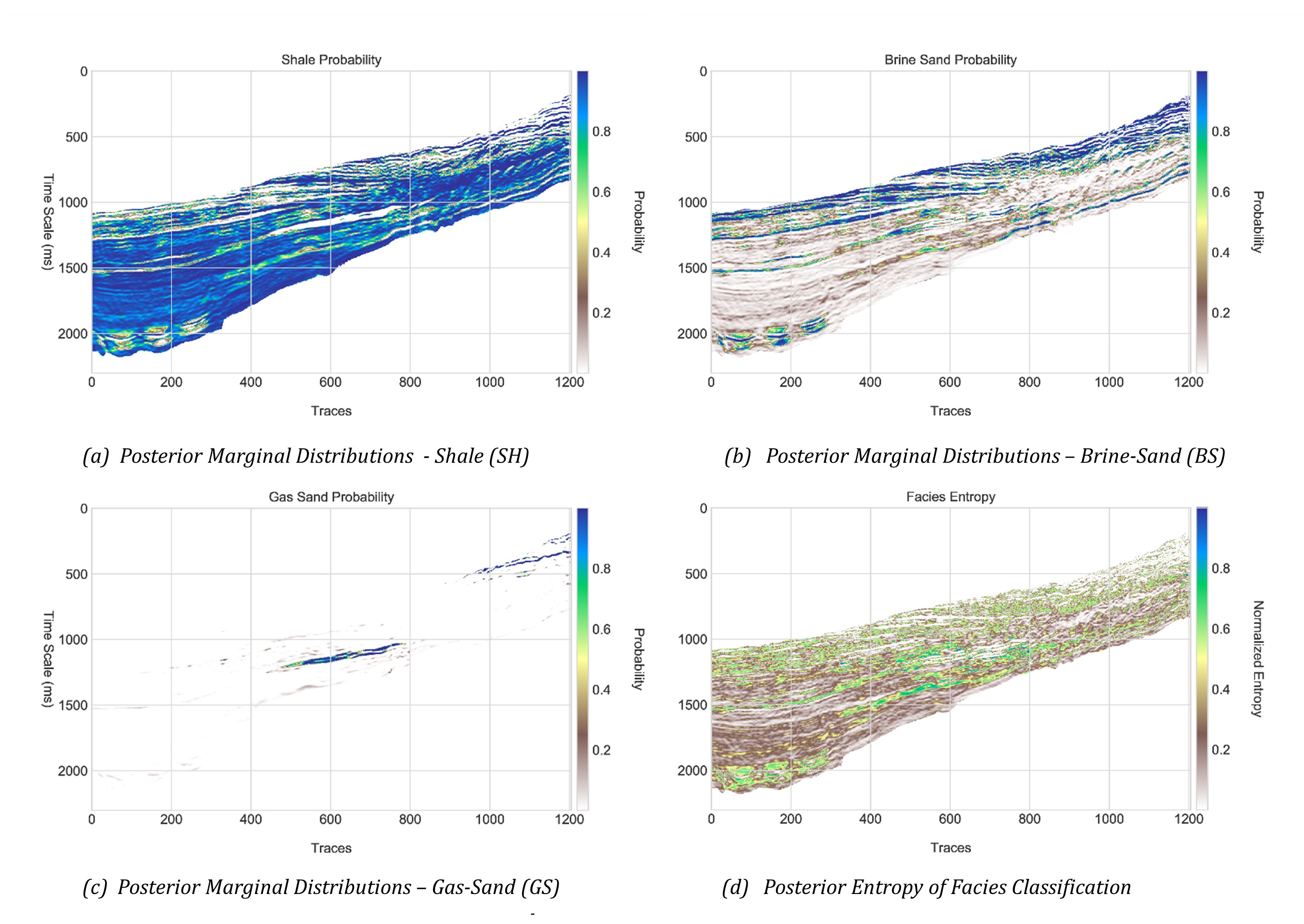}
	\caption{Cell-wise posterior marginal distributions of \textbf{(a)} shale, \textbf{(b)} brine-sand, \textbf{(c)} gas-sand, and \textbf{(d)} the posterior marginal entropy of facies classification scaled between 0.0 and 1.0. Yellow colour represents high probability or entropy (value=1.0) and dark blue colour represents low probability or entropy (value=0.0).}
	\label{fig:petro_marginals}
\end{figure}

\begin{figure}
	\includegraphics[width=1.0\linewidth]{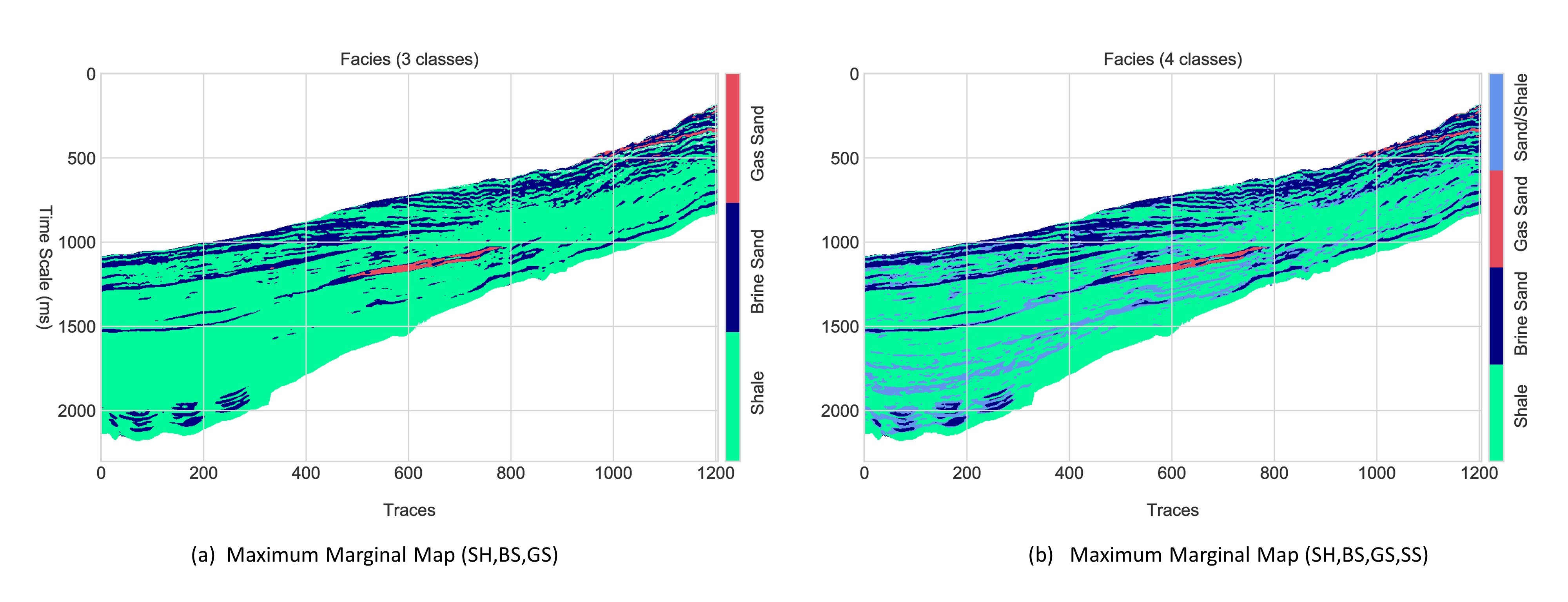}
	\caption{Cell-wise maps of facies with maximum marginal distribution. \textbf{(a)} Map of the three inverted facies: Shale (SH: shown in yellow), brine-sand (BS: blue) and gas-sand (GS: red). \textbf{(b)} Map with an additional facie “Shale/Sand” (SS: brown) identified from high entropy layers in Figure \ref{fig:petro_marginals}(d).}
	\label{fig:petro_maxmarginals}
\end{figure}

\subsubsection{Results}
We now show the application of the joint inversion method to estimate the spatial distribution of petrophysical rock properties and geological facies from well data and seismic attributes from a gas field in the North Sea. This example is based on that in \cite{nawaz2020variational}, where the available data includes vertical 2D sections of seismic attributes: $I_{p}$, $I_{s}$, and $V_{p}/V_{s}$ (Figure \ref{fig:petro_seismic_attributes}), and well logs from two wells that are located on the available 2D seismic section. The seismic attributes were available from a previous deterministic inversion of seismic waveform data. We are interested in classifying the seismic attribute data into three geological facies: shale, brine-sand and gas-sand, which are identified from the well log data (Figure \ref{fig:petro_matrix_plot}). We notice from the top-left 3x3 sub-plots in Figure \ref{fig:petro_matrix_plot} that there is a significant overlap between the shale and brine-sand elastic properties. However, we may notice from rest of the sub-plots that these shale and brine sand
may be resolved better when elastic properties are analyzed jointly with the petrophysical properties of interest: $V_{cl}$, $S_{w}$ and $\varrho$. This forms the geophysical basis for our approach to jointly invert facies and petrophysical properties from seismic attributes (elastic rock properties). Further, it may also be noticed that since well logs are recorded at a much higher resolution than seismic data, a higher number of facies could be identified from the well log data (e.g. silt, sandy-shale and shaly-sand). However, we limited our analysis to the three main facies (shale, brine-sand and gas-sand) because we hypothesized at this stage that any further sub-division of shale and sand may not be identifiable from the seismic data due to limited resolution. However, contrary to our hypothesis, we later found that seismic data could resolve at least one more facies (shaly-sand or sandy-shale) as we describe below. 

\begin{figure}
	\includegraphics[width=1.0\linewidth]{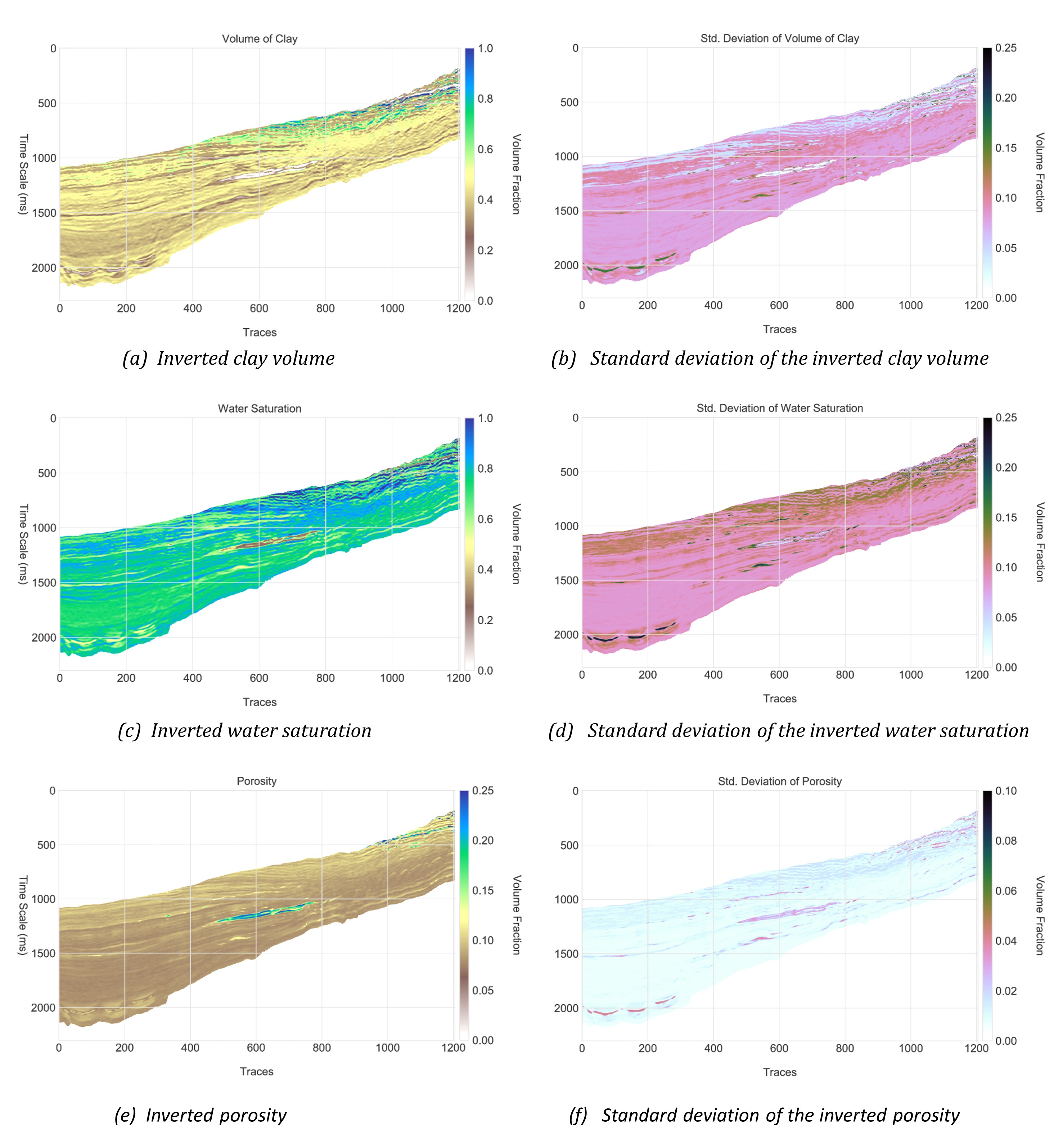}
	\caption{Cell-wise maps of petrophysical properties and their associated standard deviations (std.). \textbf{(a)} clay volume ($V_{cl}$) and \textbf{(b)} its std., \textbf{(c)} water saturation ($S_{w}$) and \textbf{(d)} its std., and \textbf{(e)} porosity ($\varrho$) and \textbf{(f)} its std. Yellow colour represents high values and dark blue colour represents low values of the respective properties.}
	\label{fig:petro_mean_std}
\end{figure}

\begin{figure}
	\includegraphics[width=1.0\linewidth]{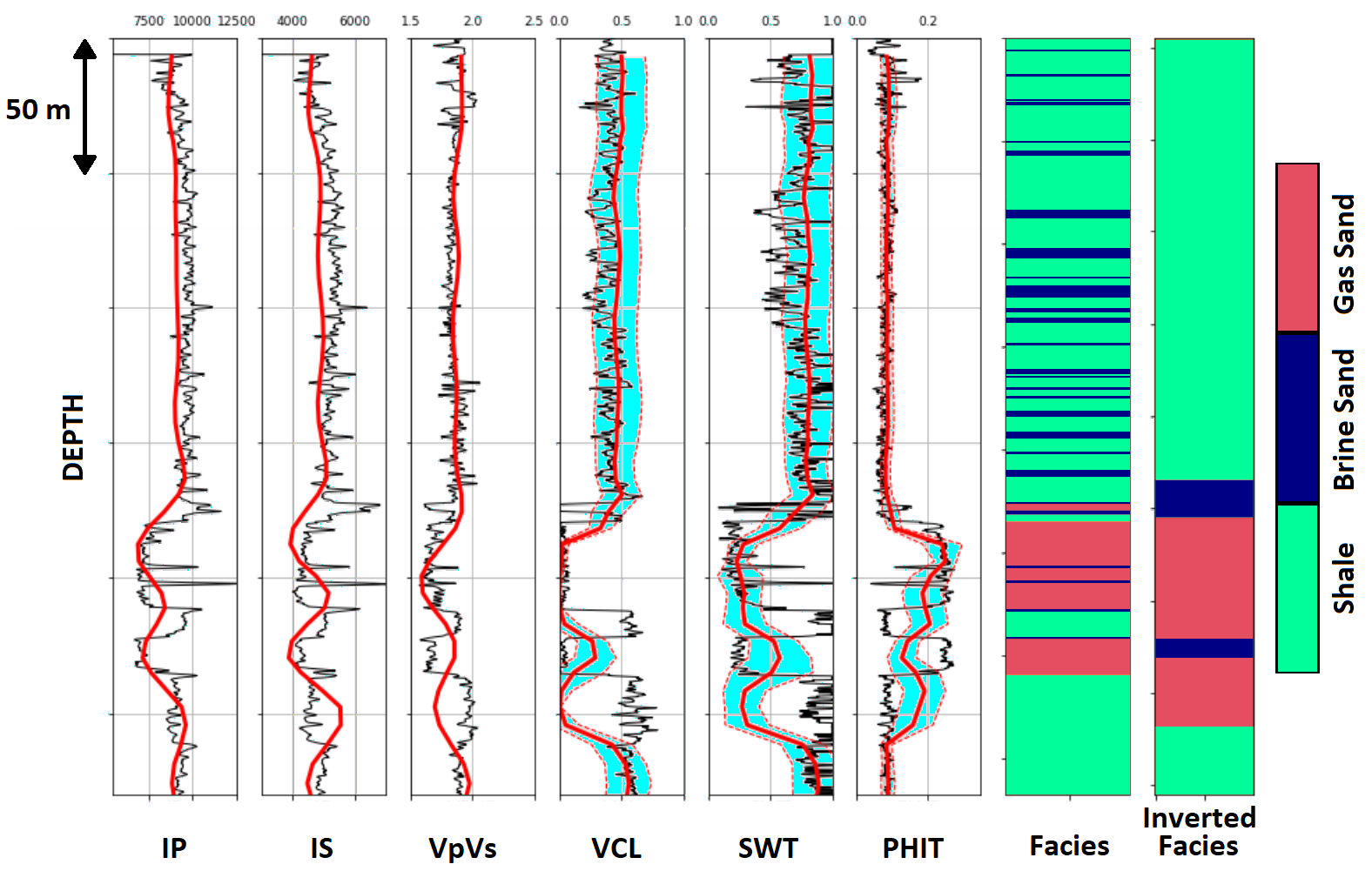}
	\caption{The inverted results compared with the well log data. Solid black lines show the measured well log data. Red lines in columns 1-3 are the input seismic attributes. Columns 4-6 show the mean (red lines) and two standard deviations (blue shaded area) of the inverted posterior distribution of petrophysical properties. The rightmost two columns show the measured facies and inverted facies, respectively.}
	\label{fig:petro_comparison_well}
\end{figure}

We used the EM method to invert the available elastic seismic attributes jointly for the spatial distributions of facies and petrophysical rock properties. The estimated marginal posterior distributions (under the mean field approximation) of the three facies and the entropy (a measure of uncertainty) of these distributions scaled between 0.0 and 1.0 is shown in Figure \ref{fig:petro_marginals}. The entropy is mostly low except at the transitions between different facies, but it appears to be high within some layers too. Since gas-sands typically have well discriminated properties, high entropy within some layers indicates the presence of a mixture of brine-sand and shale lithology that is not well discriminated. Figure \ref{fig:petro_maxmarginals}(a) shows the facies map with maximum marginal probability in each model cell for the three inverted facies: shale, brine-sand, and gas-sand. Figure \ref{fig:petro_maxmarginals}(b) shows the facies map with an additional facies defined as a combination of non-discriminated shale-sand identified to exist in the cells where entropy is greater than a cut-off value of 0.5 (i.e. 50\% of the scaled entropy range from 0.0 to 1.0). Even though we inverted for 3 facies, the entropy of the marginal posterior distributions identifies that an additional facies may also be interpreted as shaly-sand or sandy-shale shown in brown colour in Figure \ref{fig:petro_maxmarginals}(b). 

The inverted petrophysical properties along with their standard deviations are shown in Figure \ref{fig:petro_mean_std}. The seismic attribute inversion results are compared with the well data for verification and are shown in Figure \ref{fig:petro_comparison_well}. The measured well logs are shown in solid-black curves for reference. The solid-red curves are the input seismic attributes along the borehole in columns 1-3 and are means of the posterior distribution of petrophysical properties in columns 4-6. The blue shaded regions bounded by the dashed-red curves in columns 3-4 represent the two standard deviations of the posterior distribution of corresponding rock properties. The mean inverted petrophysical properties clearly identify the gas reservoir characterized by lower $V_{cl}$ and $S_{w}$, and higher $\varrho$ compared to the non-reservoir rocks.

\subsection{Travel time tomography}
\begin{figure}
	\includegraphics[width=1.0\linewidth]{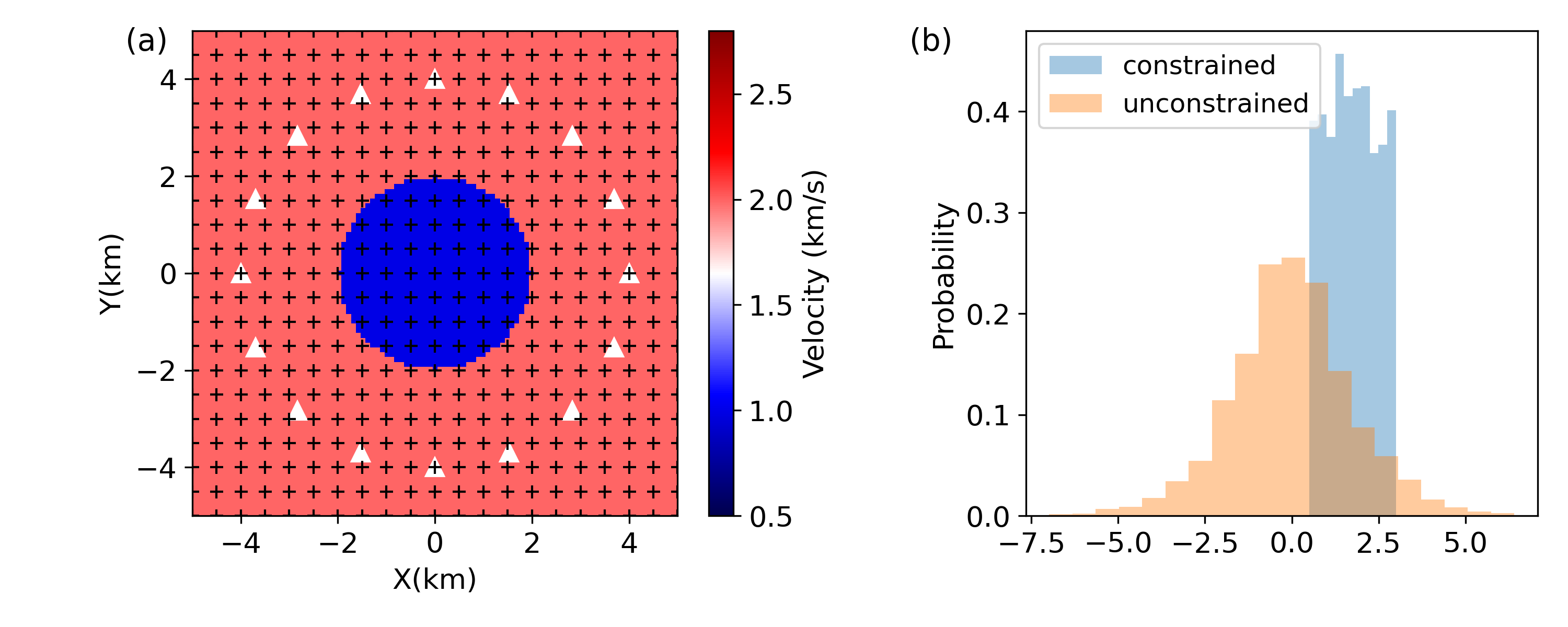}
	\caption{\textbf{(a)} The target structure and receiver geometry (while triangles). Each receiver also acts as a virtual source to simulate the scenario in ambient noise tomography \cite[e.g.,][]{shapiro2005high}. Black crosses denote the location of grid points used in inversion -- the wave velocity at each location is described by one parameter. \textbf{(b)} The prior distribution in the original space (blue histogram) and the transformed space (orange histogram) -- as described by 2000 samples.
		.}
	\label{fig:tt_true_model}
\end{figure}
In this section we explore applications of variational inference methods to seismic travel time tomography based on examples in \cite{zhang2020seismic} and \cite{zhao2020bayesian}. We image a simple 2D velocity structure that has been studied previously using Monte Carlo methods \citep{galetti2015uncertainty}. The velocity structure contains a circular low velocity anomaly with a 2 km radius and 1 km/s velocity within a homogeneous background of 2 km/s velocity (Figure \ref{fig:tt_true_model}a). 16 receivers are equally distributed around the low velocity anomaly approximating a circular acquisition geometry with a 4 km radius. Each receiver is also treated as a virtual source to simulate a typical ambient noise tomographic experiment \citep{shapiro2005high, curtis2006seismic}. Travel times between each receiver pair are calculated using the fast marching method over a 101 $\times$ 101 gridded discretisation in space of each modeled velocity structure \citep{rawlinson2004multiple}, and the travel times through the target structure are used as data to infer the velocity structure.

For inversion we use a regular 21 $\times$ 21 grid of cells to parameterize the velocity structure (black pluses in Figure \ref{fig:tt_true_model}a). The likelihood function is set to a Gaussian distribution with 0.05 s standard deviation  which represents the uncertainty on observed travel times. For each cell the prior pdf of the velocity is set to be a Uniform distribution between 0.5 km/s and 3 km/s (blue histogram in Figure \ref{fig:tt_true_model}b). To understand the characteristics of different methods we compare the posterior pdfs obtained using four methods: ADVI, Normalizing flows, SVGD and Metropolis-Hastings McMC (MH-McMC). In order to handle the hard constrains imposed by the prior information in variational methods, we transform the constrained velocity into an unconstrained space using equation \ref{eq:transform}. The orange histogram in Figure \ref{fig:tt_true_model}b shows the prior distribution in the transformed space. For all inversions, travel times are calculated using the fast marching method over a 41 $\times$ 41 grid interpolated from the lower spatial resolution properties. The gradients of the posterior pdf with respect to velocity are calculated by tracing rays backwards from each receiver to (virtual) sources using the spatial gradients of travel time fields.

In ADVI the initial Gaussian distribution in the unconstrained space is simply set to be a standard Gaussian distribution $N(\bm{\uptheta}|\mathbf{0},\mathbf{I})$, and updated using the ADAGRAD algorithm \citep{duchi2011adaptive} for 10,000 iterations using the gradients from equation \ref{eq:gradient_mu} and \ref{eq:gradient_L}. The final Gaussian distribution is transformed back to the original space, from which 5,000 samples are generated to visualize the final results.

For normalizing flows we use 6 coupling flows which each use rational quadratic splines \citep{durkan2019neural} for the bijective function. Each bijective function is parameterized by the output of a fully connected neural network, which contains 2 hidden layers each of which contains 100 hidden units with Rectified Linear Unit activation functions. The prior pdf is used as the initial distribution and is first transformed into the unconstrained space, and normalizing flows are applied in this space. The flows are updated using 3,000 iterations, and at each iteration the expectation in equation \ref{eq:gradients_normalizing_flow} is estimated using 10 samples. After the process we generate 2,000 samples from the initial (prior) distribution and transform them through the analytic flows (including the transform in equation \ref{eq:transform}) to obtain the final set of samples, whose density provides an approximation of the posterior pdf. 

For SVGD we use a RBF kernel in which the scale factor $\sigma$ is chosen to be $ \tilde{d} / \sqrt{2\mathrm{log}n}$ where $\tilde{d}$ is the median of pairwise distances between all particles. This choice is suggested by \cite{liu2016stein} based on the intuition that $\sum_{j \ne i}k(\mathbf{m}_{i}, \mathbf{m}_{j}) \approx n\mathrm{exp}(-\frac{1}{h} \tilde{d}^{2}) = 1$, such that for particle $\mathbf{m}_{i}$ the contribution from its own gradient is balanced by the influence from all other particles. We generate 800 particles from the prior distribution and first transform them into the unconstrained space. Those particles are then updated using equation \ref{eq:phi_mean} for 500 iterations and transformed back to the original space.

To demonstrate the convergence properties of these variational methods we compare the results with those obtained using the well-tested and robust method of MH-McMC \citep{metropolis1949monte}. Gaussian perturbations are used as the proposal distribution. We use a total of 6 chains, each of which contains 2,000,000 iterations with a burn-in period of 1,000,000 iterations. To reduce the correlation effects between successive samples we only retain every 50\textsuperscript{th} sample after the burn-in period. This results in a total of 120,000 samples which are used to calculate statistics of the estimated posterior pdf.

\subsubsection{Results}
\begin{figure}
	\includegraphics[width=1.0\linewidth]{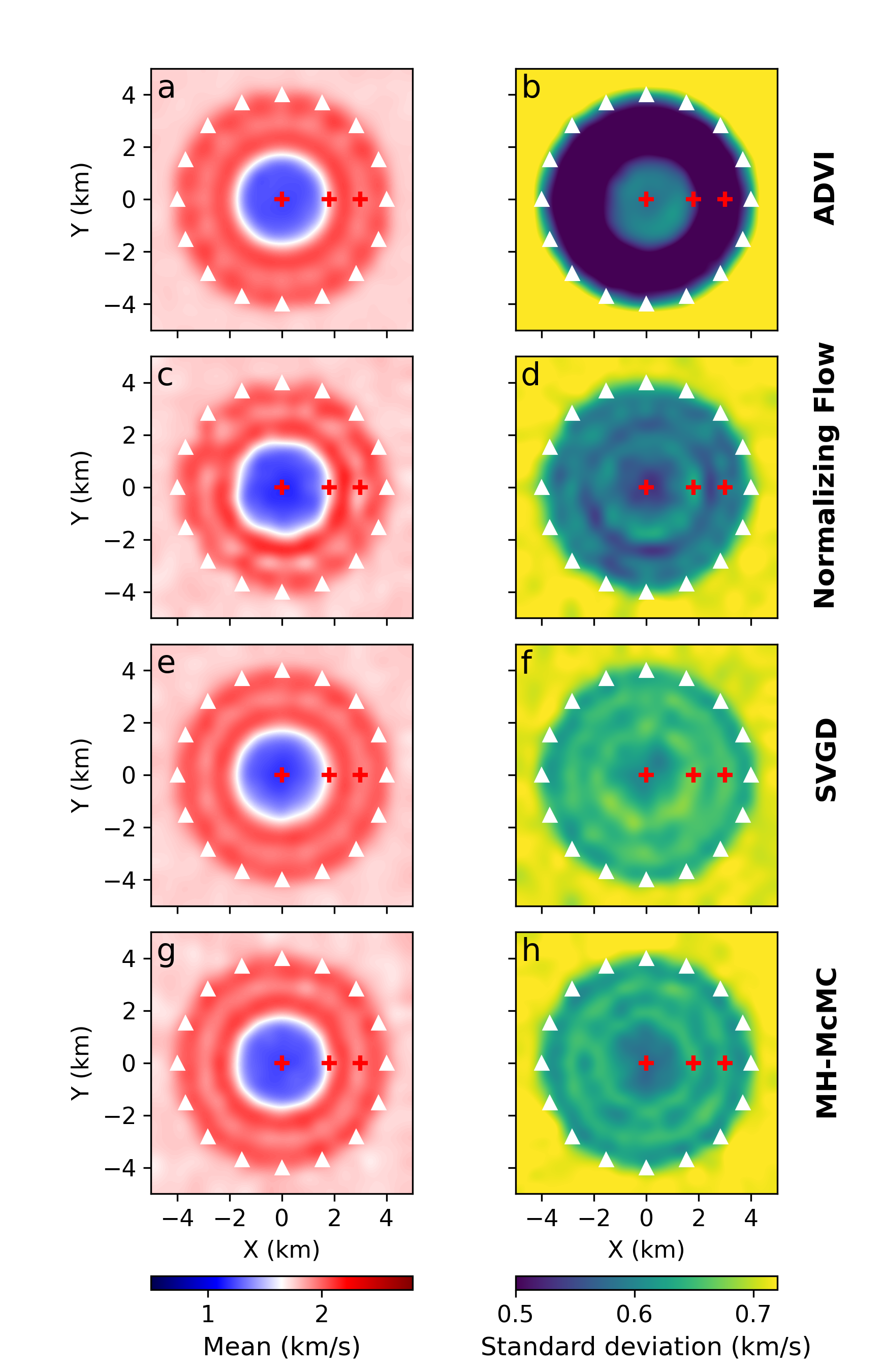}
	\caption{The mean (\textbf{left panel}) and standard deviation models (\textbf{right panel}) obtained using ADVI, Normalizing flows, SVGD and MH-McMC respectively. Red pluses are referred to in the main text and in Figure \ref{fig:tt_marginals}.}
	\label{fig:tt_mean_std}
\end{figure}
Figure \ref{fig:tt_mean_std} shows mean and standard deviation models obtained using the suite of methods. Overall the mean models obtained using different methods show similar features. For example, all models show a low velocity anomaly as in the target structure. The velocity of the mean (1.2 km/s) is slightly higher than the target value (1.0 km/s), but since this value was found by four independent methods this indicates that the mean value of the posterior pdf is genuinely lies at higher values than the target. Between the location of the receiver array and the low velocity anomaly there is a slightly lower velocity loop, and since the means from different methods show consistent features, the means probably reveal the true structure of the mean of the posterior distribution. The mean velocity structure does not necessarily need to be similar to the true velocity structure as it is the point-wise mean calculated from different samples. The circular shape of the mean velocity structure obtained from normalizing flows (Figure \ref{fig:tt_mean_std}c) is less symmetric compared to those obtained using other methods. In normalizing flows a chain of non-linear transforms are optimized to directly reshape an initial distribution towards the posterior distribution. It is highly likely that the high number of parameters in those transforms have non-unique solutions, some of which are not globally optimal. Converging to one of the latter solutions is likely to be the cause of the irregularity in the results.

The standard deviation models obtained from normalizing flows, SVGD and MH-McMC show very similar features (Figure \ref{fig:tt_mean_std}d,f and h). For example, the middle low velocity anomaly has lower standard deviation suggesting that the low velocity anomaly is well constrained. There are two high uncertainty loops: one around the middle low velocity anomaly and the other one between the low velocity anomaly and the receiver array. The inner loop has also been observed in seismic tomographic results obtained using reversible jump McMC which is due to the uncertainty caused by the trade-off between the velocity of the anomaly and its shape \citep{galetti2015uncertainty, zhang20183}. The latter high uncertainty loop is associated with the lower velocity loop in the mean velocity model. This is probably caused by the lower ray path coverage in this region, so that the mean velocity tends towards the mean of the prior (1.75 km/s) which is lower than the true value and the uncertainty is higher. In comparison the standard deviation from ADVI shows different results: higher uncertainty at the location of the middle low velocity anomaly and lower uncertainty between the low velocity anomaly and the receiver array (Figure \ref{fig:tt_mean_std}b). Instead of the double high uncertainty loops exhibited by the other results, the standard deviation only shows a slightly higher uncertainty loop around the middle low velocity anomaly. This difference is probably caused by the fact that in ADVI we use a Gaussian distribution to approximate the posterior pdf, whereas in practice the posterior pdf often assumes non-Gaussian shapes due to the nonlinear relationship between velocity structure and data. Note that outside of the receiver array all standard deviations show high uncertainties because there is no ray coverage.

\begin{figure}
	\includegraphics[width=1.0\linewidth]{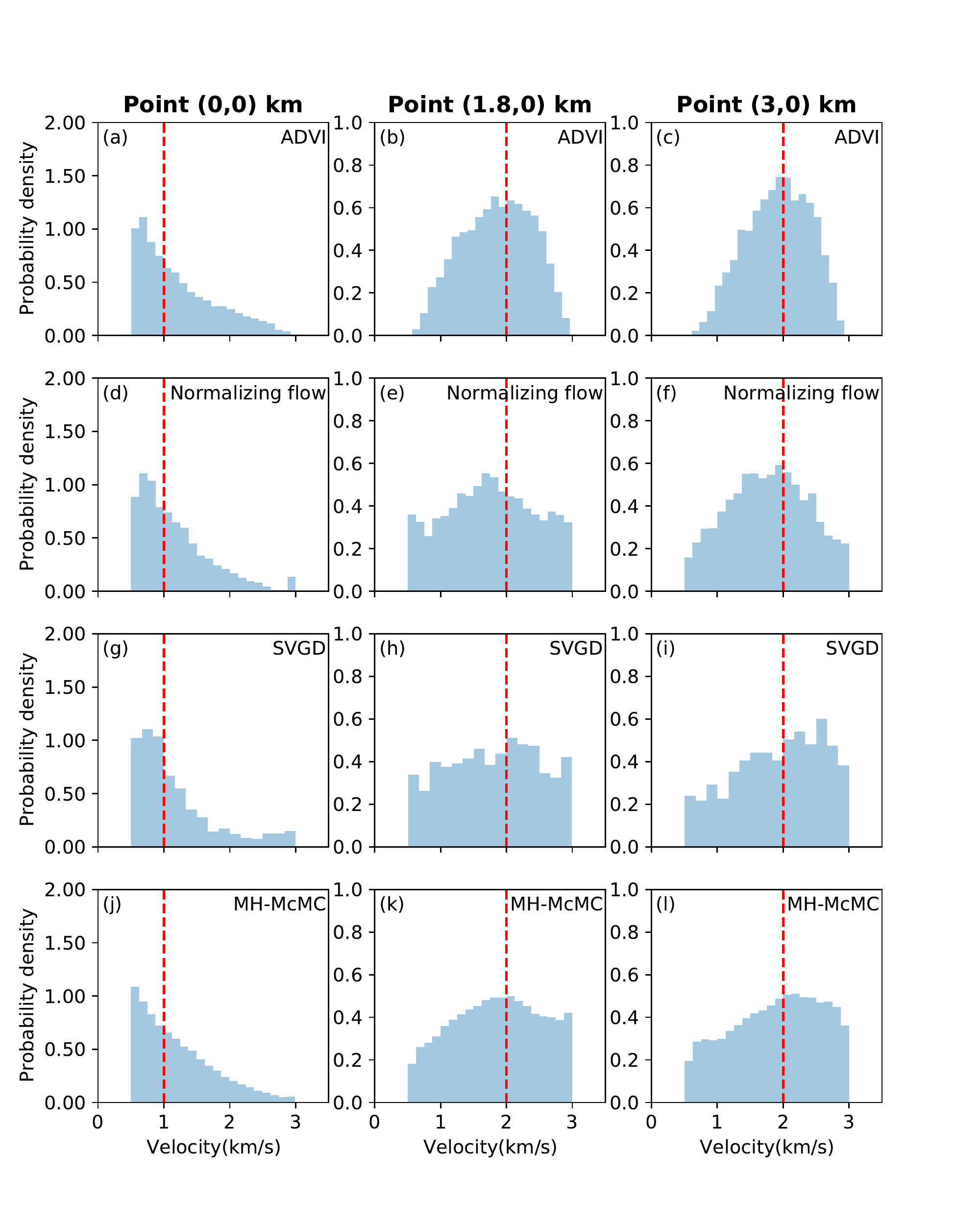}
	\caption{The marginal distributions at three locations: (0,0) km (\textbf{left panel}), (1.8,0) km (\textbf{middle panel}) and (3,0) km (\textbf{right panel}) obtained using ADVI, Normalizing flows, SVGD and MH-McMC respectively. Red line denotes the location of the true value.}
	\label{fig:tt_marginals}
\end{figure}

To further analyse the results in Figure \ref{fig:tt_marginals} we show marginal distributions obtained using different methods at three locations (red pluses in Figure \ref{fig:tt_mean_std}): point (0,0) km at the middle of the velocity structure, point (1.8,0) km and point (3.0,0) km which lie in the two high uncertainty loops. Due to symmetries of the system the marginal distributions at the three locations should reflect properties of most of the single-parameter marginal distributions. At point (0,0) km the marginal distributions are all very similar and show a distribution concentrated at one side of the prior distribution (Figure \ref{fig:tt_marginals}a, d, g and j). At point (1.8,0) km and (3.0,0) km the marginal distributions from normalizing flows (Figure \ref{fig:tt_marginals}e and f), SVGD (Figure \ref{fig:tt_marginals}h and i) and MH-McMC (Figure \ref{fig:tt_marginals}k and l) show similar features and are close to the prior distribution. This suggests that those regions are poorly constrained by the data and explains the double high uncertainty loops observed in the standard deviation structure. Note that the marginal distributions from SVGD and normalizing flows are less smooth than those obtained using MH-McMC. In SVGD this is caused by the lower number of samples used to approximate the distribution, whereas in normalizing flows it is due to the non-uniqueness of the variational optimization problem. In comparison the marginal distributions at point (1.8, 0) km and (3.0,0) km obtained using ADVI show Gaussian-like distributions due to the implicit (transformed) Gaussian assumption which fails to describe the true uncertainty structure.

\subsubsection{Computational cost}
In Table \ref{tb:tt_cost} we summarize the number of forward simulations required by each method, which provides a good metric of the computational cost since for each method the forward simulation is the most time-consuming part. The results show that ADVI is the cheapest variational method, but we demonstrated above that it may provide biased results due to the implicit Gaussian assumption. Normalizing flows is slightly less efficient than ADVI, but produced significantly more accurate results above. SVGD requires approximately ten times more simulations than normalizing flows, but provides the most accurate results among the three variational methods. In comparison MH-McMC requires far more simulations than all three variational methods; however, the comparison is not fair in this case since MH-McMC only requires forward function evaluations, whereas the variational methods also require derivatives of the (logarithm of the) posterior pdf with respect to parameters (which in turn involves calculating derivatives of forward function with respect to parameters). In these travel time examples, derivatives were calculated using ray paths, which were traced through the travel time fields calculated by the fast marching method. For each forward simulation, calculating derivatives required a computation equivalent to approximately $f=0.08$ forward simulations. A fairer comparison with the Monte Carlo method is therefore given in column 3 of Table 1 which shows the 'equivalent' number of simulations for each method, obtained by multiplying the number of simulations for the three variational methods by $1.08$. In this case because of the efficient computation of derivatives, it does not increase the computation cost of variational methods significantly. Clearly this comparison will vary for different types of problems, since factor $f$ will also vary. We demonstrate this below for waveform inversion problems for which $f$ is approximately 2 \citep{tarantola1988theoretical, liu2006finite}.

 \begin{table}
	\caption{The comparison of computational cost for all 4 methods. The third column shows the equivalent number of simulations -- these numbers are calculated as described in the main text.}
	\centering
	\begin{tabular}{l c c}
		\hline
		Method  & Number of simulations & Equivalent number of simulations\\
		\hline
		ADVI  & 10,000 & 10,870\\
		Normalizing flows & 30,000 & 32,609\\
		SVGD & {400,000} & 434,782\\
		MH-McMC & 12,000,000 & 12,000,000 \\
		\hline
	\end{tabular}
	\label{tb:tt_cost}
\end{table}
Note that the above comparison is only valid for this specific example and does not necessarily provide general guidance for the practical choice of algorithms. For example, although ADVI provides biased results, it can still be useful for weakly nonlinear problems in scenarios where efficiency is important and a Gaussian distribution is sufficient for uncertainty analysis. For very high dimensional problems such as 3D tomography and full-waveform inversion, ADVI can become inefficient as the full covariance matrix may require extremely large memory. In the above example, normalizing flows would be a good choice given that it produces reasonably accurate results yet requires the same order of computational cost as ADVI. However we note that normalizing flows may require more human interaction as it has many hyperparameters to tune -- which specific flow to use, how many flows to use, and if invertible neural networks are used  then the structure of the neural network needs elaborate design. For very high dimensional problems we may require large neural networks, so the training time cannot be neglected and may even dominate the whole calculation. SVGD solves variational inference problem using a set of samples, which provides a flexible way to approximate complex probability distributions but at the price of an increased number of forward function evaluations. The method is fully parallelizable which makes it more efficient in real time when combined with modern parallel computer architecture. However it remains unclear how the method performs in very high dimensional space, as it might be impossible to use hundreds of samples to approximate the posterior pdf meaningfully in high dimensional problems.

In this example we only compared the computational cost of variational methods with MH-McMC. In practice there are many ways to make Monte Carlo methods more efficient, for example reversible-jump McMC \citep{green1995reversible, malinverno2002parsimonious, bodin2009seismic}, Hamiltonian Monte Carlo \citep{duane1987hybrid, neal2011mcmc, fichtner2018hamiltonian}, Langevin Monte Carlo \citep{roberts1996exponential, girolami2011riemann}, Sequential Monte Carlo \citep{liu1998sequential, smith2013sequential}, slice sampling \citep{neal2003slice},  physics informed Monte Carlo \citep{khoshkholgh2020informed} and parallel tempering \citep{hukushima1996exchange, earl2005parallel, sambridge2013parallel}. Nevertheless, Monte Carlo methods cannot be parallelized within a Markov chain, several of these Monte Carlo methods require calculation of gradients of the forward function which introduces an additional factor $f$ to the cost as described above, and the methods often become intractable for large datasets which are usually expensive to simulate. In contrast, variational methods can be parallelized at the sample level in each iteration -- for example gradient calculation in ADVI, normalizing flows and SVGD can be fully parallelized. In addition variational methods can be applied to large datasets by using stochastic optimization \citep{robbins1951stochastic, kubrusly1973stochastic} and distributed optimization, which is likely to make variational methods more efficient in practice for some types of problems. 

In travel time tomography the gradients of posterior pdf with respect to model parameters can be calculated efficiently using the travel time field obtained in the forward simulation. In the case that gradients are difficult to calculate, MH-McMC may be more efficient than both variational methods and many other Monte Carlo methods since MH-McMC does not require gradient information. We also note that our comparison above depends on subjective assessments of the point of convergence of each method, so the absolute number of simulations required by each method may not be accurate. Nevertheless they at least provide a reasonable insight into the computational efficiency of each method.

\subsection{Full waveform inversion}
Full waveform inversion (FWI) uses filtered versions of full seismic recordings to characterize properties of the subsurface, and can produce high resolution images of the Earth's interior \citep{tarantola1984inversion, tarantola1988theoretical, gauthier1986two, pratt1999seismic, tromp2005seismic}. The method has been used at industrial scale \citep{prieux2013multiparameter, warner2013anisotropic}, regional scale \citep{chen2007full, tape2009adjoint, fichtner2009full} and global scale \citep{french2014whole, bozdaug2016global, fichtner2018collaborative}. Due to the high nonlinearity and nonuniqueness of the problem, in traditional optimization-based methods a good starting model is required to avoid converging to incorrect solutions. A variety of misfit functions that can reduce multimodalities in the posterior pdf have also been proposed \citep{luo1991wave, gee1992generalized, fichtner2008theoretical, brossier2010data, van2010correlation, bozdaug2011misfit, metivier2016measuring, warner2016adaptive}. In addition, to quantify uncertainties in the solution Monte Carlo methods have recently been used to solve FWI problems \citep{ray2016frequency, ray2017low, biswas20172d, zhao2019gradient, gebraad2020bayesian, guo2020bayesian}. We now use variational inference methods, specifically SVGD to solve FWI problems probabilistically, which we refer to as variational full waveform inversion or VFWI, based on examples in \cite{zhang2020variational, zhang2021bayesian}.

\subsubsection{Transmission seismic FWI with strong prior information}
\begin{figure}
	\includegraphics[width=1.0\linewidth]{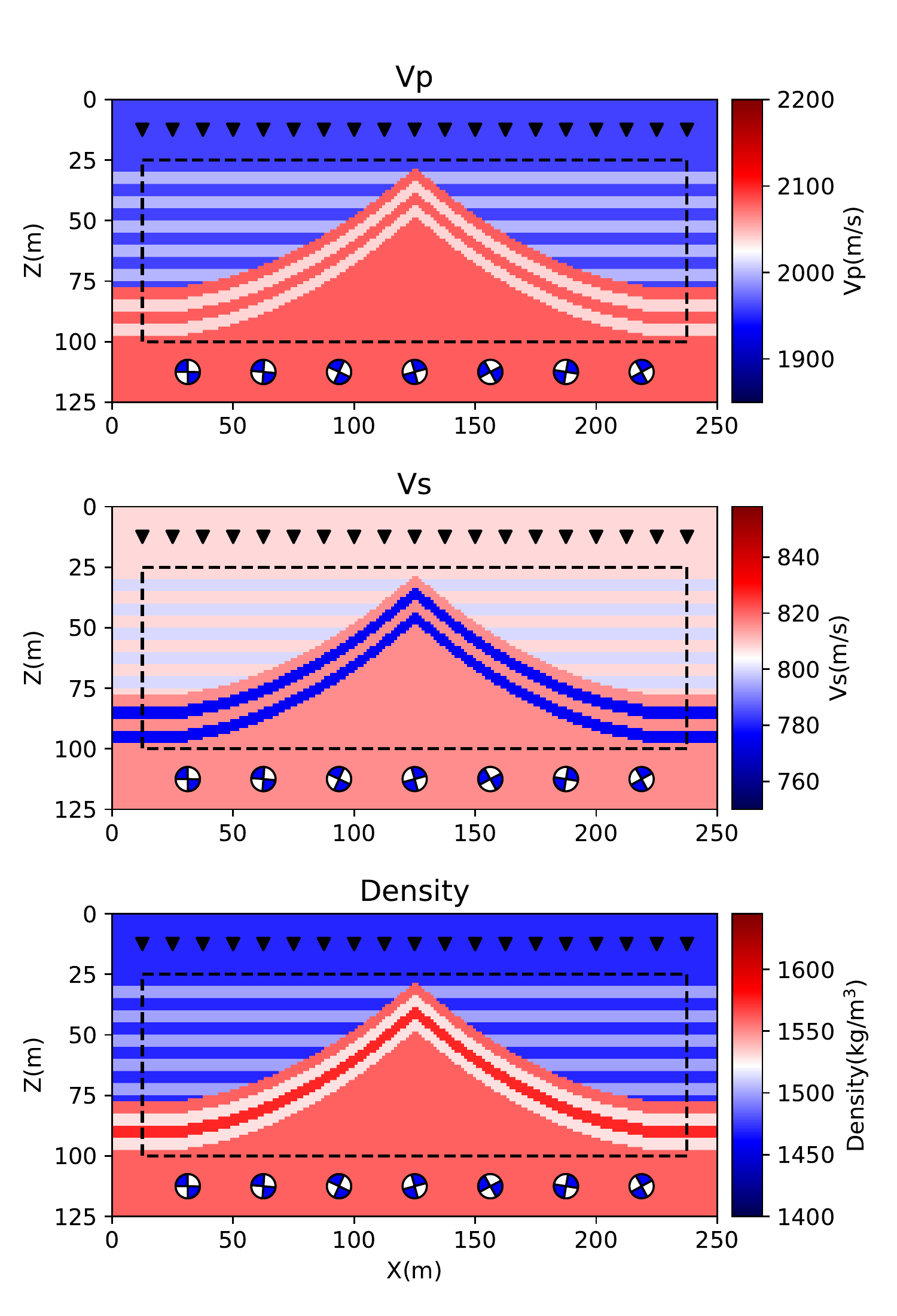}
	\caption{The target structure of Vp, Vs and density. Sources are located at the bottom of the model with random moment tensors and receivers are located at the near surface (black triangles). The black dashed line indicate the area that has free parameters. This inverse problem setup is identical to that in \cite{gebraad2020bayesian}.}
	\label{fig:tfwi_true_model}
\end{figure}

We first apply SVGD to a transmission FWI problem in which seismic data are recorded on a receiver array that lies above the structure to be imaged given earthquake-like sources located underneath the structure. We use a 2D fully elastic target structure and data acquisition setup that is identical to that used by \cite{gebraad2020bayesian} such that the results obtained by SVGD can be fairly compared to those that \cite{gebraad2020bayesian} obtained using Hamiltonian Monte Carlo (HMC). Figure \ref{fig:tfwi_true_model} shows the target Vp, Vs and density model. 7 sources with random moment tensors are located at the bottom of the region. Similarly to \cite{gebraad2020bayesian} we use a Ricker wavelet source-time function with a dominant frequency of 50 Hz. 19 receivers are located at the depth of 10 m with a regular spacing of 12.5 m. The model is discretised using a regular $200 \times 100$ grid of cells, within which a $180 \times 60$ sub-grid of cells have free parameters (black dashed box in Figure \ref{fig:tfwi_true_model}). This leads to a total of $180 \times 60 \times 3 = 32,400$ free parameters. The waveform data are modelled using a fourth-order variant of the staggered-grid finite difference scheme \citep{virieux1986p, gebraad2020bayesian}. The gradients of the likelihood function with respect to velocities and density are computed using the adjoint method \citep{tarantola1988theoretical, liu2006finite, fichtner2006adjoint, plessix2006review}.  

To reduce the complexity of the inverse problem and guide both methods towards to correct solution we use a strong prior information as in \cite{gebraad2020bayesian}: Uniform distributions in the interval of 2000 $\pm$ 100 m/s for Vp, 800 $\pm$ 50 m/s for Vs and 1500 $\pm$ 100 kg/m\textsuperscript{3} for density. For the likelihood function, we assume a Gaussian distribution with a diagonal covariance matrix:
\begin{equation}
	p(\mathbf{d}_{\mathrm{obs}}|\mathbf{m}) \propto \mathrm{exp}
	[-\frac{1}{2} \sum_{i}(\frac{d_{i}^{\mathrm{obs}}-d_{i}(\mathbf{m})}{\sigma_{i}})^{2}]
	\label{eq:fwi_likelihood}
\end{equation} 
where $i$ is the index of time samples and $\sigma_{i}$ is the standard deviation of that data point. To keep the inverse problem identical to that in \cite{gebraad2020bayesian}, we set $\sigma_{i}$ to be 1 $\mu$m\textsuperscript{2} and did not add any noise to the waveform data. The effects of different values of $\sigma_{i}$ on the solution are analysed in \cite{gebraad2020bayesian}.

Similarly to the previous section we use a RBF kernel for SVGD whose scale factor is determined from the median of pairwise distances between all particles. We generated 600 particles from the prior distribution and transformed them into an unconstrained space using equation \ref{eq:transform}. Those particles are then updated using equation \ref{eq:phi_mean} for 600 iterations, and are finally transformed back to the original space.

\begin{figure}
	\includegraphics[width=1.0\linewidth]{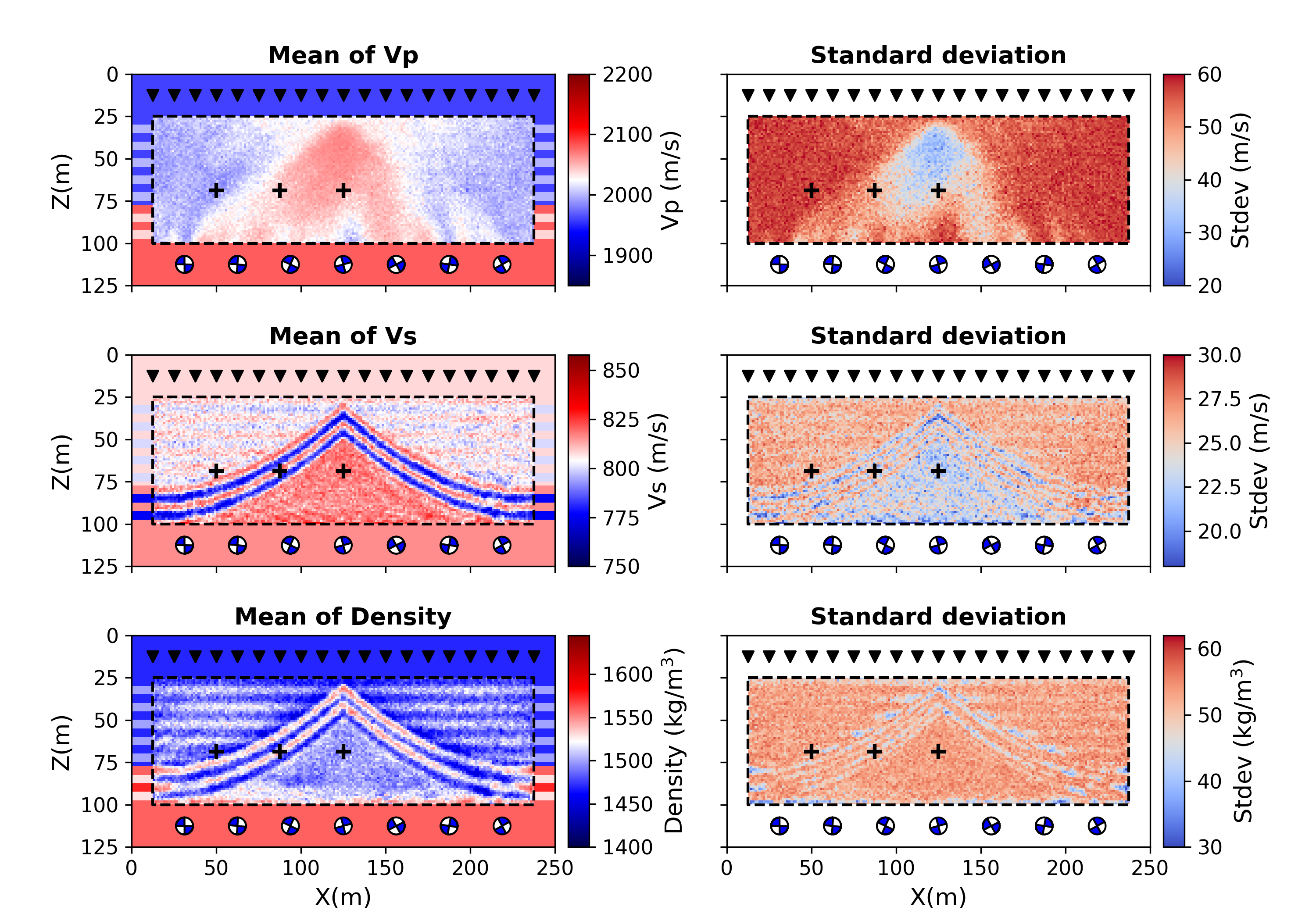}
	\caption{The mean and standard deviation of Vp, Vs and density obtained using SVGD. Black crosses are referred in the main text.}
	\label{fig:tfwi_svgd_model}
\end{figure}

\begin{figure}
	\includegraphics[width=1.0\linewidth]{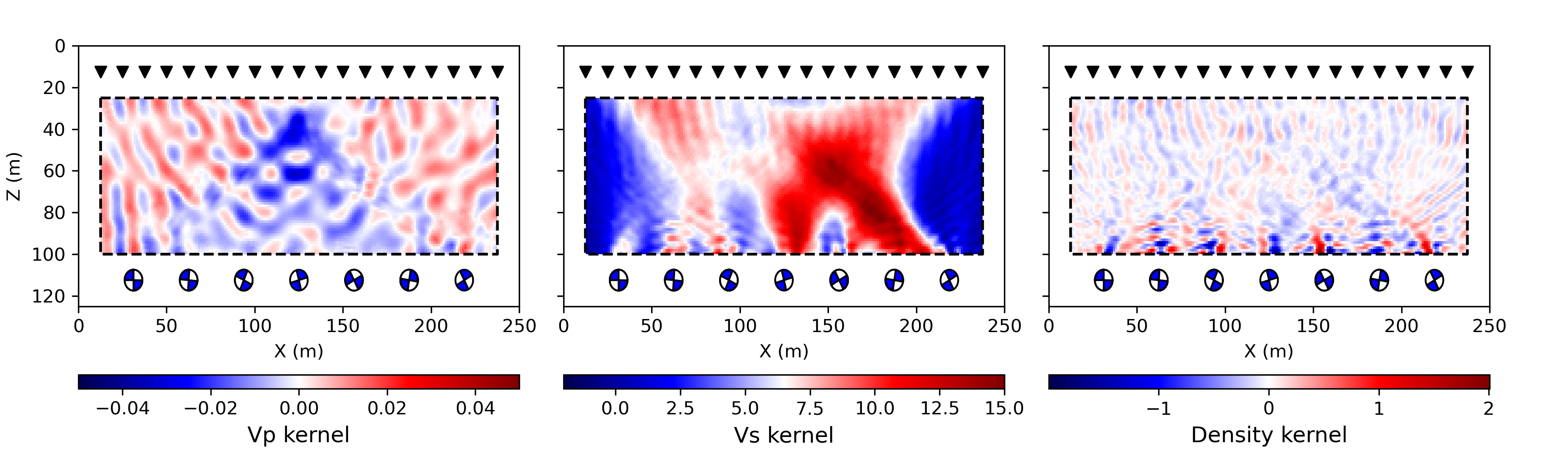}
	\caption{The kernel (gradients of the misfit function) of Vp, Vs and density calculated using mean structures in Figure \ref{fig:tfwi_svgd_model}. The magnitude of Vs and density kernels are significantly higher than that of Vp kernel.}
	\label{fig:tfwi_kernel}
\end{figure}

Figure \ref{fig:tfwi_svgd_model} shows the mean and standard deviation structures obtained using SVGD. The mean Vs model shows similar features to the true velocity structure, for example the bottom high velocity structure and tilted layers above that structure. The horizontal layers at the shallow part ($<$ 80 m) are not as clearly observable as those in the true velocity structure, which probably reflects the limits of the resolution of the data. By contrast, the mean Vp model only recovers the bottom large scale structure. This is probably because when a simple unweighted L2 norm misfit function is used, seismic waveforms are more sensitive to Vs than to Vp due to the higher amplitudes of shear waves. Figure \ref{fig:tfwi_kernel} shows kernels (gradients of the misfit function) of Vp, Vs and density calculated using the mean models in Figure \ref{fig:tfwi_svgd_model}. The magnitude of Vs and density kernels are significantly higher than that of Vp. As a result, the Vp structure is not well constrained by the data. The mean density model clearly shows horizontal and tilted layers except that the value of the lower density titled layers is smaller than the true value. In comparison the bottom high density structure is not present in the mean model which is probably because seismic waveforms are mainly sensitive to spatial gradients of density.

Overall the standard deviation models show similar features to their associated mean structure. For example, the standard deviation model of Vp shows lower uncertainty at the location of the large scale high velocity structure. The Vs standard deviation model shows lower uncertainties at the location of the horizontal high velocity layers and the bottom high velocity structure. There are high uncertainties at the boundaries of tilted layers, which suggests that the location of velocity layers are not well-constrained. Note that a similar phenomenon observed in the travel time tomography examples in the previous section. Similarly there are high uncertainties at those boundaries in the standard deviation model of density. Due to the fact that seismic waveforms are mainly sensitive to density spatial gradients, the bottom high density structure has high uncertainty.  

\begin{figure}
	\includegraphics[width=1.0\linewidth]{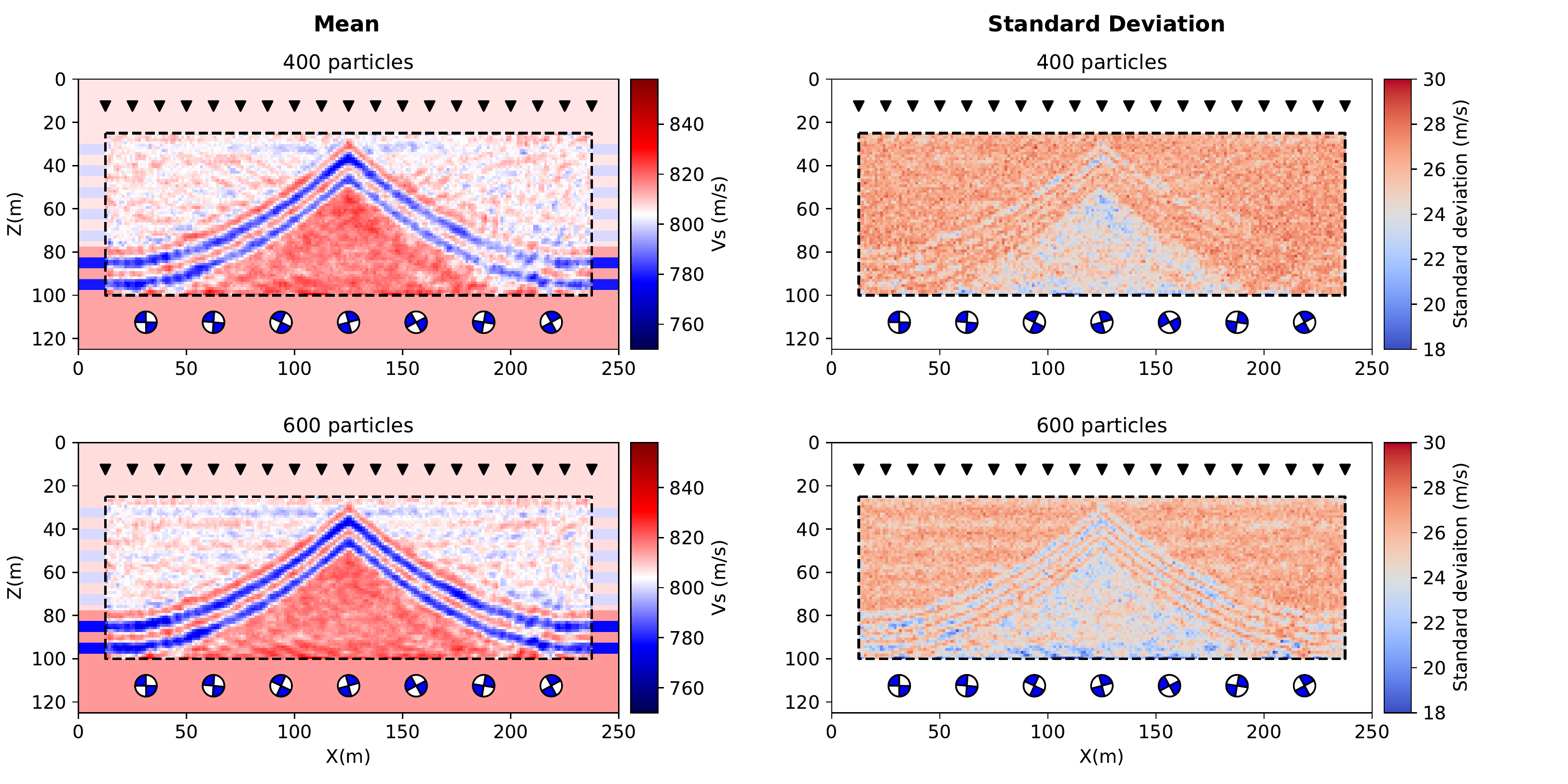}
	\caption{The mean (\textbf{left}) and standard deviation (\textbf{right}) of Vs obtained using SVGD with 400 and 600 particles respectively.}
	\label{fig:tfwi_svgd_particle_comparison}
\end{figure}

To explore the effects that the number of particles have on the results, in Figure \ref{fig:tfwi_svgd_particle_comparison} we show the mean and standard deviation models of Vs obtained using 400 particles and 600 particles respectively. As expected, the results show that when using 600 particle, we can obtain more accurate results. For example, the mean Vs model obtained using 400 particles only shows the bottom high velocity structure and the tilted layers. The shallow horizontal layers are smeared into each other. Similarly the standard deviation model does not show much structure in the shallow part compared with that obtained using 600 particles. This shows that the accuracy of the results of SVGD improves with the number of particles \citep{liu2017stein}.

\begin{figure}
	\includegraphics[width=1.0\linewidth]{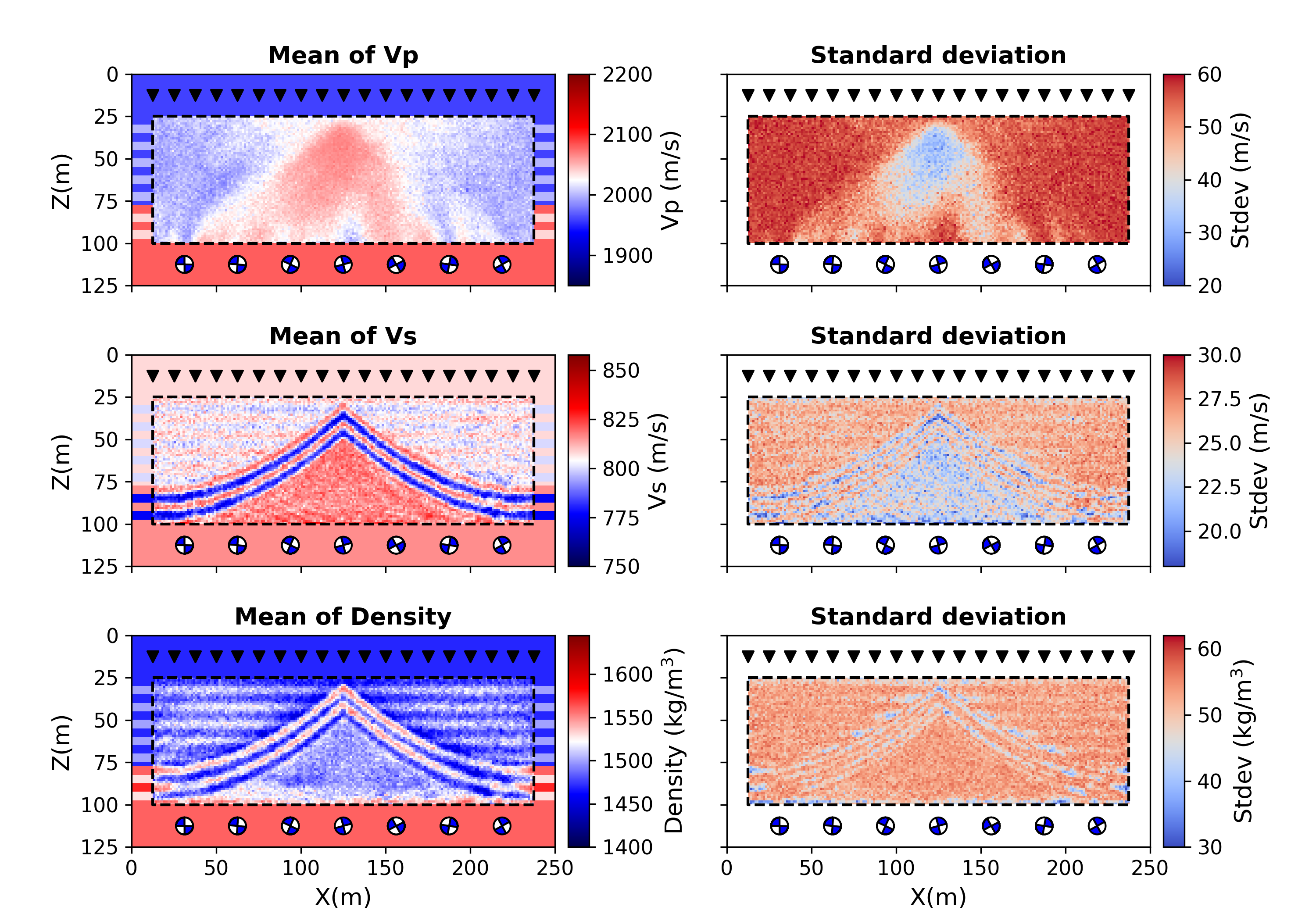}
	\caption{The mean and standard deviation of Vp, Vs and density obtained using HMC from \cite{gebraad2020bayesian}.}
	\label{fig:tfwi_hmc_model}
\end{figure}

To validate the results obtained using SVGD, we compared the results with those obtained using HMC by \cite{gebraad2020bayesian} (Figure \ref{fig:tfwi_hmc_model}). The mean and standard deviation structures obtained using HMC are very similar to those obtained using SVGD. For example, the mean Vp only shows the bottom large scale structure whereas the mean Vs successfully recovers the true structure. The mean density shows the horizontal and tilted layers and fails to find the bottom high density structure. The standard deviations also show similar features to associated mean structures. Since the two methods are completely different, it is highly likely that these results represent the true solution to this specific FWI problem. Note that the results from SVGD are smoother than those from HMC, which is probably caused by undersampling of both methods and lack of convergence of HMC \citep{gebraad2020bayesian}.

\begin{figure}
	\includegraphics[width=1.0\linewidth]{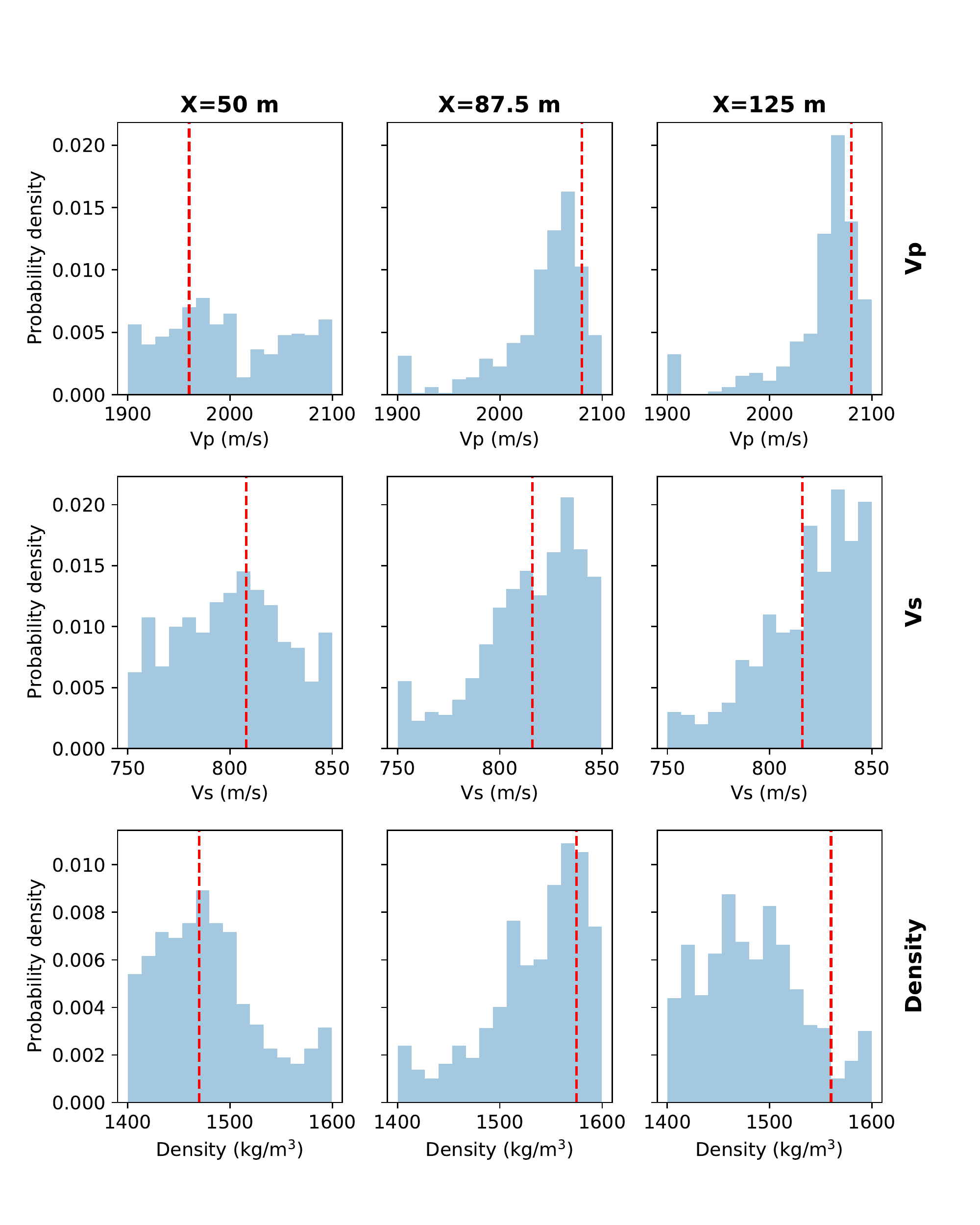}
	\caption{The marginal distributions of Vp, Vs and density obtained using SVGD at the depth of 68.75 m and at X=50 m, 87.5 m and 125 m respectively. Red lines denote the true values.}
	\label{fig:tfwi_marginals}
\end{figure}

To further analyse the results, in Figure \ref{fig:tfwi_marginals} we show marginal distributions of Vp, Vs and density obtained using SVGD at three points (black crosses in Figure \ref{fig:tfwi_hmc_model}): (50, 68.75) m, (87.5, 68.75) m and (125, 68.75) m. Overall the results show high probability around the true value. At X=50 m the marginal distributions are wider than those at the other locations, which indicates high uncertainties at this location. At X=125 m the true value of density deviates from the values with highest probability as we have observed in the mean model due to the fact that seismic waveforms are mainly sensitive to density spatial gradients. Note that the marginal distributions show nonsmoothness due to the undersampling of the posterior pdf (a small number of particles).

Since SVGD is based on particles, the method can be computationally expensive. For example, the above example requires 600$\times$600 = 360,000 forward and adjoint simulations; whereas HMC took approximately 130,000 forward and adjoint simulations. Although in this case it appears that HMC is slightly more efficient, in the above example HMC has clearly not fully converged. While SVGD can be easily parallelized, it is difficult to parallel a Markov chain due to the dependence between successive samples \citep{neiswanger2013asymptotically}. Also in practice HMC often requires deliberate and tedious tuning to construct an efficient Markov chain \cite[see discussions in ][]{gebraad2020bayesian} so the actual computational cost may be significantly higher than the number of samples reported above. In contrast SVGD is much easier to tune by using adaptive gradient ascent methods \citep{duchi2011adaptive, liu2016stein}. In addition SVGD can be performed on large datasets by using stochastic optimization by dividing large datasets into minibatches \citep{liu2016stein}. The same technique cannot be used in McMC methods because it breaks the detailed balance required by McMC. To give an idea about the overall computational cost required by SVGD, the above example took 6 days of computation parallelized across 16 Intel Xeon cores. 
		  
\subsubsection{Reflection seismic FWI with realistic prior information}

\begin{figure}
	\includegraphics[width=1.0\linewidth]{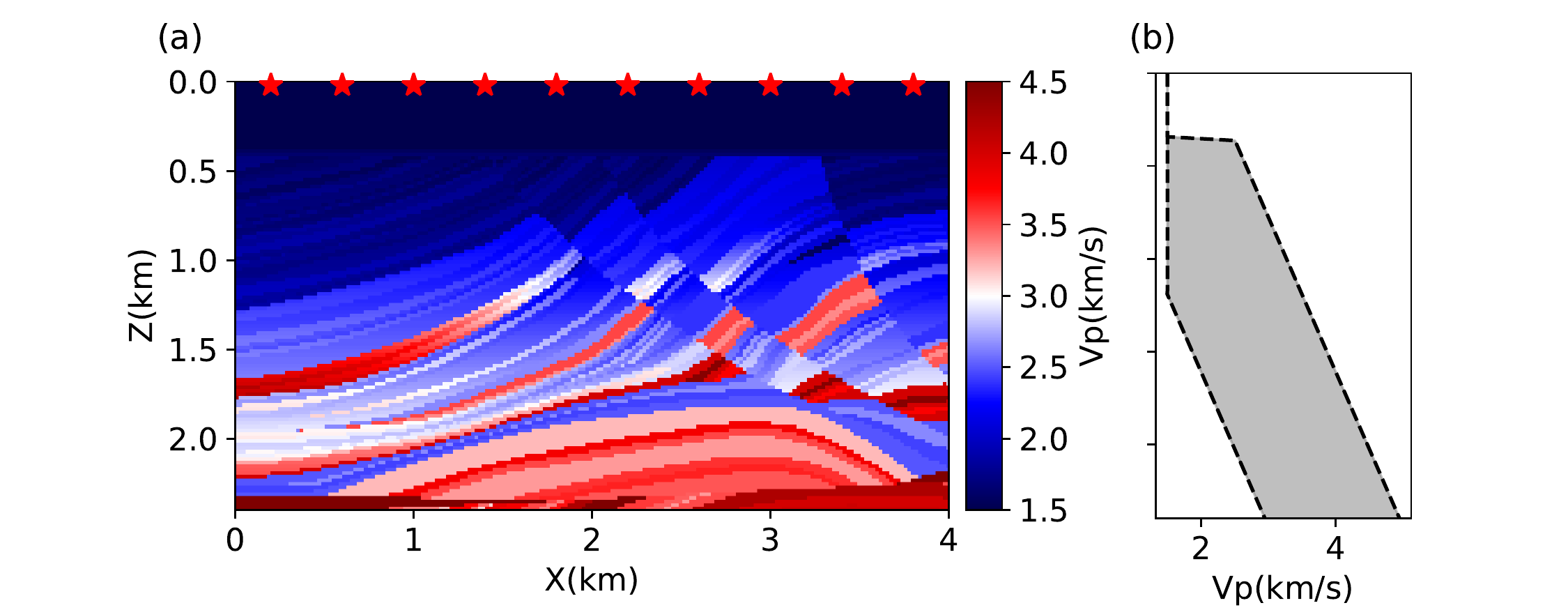}
	\caption{\textbf{(a)} Part of the Marmousi model \citep{martin2006marmousi2} which is used as the target velocity structure. 10 sources (stars) are located at the depth of 20 m and 200 receivers (not shown) are equally spaced at a depth of 360 m across the horizontal extent of the model (this depth represents the seabed). \textbf{(b)} Prior distribution used in the inversion: a Uniform distribution with an width of 2 km/s at each depth. Note that an extra lower bound is also imposed to the velocity to ensure that the rock velocity is higher than the velocity in water (1.5 km/s).}
	\label{fig:rfwi_true_model}
\end{figure}
In the previous section we applied SVGD to a transmission FWI problem with known, double-couple (earthquake-like) sources and strong prior information on parameters. Unfortunately such strong prior information about sources and parameters is never available in practice. To explore the applicability of the method in practice, in this section we apply SVGD to seismic reflection data generated by known near-surface sources with more practically realistic prior information.

We solve a 2D acoustic FWI problem using the waveform data generated from a part of the Marmousi model \citep{martin2006marmousi2}. The model is discretised in space using a 200$\times$100 regular grid of cells. 10 sources are located at 20 m depth and 200 receivers are located at the 360 m depth (which represents the seabed) across the full horizontal extent of the model with a regular spacing of 20 m (Figure \ref{fig:rfwi_true_model}). Similarly to the previous section, the waveform data are generated using the finite difference method and the gradients of the posterior pdf with respect to velocity parameters are computed using the adjoint method.
 
\begin{figure}
	\includegraphics[width=1.0\linewidth]{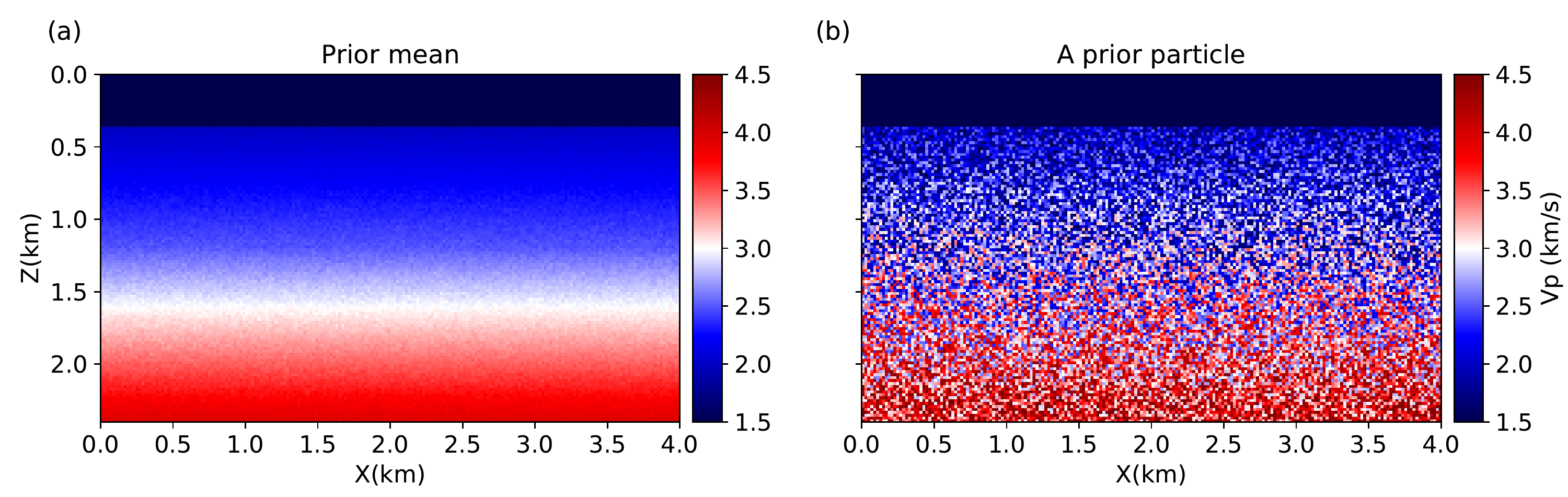}
	\caption{\textbf{(a)} The mean of the prior distribution and \textbf{(b)} a random sample generated from the prior pdf.}
	\label{fig:rfwi_initial_model}
\end{figure}

Instead of using strong prior information (a Uniform distribution over an interval of 0.2 km/s) as in the previous section, we impose ten times weaker prior information to the velocity: a Uniform distribution over an interval width of 2 km/s at each depth (Figure \ref{fig:rfwi_true_model}b). We also impose a lower bound on the velocity to ensure that the rock velocity is higher than that in the water (1.5 km/s). The velocity of the water layer is fixed to be the true velocity (1.5 km/s) in the inversion as is standard in practical marine seismic FWI. Figure \ref{fig:rfwi_initial_model} shows the mean of the prior distribution and a random particle generated from the prior distribution. We simulate waveform data using a Ricker wavelet with a dominant frequency of 10 Hz. Uncorrelated Gaussian noise with a standard deviation of 0.1 amplitude units is added to the data. For the likelihood function we use the same Gaussian distribution as described in equation \ref{eq:fwi_likelihood} where $\sigma_{i}$ is set to be the true value. For standard optimisation-based FWI this problem is difficult because the reference parameter values from which the inversion begins (which in practice would normally be the mean structure) is very different from the target. 

\cite{zhang2021bayesian} showed that one can improve accuracy of the inversion results by performing an inversion using low frequency data first, and using the results of the low frequency inversion as the starting distribution for high frequency inversions. Therefore, we first perform SVGD on low frequency data generated by a Ricker wavelet with a dominant frequency of 4 Hz with the same Gaussian noise as above added to the data (a standard deviation of 0.1). The inversion is conducted using 600 particles that are initially generated from the prior distribution (e.g., Figure \ref{fig:rfwi_initial_model}b) and the matrix kernel described in equation \ref{eq:phi_matrix_kernel} where $\mathbf{Q}^{-1}=diag(var(\mathbf{m}))$ and $var(\mathbf{m})$ is the variance computed across those particles. For parameters with higher variance this kernel applies higher weights to the posterior gradients, and also enables more distant interactions with other particles. As in the previous section we first transform those particles to an unconstrained space using equation \ref{eq:transform} and update them using equation \ref{eq:phi_mean} for 600 iterations. Those particles are then used as the starting particles for the high frequency inversion and are updated for another 300 iterations. The mean and standard deviation are calculated after transforming those particles back to the original parameter space.

\begin{figure}
	\includegraphics[width=1.0\linewidth]{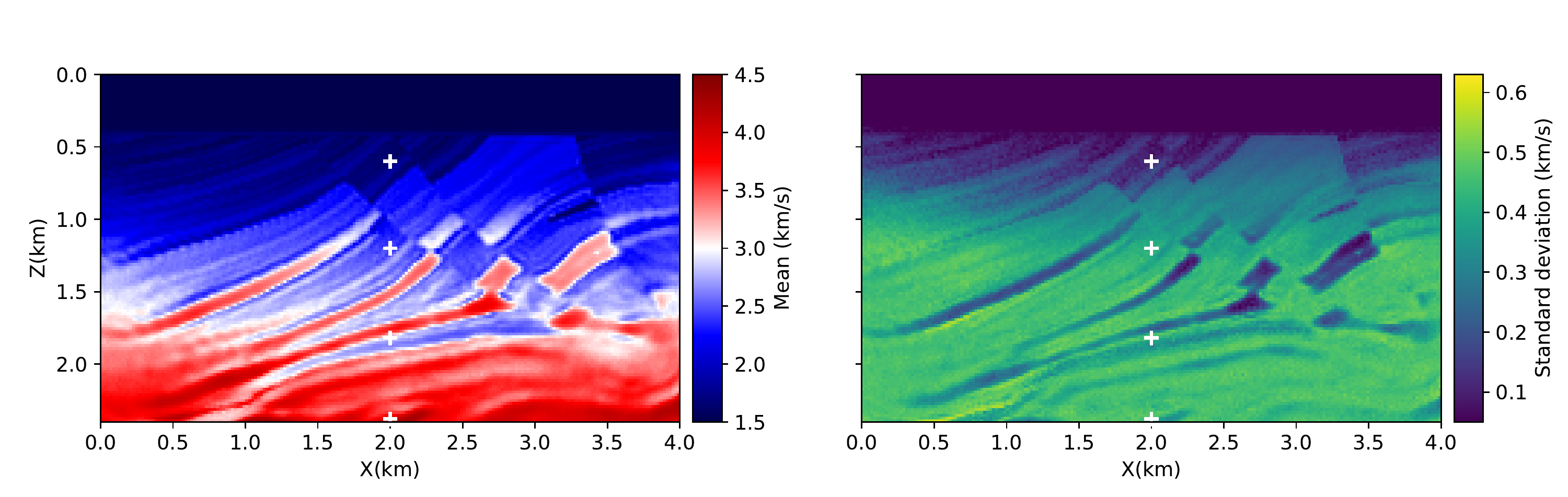}
	\caption{The mean structure and its point-wise standard deviation obtained using SVGD given high frequency data and using particles from low frequency inversion as the starting distribution.}
	\label{fig:rfwi_boost_results}
\end{figure}

Figure \ref{fig:rfwi_boost_results} shows the mean and standard deviation structures obtained using the above strategy. A comparison of the results to those obtained using only low frequency data and only high frequency data is discussed in \cite{zhang2021bayesian}. Overall the mean exhibits similar features to the target structure, except that the deeper part ($>$ 2 km) is slightly different from the target structure because of the poor illumination. The standard deviation has qualitatively similar features to the mean as we observed in the previous section. For example, in the near surface ($<$ 1 km) the low velocity anomalies are associated with lower uncertainty, and in the deeper part ($>$ 1 km) there are lower uncertainties at the location of high velocity anomalies. This phenomenon probably reveals the fact that waves spend comparatively longer in low velocity area which results in higher sensitivity. Note that due to the stronger prior information and better data coverage, the shallower part ($<$1 km) has lower uncertainty compared to the deeper part. 


%
%

\begin{figure}
	\includegraphics[width=1.0\linewidth]{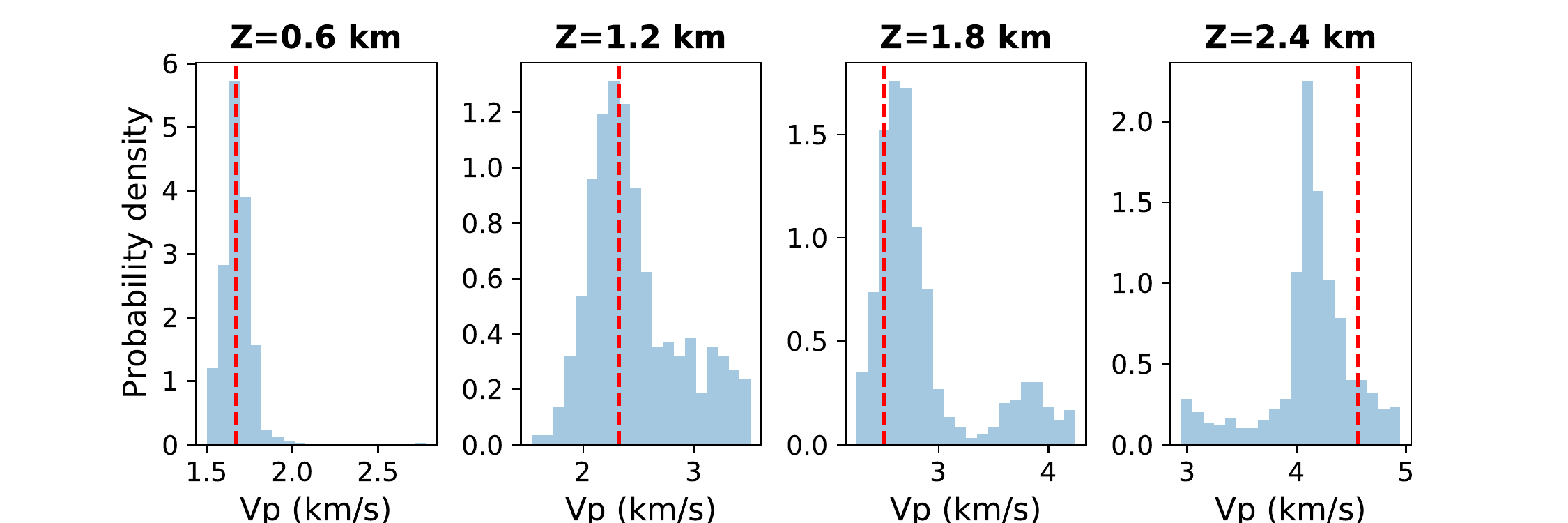}
	\caption{The marginal distributions at the horizontal location of 2 km and the depth of 0.6 km, 1.2 km, 1.8 km and 2.4 km obtained using SVGD given high frequency data and using particles from low frequency inversion as the starting distribution. Red lines denotes the true values.}
	\label{fig:rfwi_marginals}
\end{figure}

To further analyse the results, in Figure \ref{fig:rfwi_marginals} we show marginal distributions at four locations (white pluses in Figure \ref{fig:rfwi_boost_results}): (2.0, 0.6) km, (2.0, 1.2) km, (2.0, 1.8) km and (2.0, 2.4) km. Overall the true velocity values are around the high probability area, except that at the depth of 2.4 km the true value slightly deviates from the value with highest probability because of the poor illumination. At the deeper locations (1.8 and 2.4 km) the marginal distributions show complex, multimodal distributions which reflects the complexity of this inverse problem.
 
The above inversion took about 10,055 CPU hours for the total 900 iterations and required approximately 111.7 hours to run on 90 Intel Xeon CPU cores. In practice for larger datasets the method can be implemented using stochastic minibatch optimization. In addition, since the method does not require strong prior information, it could also be used to provide a good starting model for standard linearised FWI by using a small part of a large dataset. In addition, one may be able to perform the method on data types that require lower computational cost first, e.g. travel time tomography, and use those results as the starting distribution for VFWI to improve efficiency.

\section{Discussion}

In this study prior probabilities are simply set as Uniform distributions. While Uniform prior probabilities are simple to impose and are useful to explore properties of different methods, they may cause complex posterior pdfs which are hard to explore. In practice in cases where we have more knowledge about the subsurface, a more informative prior distribution should ideally  be used. For example some prior regularization can be used to produce smoother models \citep{mackay2003information}, or a Gaussian process may be used to inject prior information with adaptable complexity into inference scheme \citep{ray2019bayesian}. Neural networks can also be used to encode geological information into prior distributions \citep{laloy2017inversion, mosser2020stochastic}. For likelihood functions we simply used Gaussian distributions with a known, fixed data noise level. In practice this noise level might be estimated from data using the maximum likelihood method \citep{sambridge2013parallel, ray2016frequency} or a variety of other methods \citep{bensen20093, yao2009analysis, weaver2011precision, nicolson2012seismic, nicolson2014rayleigh}. It may also be possible to estimate the noise level in variational methods using a hierarchical Bayesian formulation  \citep{ranganath2016hierarchical}. To further improve the results a non-Gaussian likelihood function might also be used at little or no additional cost to the method.

In above examples we used a fixed regular grid of cells to parameterize the subsurface which can cause overfitting or underfitting of the data. For instance, in the travel time tomography example we observed a lower velocity loop with high uncertainty between the middle velocity anomaly and the receiver array (Figure \ref{fig:tt_mean_std}), which may be caused by overfitting as there is no such structure in the true model. To resolve this issue, an optimal grid might be sought. This can be achieved by applying a series of different grids and selecting the best one based on Bayesian or other model selection theories \citep{walter1997identification, curtis1997reconditioning, arnold2018interrogation}. For example, the ELBO calculated implicitly in variational methods can be used as a model selection criterion \citep{sato2001online, bernardo2003variational, mcgrory2007variational}. However, we note that the statistical theory behind such a design criterion is currently under explored, especially compared to McMC methods: in McMC a variety of well-established methods are available to perform model selection, for example reversible-jump McMC \citep{green1995reversible}, sequential Monte Carlo \citep{smith2013sequential} and nested sampling \citep{skilling2004nested, feroz2008multimodal}. Further research is required to develop appropriate model selection in variational inference. Apart from regularly gridded cells, we note that other more advanced parameterizations can be used in variational methods to provide more flexibility, such as Voronoi cells \citep{bodin2009seismic, zhang20183}, wavelet parameterization \citep{fang2014wavelet, zhang2015wavelet, hawkins2015geophysical}, Johnson-Mehl tessellation \citep{belhadj2018new} and Delaunay
and Clough-Tocher parameterizations \citep{curtis1997reconditioning, hawkins2019trans}.

While we focused on variational inference using KL-divergence to measure difference between two distributions, it is also possible to use other measures of divergence. For example, \cite{minka2013expectation} proposed the expectation propagation method by using KL-divergence in the other direction, that is $\mathrm{KL}[p||q]$ rather than $\mathrm{KL}[q||p]$. Other more general divergences, such as $\alpha$-divergence \citep{amari1985alpha} and $f$-divergence \citep{ali1966general} have also been used employed within variational inference \citep{li2016r, hernandez2016black, bamler2017perturbative, wang2018variational}. Stein's discrepancy provides another measure of difference between two distributions \citep{stein1972bound, gorham2015measuring, liu2016kernelized} and can also be used to develop variational methods \citep{ranganath2016operator, liu2017steinpolicy}.

Since the ELBO is a nonconvex objective function, variational inference can converge to a local optimum. For instance, in our travel time tomography example the results obtained using normalizing flows show irregularities and non-somoothness, which likely reflects convergence to a local optimum. To reduce this issue, more advanced optimization methods can be used, for example variational tempering \citep{mandt2016variational}, the trust-region method \citep{theis2015trust} or population variational inference \citep{kucukelbir2014population}. Different variational methods may also be combined together to increase robustness. For example, the probability distribution obtained using ADVI can be used as a starting distribution for normalizing flows and SVGD.

Monte Carlo sampling methods and variational inference are different methods that can be used to solve similar problems. Monte Carlo methods are usually applied using Markov chains, which generate a chain of samples that approximately follow the posterior pdf; variational inference seeks an optimal approximation to the posterior pdf within a predefined family of probability distributions. Monte Carlo methods are well-understood and are guaranteed to converge to the true posterior pdf asymptotically as the number of samples tends to infinity \citep{robert2013monte}, whereas the theoretical aspects of accuracy and convergence of variational inference are still unknown. The two methods can be used together to combine the merits of both. For example, a variational approximation can be used to build proposal distributions for  Metropolis-Hastings algorithms to improve their efficiency \citep{de2001variational}, or McMC steps can be incorporated into variational inference to improve accuracy \citep{salimans2015markov}. Further research on the interface between the two methods is certainly an interesting topic.

We have applied variational inference methods to petrophysical inversion, 2D travel time tomography and 2D FWI, and demonstrated their efficiency in solving these problems. However, it remains a challenge to apply variational methods to very high dimensional inverse problems, e.g. 3D FWI. In such cases the forward modelling itself is usually computationally extremely expensive. For methods like normalizing flows we may end up with very large neural networks, which can occupy huge memory and become very difficult to train. For SVGD we are likely to need many more particles than used herein, which may demand more resources than one can afford. In addition kernel metrics used in SVGD may become inefficient in high dimensional space due to the curse of dimensionality \citep{wainwright2019high}. Therefore further work is required to explore the properties of variational methods in a range of high dimensional, practical applications. 
   
\section{Conclusion}
In this chapter we reviewed the basic concepts of variational inference, and discussed four specific methods: mean-field approximation, automatic differential variational inference (ADVI), normalizing flows and Stein variational gradient descent (SVGD). Mean-field approximations can provide very efficient methods, but they assumes mutually independent parameters. ADVI uses a Gaussian distribution to approximate the posterior distribution, again leading to a reasonably efficient method but results that may be biased. Both normalizing flows and SVGD use a series of invertible transforms to transform an initial distribution to an approximation to the posterior distribution. Normalizing flows use a series of analytical invertible transforms, whereas SVGD uses an implicit transform to rearrange a set of particles from an initial distribution to represent the posterior distribution. We reviewed previous applications of the methods to a range of different examples: petrophysical inversion, travel time tomography and full-waveform inversion (FWI). In travel time tomography example we compared the results from ADVI, normalizing flows and SVGD with those obtained using Monte Carlo methods. The results show that ADVI is the cheapest method but provides biased results due to the implicit Gaussian assumption. In comparison, normalizing flows and SVGD can provide more accurate approximations to the results from the Monte Carlo method. Normalizing flows further improved efficiency of the inversion compared with SVGD. To further demonstrate variational methods, we applied SVGD to full-waveform inversion (FWI) problems and demonstrated that SVGD can produce accurate results to FWI problems, similar to those from Monte Carlo where the comparison has been made. We conclude that variational inference is an efficient and valuable tool to solve Geophysical inverse problems. We also note that variational inference is still in a phase of rapid development, for example, to solve the variational optimization problem more efficiently and to make the method more feasible to large scale inverse problems, so the method may become more accurate and more efficient in the near future.

  \begin{frontmatter}
\setglossarysection{chapter}
\glsaddall
\printglossary[nonumberlist]
\end{frontmatter}


\Backmatter
  \begin{frontmatter}
\chapter*{References}
\end{frontmatter}

\bibliographystyle{apacite}
\bibliography{bibliography}
%

\end{document}